# The International X-ray Observatory
## Activity submission in response to the Astro2010 RFI#2

Jay Bookbinder
Smithsonian Astrophysical Observatory
1-617-495-7058
jbookbinder@cfa.harvard.edu

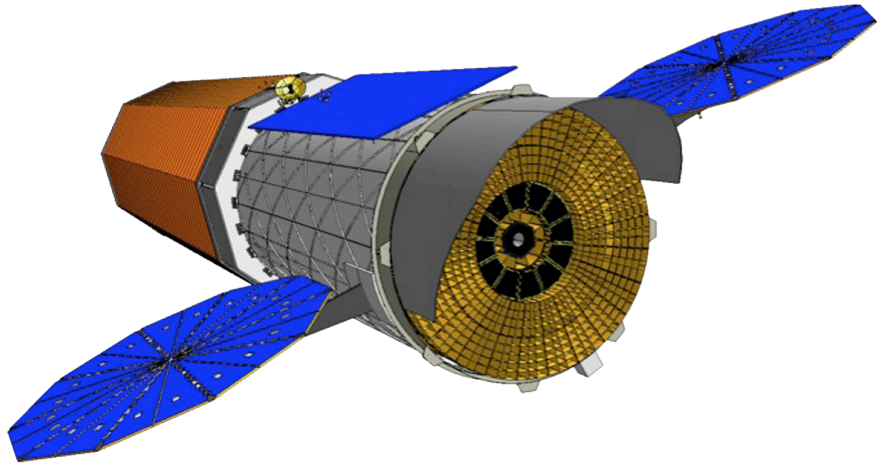

**Submitted on behalf of the IXO Study Coordination Group, whose members are**

Didier Barret (CESR, Toulouse)
Mark Bautz (MIT, Cambridge)
Jay Bookbinder (SAO, Cambridge)
Joel Bregman (University of Michigan, Ann Arbor)
Tadayasu Dotani (ISAS/JAXA, Sagamihara) – JAXA Project Manager
Kathryn Flanagan (STScI, Baltimore)
Philippe Gondoin (ESA, Noordwijk) – ESA Study Manager
Jean Grady (GSFC, Greenbelt) – NASA Project Manager
Hideyo Kunieda (Nagoya University, Nagoya) – SCG Co-Chair
Kazuhisa Mitsuda (ISAS/JAXA, Sagamihara)
Kirpal Nandra (Imperial College, London)
Takaya Ohashi (Tokyo Metropolitan University, Tokyo)
Arvind Parmar (ESA, Noordwijk) – ESA Study Scientist, SCG Co-Chair
Luigi Piro (INAF, Rome)
Lothar Strüder (MPE, Garching)
Tadayuki Takahashi (ISAS/JAXA, Sagamihara)
Takeshi Go Tsuru (Kyoto University) – JAXA Study Scientist
Nicholas White (GSFC, Greenbelt) – NASA Project Scientist, SCG Co-Chair

**and on behalf of the 69 members of the IXO Science Definition Team, Instrument Working Group, and Telescope Working Group, whose membership is listed at**
**http://ixo.gsfc.nasa.gov/people/**

# IXO QUICK REFERENCE GUIDE

- **Launch Date:** 2021

- **Orbit:** L2

- **Launch Vehicle:** EELV or Ariane V

- **Payload:** Five Instruments and Flight Mirror Assembly (FMA)

- **Observatory Wet Mass:** 4374 kg

- **Power Load:** 3.7 kW Max

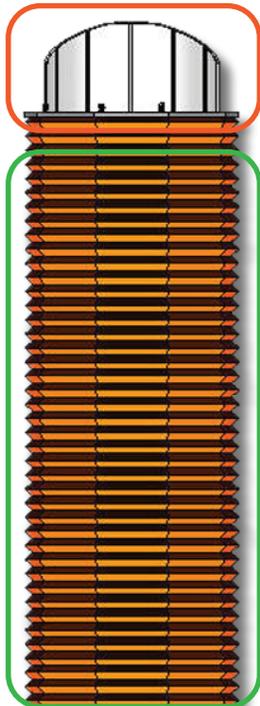

### Instrument Module (IM)
Mass: 736 kg

Length Scale 3 m

### Deployment Module (DM)
Mass: 439 kg

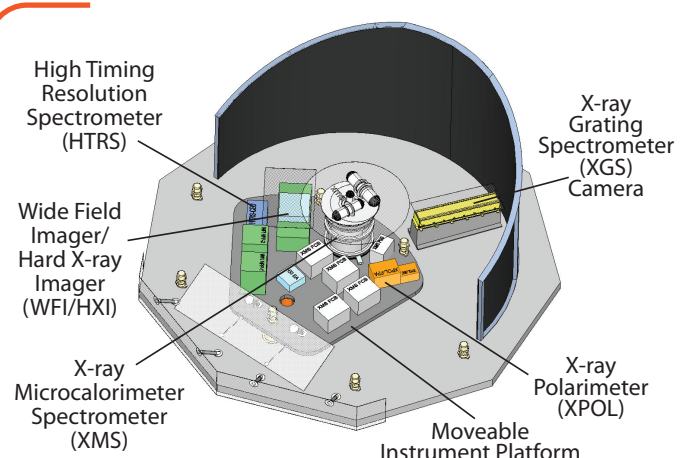

High Timing Resolution Spectrometer (HTRS)

Wide Field Imager/ Hard X-ray Imager (WFI/HXI)

X-ray Grating Spectrometer (XGS) Camera

X-ray Microcalorimeter Spectrometer (XMS)

X-ray Polarimeter (XPOL)

Moveable Instrument Platform

| Instrument | Bandpass [keV] | FOV [arcmin] | Energy Resolution [eV@keV] |
|---|---|---|---|
| XMS core | 0.3–12 | 2 x 2 | 2.5@6 |
| XMS outer | | 5.4 x 5.4 | 10@6 |
| WFI/HXI | 0.1–15/10–40 | 18 diam/8 x 8 | 150@6/1000@30 |
| XGS | 0.3–1 | N/A | E/ΔE = 3000 |
| HTRS | 0.3–10 | N/A | 150@6 |
| XPOL | 2–10 | 2.5 x 2.5 | 1200@6 |

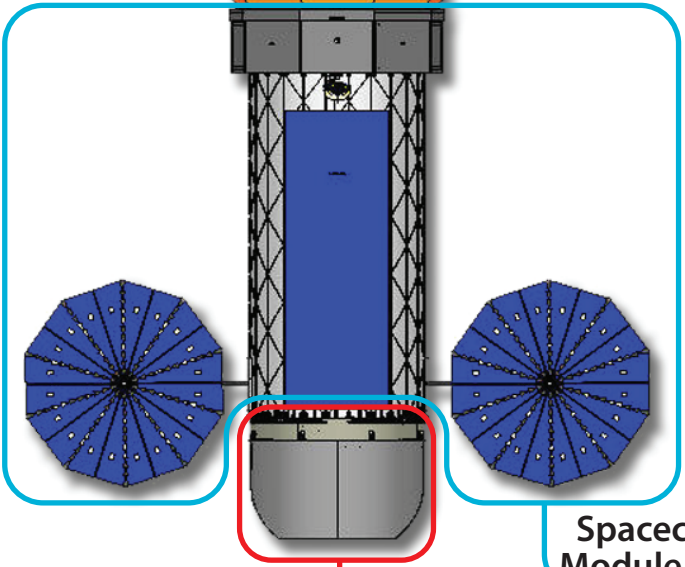

Three Deployable Masts (12m)

Shroud: 2 Concentric Pleated MLI Blankets (Whipple Shield)

Shroud Stowage Ring and Bus Interface Panel

Masts similar to those on the International Space Station's Shuttle Radar Topography Mission (SRTM) and NuSTAR

### Spacecraft Module (SM)
Mass: 1084 kg

### Optics Module (OM)
Mass: 1952 kg

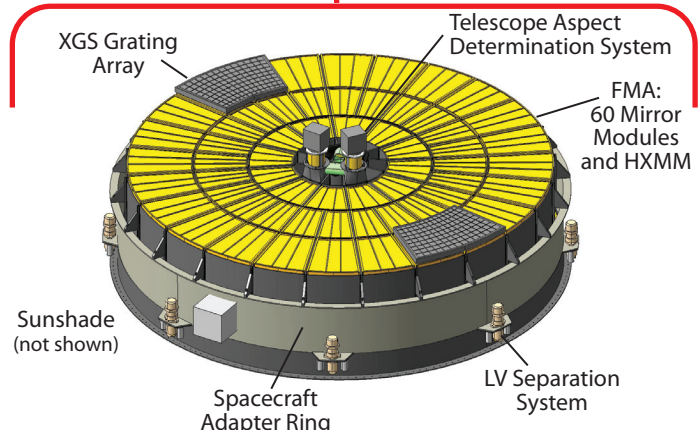

XGS Grating Array

Telescope Aspect Determination System

FMA: 60 Mirror Modules and HXMM

Sunshade (not shown)

Spacecraft Adapter Ring

LV Separation System

- **Eff Area:** 3 m² @ 1.25 keV, 0.65 m² @ 6 keV, 150 cm² @ 30 keV
- 4 arcsec Half-Power Diameter Angular Resolution (FMA only)
- 5 arcsec Half-Power Diameter Angular Resolution (Full System)

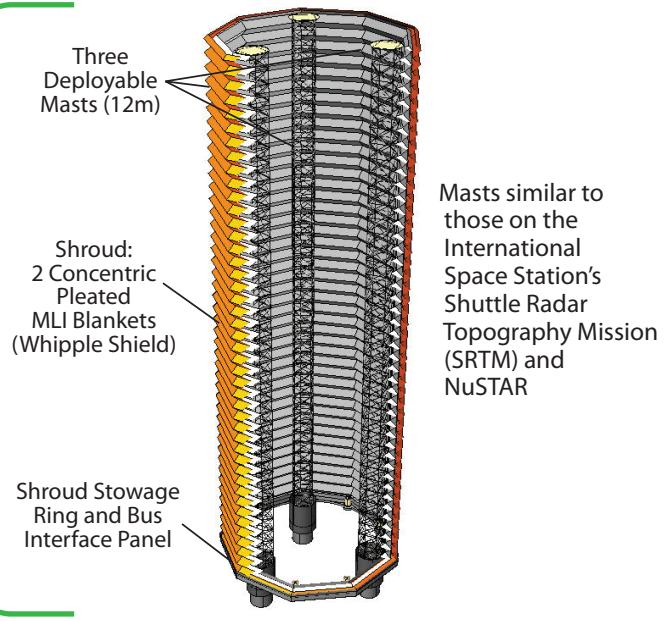

9-sided Bus Frame with Honeycomb Equipment Panels

Avionics

Bi-Prop and Monoprop Propulsion System

CFRP Isogrid Tube Fixed Metering Structure

2.4 m Diameter Hole for X-ray Beam

High Gain Antenna

Fixed Solar Array, 3.2 kW

2 Deployable Solar Arrays, 3.4 kW Total

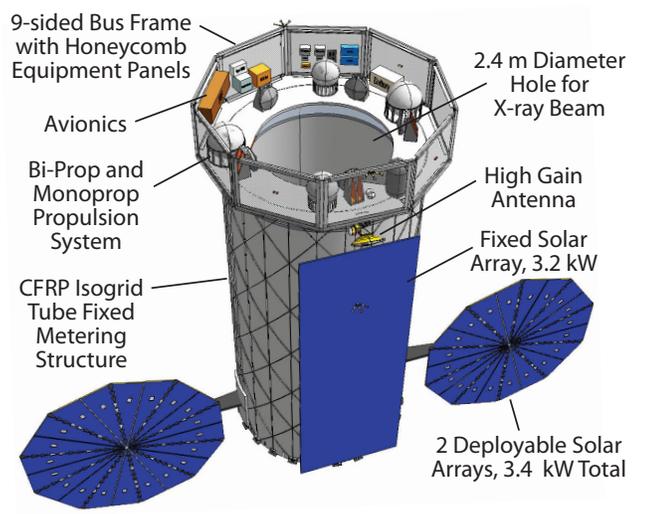

*All mass and power values are CBE*

# TABLE OF CONTENTS

## LIST OF TABLES

# QUESTIONS MAP

## Executive Summary

The International X-ray Observatory (IXO) is a joint NASA-ESA-JAXA effort. X-ray observations will resolve pressing astrophysical questions such as: *What happens close to a black hole? How do supermassive black holes grow? How does large scale structure form? What is the connection between these processes?* To address these questions requires dramatic increases in collection area (see figure) combined with sensitive new instrumentation.

IXO's spectroscopic, timing, and polarimetric capabilities will probe close to the event horizon of super-massive black holes (SMBH) where strong gravity dominates. IXO will determine the evolution and origin of SMBH by measuring their spin to understand their merger history, surveying them to find their luminosity distribution out to high redshift (z~8), and spectroscopically characterizing their outflows during peak activity. IXO will revolutionize our understanding of galaxy clusters by mapping their bulk motions and turbulence. IXO will observe the process of cosmic feedback where black holes inject energy on galactic and intergalactic scales, and characterize the missing baryons in the cosmic web. Meanwhile, surveys of distant clusters will constrain cosmological models.

IXO will be available to all astronomers, taking X-ray astrophysics from an era where high-resolution spectra are a rarity to one with vast numbers of spectra from all types of sources. Powerful spectral diagnostics and large collecting areas will reveal unexpected discoveries, with IXO studying new phenomena as they appear—a key feature of great observatories (**Sembach et al. 2009**).

A single mirror assembly with a 3 m diameter and a 20 m focal length provides IXO's required 3 m² collecting area, and a deployable optical bench is employed to fit the optics and the science instruments within the launcher shroud. To reduce risk in achieving the required 5 arcsec angular resolution, two independent optics technologies are under development. In the US, segmented glass technology has demonstrated ~15 arcsec performance with a path to achieve 5 arcsec and TRL 6 by early 2012. In Europe, silicon pore optics uses infrastructure from the microprocessor industry and has achieved ~15 arcsec with a path to reach 5 arcsec and TRL 6 by early 2012. The

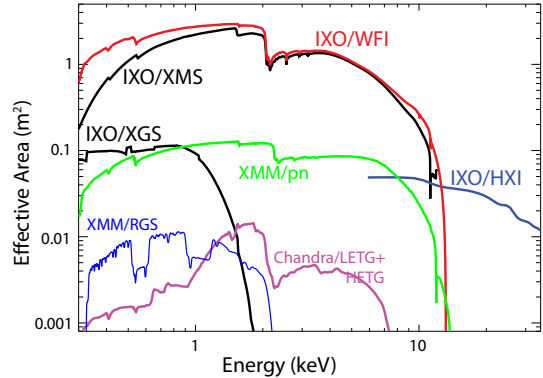

*The IXO effective area, which will be more than an order of magnitude greater than current imaging X-ray missions. Coupled with the large spectral resolving power, IXO will open a vast discovery space.*

IXO optics technology selection will be made in 2012 based on performance, cost, and schedule considerations.

The X-ray Microcalorimeter Spectrometer (XMS) imager with ΔE = 2.5 eV provides unprecedented spectro-imaging capability. The Wide Field Imager/Hard X-ray Imager (WFI/HXI) provides deep imaging over an 18 arcmin field of view in the 0.1–15 keV band with the WFI while the HXI extends the imaging bandpass to 40 keV. An X-ray Grating Spectrometer (XGS) provides resolving power of 3000 below 1 keV. A High Time Resolution Spectrometer (HTRS) provides microsecond spectroscopic timing at high count rates. An imaging X-ray polarimeter (XPOL) will enable sensitive searches in this new parameter space. These instruments are currently at TRL 3–6, with plans to achieve TRL 6 by 2013.

The IXO architecture provides well-defined interfaces to simplify development and I&T, as well as facilitating shared development between international partners. A final division of responsibilities will be made during Phase A when the optics technology has been selected. Costs are shown assuming NASA leads the mission and provides the optics module. IXO will be placed in L2 orbit using either a NASA EELV or an ESA Ariane V. The five-year mission lifetime has consumables for at least 10 years. The total Phase A-E mission cost in FY09 dollars is $3.3B, with $1.8B from NASA. This includes $419M for Phase E, with $24M per year in grants to the US community. If Phase A starts in 2011, the mission schedule allows launch as soon as 2021.

# 1. Science Overview

The driving science goals of IXO are to understand evolution of black holes and the properties of their extreme environments, measure the energetics and dynamics of hot gas in large cosmic structures, and reveal the connections between these phenomena. IXO will also constrain the equation of state of neutron stars and track the dynamical and compositional evolution of interstellar and intergalactic matter. IXO measurements of virtually every class of astronomical object will also return serendipitous discoveries, as with all major advances in astronomical capabilities.

This section includes only selected science that drives key requirements. A more complete list of IXO science goals, including topics such as measuring Galactic black hole spins and studying the formation and distribution of the elements, is available in RFI#1 and the **IXO Decadal White Papers** (WP).

## 1.1 Science Objectives and Required Measurements

*Q1. Describe the measurements required to fulfill the scientific objectives expected to be achieved by your activity.*

*Q2. Describe the technical implementation you have selected, and how it performs the required measurements.*

*Q4. Present the performance requirements (e.g., spatial and spectral resolution, sensitivity, timing accuracy) and their relation to the science measurements.*

*Q6. For each performance requirement, present as quantitatively as possible the sensitivity of your science goals to achieving the requirement. For example, if you fail to meet a key requirement, what will the impact be on achievement of your science objectives?*

The answers to RFI#2 Questions 1, 4, 6, and the second part of 2 are broken out by science topic. The first part of Question 2, including descriptions of all instruments and their acronyms, is addressed in Section 2.1 and Questions 3 and 5 are addressed at the end of this section.

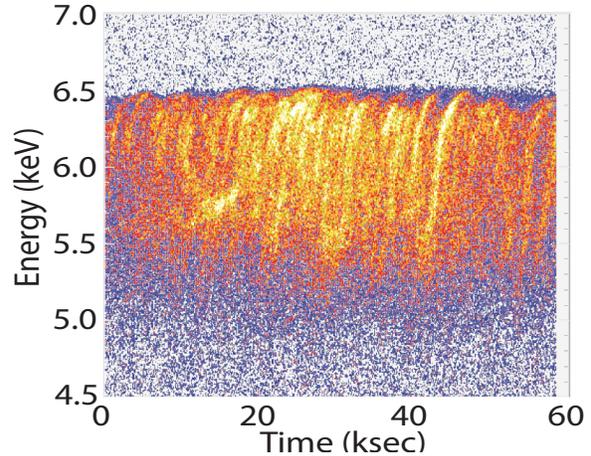

**Figure 1-1.** *IXO will resolve "hot spots" in energy and time as they orbit the SMBH. In the time-energy plane, the emission from these hot spots appears as "arcs," each corresponding to an orbit of a given bright region.*

A summary of the science objectives and the flowdown to the mission requirements is given in Table 1-1 at the end of this section.

### 1.1.1 Studies of Strong Gravity
**(Key Requirement: Effective Area at 6 keV)**

Observations of accretion flows around supermassive black holes (SMBH) can probe General Relativity's (GR) spacetime metric due to the geometric and dynamic simplicity of accretion disks. Each parcel of gas has an orbit around the black hole that closely approximates a circular test-particle orbit, with typical deviations less than 1% in such thin accretion disks (Armitage & Reynolds 2003). IXO will add a new dimension—time—to the study of iron lines, with "hot spots" of iron Kα emission in the disk appearing as "arcs" in a time-energy plane (See Fig 1-1). GR predicts the form of these arcs, and the ensemble of arcs can be fitted for the mass and spin of the black hole and the inclination of the accretion disk.

**Technical Implementation and Performance Requirements for Measurement (Q2,Q4):** The centroid of narrow but varying iron lines must be measured in at least 10 phase bins throughout the orbit (order of hours) for a range of SMBH masses. A mirror area of 0.65 m² at 6 keV will ensure at least 100 photons in the Fe line per orbital bin, enough for an accurate energy centroid, for about 10 SMBH targets. The XMS is the preferred instrument as it allows accurate centroiding of emission lines with ~100 photons and can detect other narrow features.

**Sensitivity of Science to Requirement (Q6):** These measurements are signal dominated, as the sources are bright. The number of sources sufficiently bright to resolve the emission-line arcs will scale with a change $\Delta$ to the 6 keV effective area roughly as $\Delta^{3/2}$.

## 1.1.2 Measuring Black Hole Spin
### (Key Req: Effective Area and Resolution at 30 keV, Polarization Sensitivity)

Despite their immense potential for energy generation and consequent impact on cosmic evolution, black holes have only two measurable parameters: mass and spin. The spin of a SMBH depends upon its growth history: an accretion-dominated history leads to high spin and a merger-dominated one to low spin (Berti & Volonteri 2008). By determining the spins of a few hundred SMBH, using multiple approaches, IXO will determine how SMBH grow.

**Technical Implementation and Performance Requirements for Measurement (Q2,Q4):** The primary method is measurement of broad orbitally-averaged iron lines, which unlike the strong gravity study requires only moderate energy resolution (150 eV) but broad energy coverage to determine the continuum both below and above the 6.4 keV iron line. The WFI/HXI will enable measurements covering 0.1–40 keV with adequate effective area and resolution at 30 keV from the HXI to measure the hard continuum.

Another method determines spin by measuring the polarization properties of X-rays reflected from the disk, which depend upon the inner disk radius, a spin-dependent property (Miniutti & Fabian 2004; Dovciak, Karas & Matt 2004). The expected polarization degree ranges from ~1-30%. Measuring this effect for ~10 SMBH leads to the requirement of 1% minimum detectable polarization (MDP) for a 1 mCrab source in 100 ksec. The GEMS X-ray polarimetry SMEX would require 2.5 Msec to reach the same MDP for each source.

**Sensitivity of Science to Requirement (Q6):** The 150 cm$^2$ mirror area at 30 keV provides the lever arm to determine the continuum. Areal reductions will require longer measurements, reducing the number of sources that can be sur-veyed. The expected SMBH polarization is a few percent, above the 1% MDP requirement. If this requirement is not met, the observing time would increase quadratically with the MDP to reach the same sensitivity, since the observations will be signal-dominated.

## 1.1.3 Neutron Star Equation of State
### (Key Req: High Count Rate Timing and Spectral Resolution)

Neutron stars have the highest known matter densities in nature, utterly beyond the densities produced in terrestrial laboratories. At these densities, the uncertainties in the underlying physics lead to widely differing equations of state, each of which imply different neutron star radii as a function of mass. IXO will determine the equation of state for neutron stars via their mass-radius relationship for approximately a dozen neutron stars of various masses with several distinct methods (**Paerels et al.** WP). The most robust method will measure energy dependent pulsations present during thermonuclear X-ray bursts from fast spinning neutron stars in low mass X-ray binaries (LMXB)(Strohmayer 2004). The same modeling technique will allow mass–radius measurements for several rotation-powered millisecond X-ray pulsars (Bogdanov et al. 2008).

**Technical Implementation and Performance Requirements for Measurement (Q2,Q4):** Measuring the mass and radius of LMXB bursters requires fast relative timing (10 μs), high throughput (>10$^5$ cts/s), low dead-time and modest spectral resolution (150 eV). These capabilities are provided by the combination of IXO's large collecting area and the HTRS detector. Modeling of simulated pulse profiles indicates that an ~8% measurement of mass and radius is statistically achievable using bright bursts which have pulsations present during burst rise. Another method uses the spectral resolution of the HTRS to detect rotationally-broadened absorption lines from these and other LMXB sources. This approach will provide an independent measure of the mass and radius, if the lines are sufficiently strong.

**Sensitivity of Science to Requirement (Q6):** The statistical precision scales approximately as the square root of the counts present in the pulse profile. Combining pulsation data from several bursts would improve the precision according to the same relation. The burst oscillations have 200-

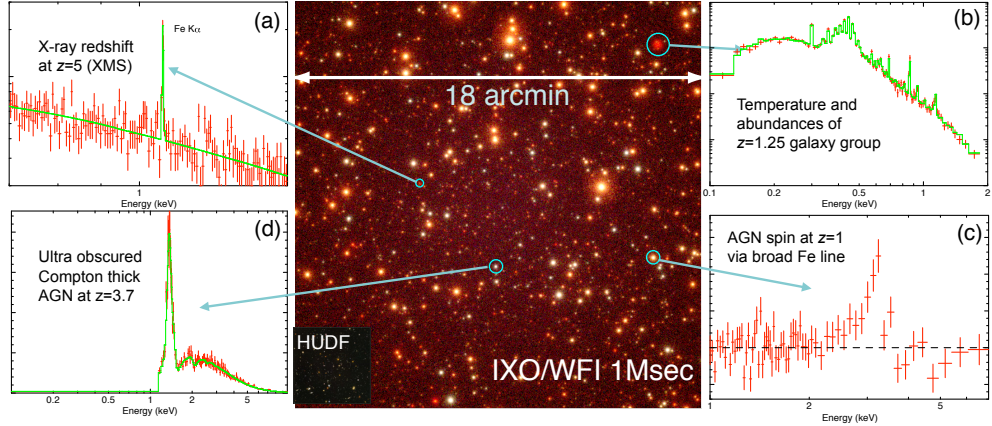

*Figure 1-2. WFI Simulation of the Chandra Deep Field South with Hubble Ultra Deep Field (HUDF) in inset. Simulated spectra of various sources are shown, illustrating IXO's ability (clockwise from top left) to: a) de-* termine redshifts in the X-ray band, b) determine temperatures and abundances even for low luminosity groups to z>1, c) make spin measurements of AGN to a similar redshift, and d) uncover the most heavily obscured, Compton-thick AGN.

600 Hz periods, requiring sub-ms relative timing. Degradation in the HTRS resolving power R will reduce sensitivity to the absorption lines by $R^{-1/2}$.

### 1.1.4 Growth of Supermassive Black Holes

**(Key Req: Effective Area at 1.25 keV, FOV, PSF, Astrometry)**

SMBHs are a critical component in the formation and evolution of galaxies, and IXO can detect accretion power from embedded high-red-shift SMBHs ($10^7$–$10^9 M_\odot$), even when obscured (**Nandra et al.** WP; see Fig 1-2). IXO will determine the luminosity function of SMBHs out to z ~ 8, exploring the early growth phase of SMBHs.

**Technical Implementation and Performance Requirements for Measurement (Q2,Q4):** This science can be achieved using a combination of large mirror effective area (3 m² at 1.25 keV), good angular resolution (5 arcsec) and large field of view (FOV; 18 arcmin diameter) with moderate spectral resolution ($\Delta E$ ~ 150 eV @ 6 keV). These capabilities are provided by the WFI/HXI and will allow IXO to carry out a multi-tiered survey in a manageable amount of observing time (~10 Msec). IXO can efficiently survey significant areas of the sky an order of magnitude faster than Chandra and to a limiting depth that surpasses the 2 Msec Chandra deep field. The observational approach is a survey of increasing solid angle with decreasing exposure time (1000, 300, 100, 30, and 10 ksec, with solid angles increasing from 0.3 to 3.5 sq. degrees). This survey will need to be complemented by optical and IR surveys, so the point spread function (PSF) must be small enough that optical-IR counterparts can be identified. Combined with 5 arcsec resolution and a 50 photon source, the statistical error circle of the source centroid has a radius of 0.4 arcsec, which is smaller than the mean source separation even in the Hubble Ultra Deep Field (1.5 arcsec). Systematic errors on registration will be of a similar scale if at least three known sources are in the field to allow for post-observation astrometric correction (improving on the absolute astrometric accuracy from the attitude reconstruction of 1 arcsec). The IXO 0.5–2 keV confusion limit is ~6 × 10⁻¹⁸ erg/cm²/s (PSF/5 arcsec)²·⁹, well-matched to the sensitivity of the 1 Msec survey component.

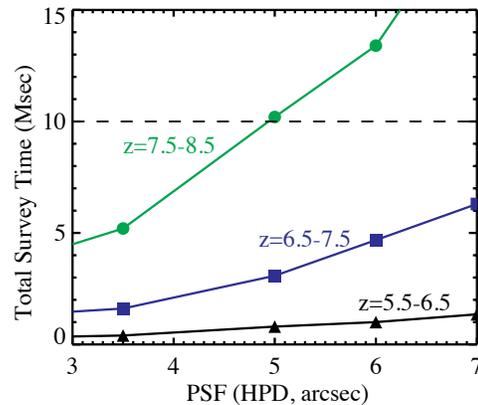

*Figure 1-3. Effect of PSF on total observing time for survey of $L_x > 10^{43}$ erg/s SMBH at z~6, 7, and 8. The dashed line shows the total time required for the multi-tiered survey.*

**Sensitivity of Science to Requirement (Q6):** A modest decrease in the collecting area or the field of view could be compensated for linearly by longer observing times. The size of the PSF affects the detection threshold (due to a background increase), the source confusion limit, and the ability to identify optical counterparts. Our simulations using current knowledge of SMBH space densities indicate that a 5 arcsec PSF is required to achieve our scientific goals in a 10 Msec survey. The dependence of total observing time on the PSF is shown in Fig 1-3.

### 1.1.5 Evolution of Galaxy Clusters & Feedback

**(Key Req: Effective Area at 1.25 keV, Energy Resolution, FOV, PSF)**

Galaxy formation depends on the physical and chemical properties of the intergalactic medium (IGM), which, in turn, is affected by energy and metal outflows from galaxies (feedback). Detailed studies of the IGM in galaxy clusters are now limited to the nearby Universe ($z < 0.5$). IXO will measure the dynamical and thermodynamic properties as well as the metal content of the first low-mass clusters emerging at $z \sim 2$ and directly trace their evolution into today's massive clusters (**Arnaud et al.** WP, **Fabian et al.** WP).

**Technical Implementation and Performance Requirements for Measurement (Q2,Q4):** A resolving power of 2400 at 6 keV ($\Delta E=2.5$ eV, $\Delta v=125$ km/s) of the XMS enables determination of the state of galaxy cluster evolution from velocity structures (including feedback from SMBHs), precise redshifts directly from X-ray observations, and allows turbulence to be measured in the intracluster medium (Fig 1-4). The resolving power also allows precise abundance measurements down to the photon-limited detection limit. The FOV and mirror effective area at 1.25 keV are needed to observe both feedback in cluster cores (~4 Msec) and the metallicity and dynamics of clusters across cosmic time (~10 Msec).

**Sensitivity of Science to Requirement (Q6):** The total observing time for the cluster program scales linearly with collecting area. With only the inner core of the XMS, the field of view would decrease by a factor of ~6. For large clusters that require mapping (at fixed S/N, nearby clusters), this FOV decrease would increase the observing time by a factor of 6. The study of feedback in low-$z$ cluster cores is largely unaffected, as these regions are typically <2 arcmin in size. Measurement of metallicity in distant ($z >1$) clusters is also largely unaffected by field of view. The spectral resolving power, which scales inversely with velocity resolution, impacts sensitivity to turbulence and mixing velocities. A PSF of 5 arcsec is important both for studies of feedback in nearby clusters (to resolve the important structures) and distant clusters (where the core is 6–7 arcsec and must be resolved).

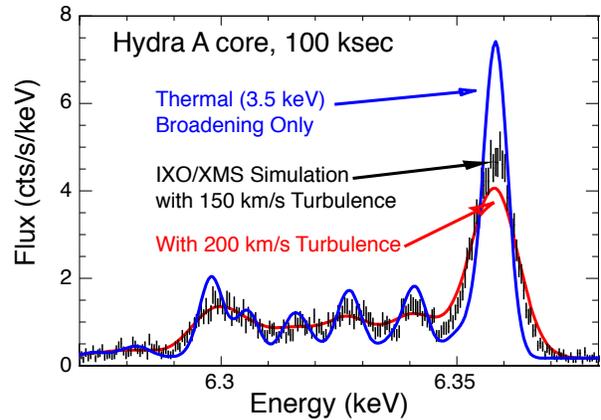

*Figure 1-4. IXO spectrum of Fe XXV lines shows that turbulence as low as 150 km/s may be distinguished from thermal broadening alone. Simulated IXO XMS data in black, models in color.*

### 1.1.6 Cosmology

**(Key Req: Energy Resolution, FOV)**

The growth of galaxy clusters, the largest virialized systems, is fundamental. In particular, the evolution of the mass function of galaxy clusters and a measurement of the distance-redshift relation [$d(z)$] places strong constraints on cosmology including the properties of Dark Energy. IXO observations of galaxy clusters will provide both tests, complementing other cosmological experiments (**Vikhlinin et al.** WP).

**Technical Implementation and Performance Requirements for Measurement (Q2,Q4):** Observations of 1000 clusters at $z$=1-2 together with existing low-$z$ data will constrain the growth of structure independent of other methods. Precise temperature measurements and surface brightness distributions are essential to determining the cluster masses, and in the outskirts of galaxy clusters these are done with the outer pixels of the XMS, with a resolving power of 150-300 in the redshifted Fe Kα line (10 eV @ 6 keV). A characteristic cluster diameter at 500 times the critical density ($2R_{500}$) is 3 arcmin at $z = 1$ (5 keV cluster). Sky background near the galaxy cluster must also be measured, increasing the FOV to be observed. With the XMS, distant clusters will require only 1–4 pointings.

**Sensitivity of Science to Requirement (Q6):** If only the inner core of the XMS were built, it would decrease the FOV to 2×2 sq. arcmin and for typical clusters at $z$=1-2 (with diameter ~2-4 arcmin) would increase by a factor of 2-3 the amount of observing time required. Reductions in the spectral resolving power would increase the systematic uncertainty in the temperature and metal abundances due to modeling uncertainties in low-count spectra.

### 1.1.7 Cosmic Web of Baryons

**(Key Requirement: Energy Resolution)**

In addition to determining whether half the baryons in the Universe lie in the $0.3–10 \times 10^6$ K range, IXO will discover whether this hot gas is enriched by galactic superwinds and if it has the anticipated web-like topology. Key observations are the equivalent width measurements of He-like and H-like lines of O, N, and C, as seen against the continuum of bright background AGNs; detection of O VII and O VIII absorption lines are key. The cosmic web should contain numerous O VII and O VIII absorption lines with equivalent widths of 2–8 mÅ (Bregman 2007). These lines may have velocity structures imposed by galactic superwinds and the absorption may be associated with individual galaxies.

**Technical Implementation and Performance Requirements for Measurement (Q2,Q4):** Addressing these two issues requires near-Doppler width resolving power, about R=3000 (v~100 km/s). This resolution is needed in the 0.3–1 keV range where the lines will occur, and it is achievable with the XGS. To determine the gas mass contribution, we need to define the differential equivalent width distribution for O VII, which will require about 200 absorption systems; ratios of other ions to O VII are valuable but needed for fewer cases. These goals can be met by measuring absorption lines toward 30 bright AGNs with a 0.1 m² effective area for the XGS.

**Sensitivity of Science to Requirement (Q6):** For a known absorption feature, the minimum detectable equivalent width (EqW) is proportional to the width of a resolution element divided by the square root of the counts in that resolution element, or (R × Area × Time)$^{-1/2}$. With the XGS, an O VII absorption line at $z = 0.15$ observed toward a source with soft X-ray flux $5 \times 10^{-12}$ erg/cm²/s can be detected to a minimum EqW of ~2 mÅ at 3σ in 200 ksec. The measurement is signal-dominated, so a modest decrease in the effective area or resolution could be compensated for by longer observations. However, reducing the spectral resolution will result in a loss of sensitivity to weak absorption lines and velocities, and a significant degradation will render the distinction between intervening WHIM lines from those arising from galaxy superwinds impossible.

### 1.1.8 The Most Demanding Measurements

*Q3. Of the required measurements, which are the most demanding? Why?*

Each program above defines at least one key requirement and usually requires substantial allocations of telescope time (10% or more of the entire emission over five years). However, one program, The Growth of Supermassive Black Holes, (see Section 1.1.4) sets four key requirements, including three on the FMA. For these reasons, this is the most demanding program.

## 1.2 Science Flowdown

*Q5. Present a brief flow down of science goals/requirements and explain why each payload instrument and the associated instrument performance are required.*

Table 1-1 shows the selected science objectives, including typical targets and their fluxes. The observational requirements for each are included, with driving requirements marked in ***bold green italics***. The majority of observations have both a primary and a secondary instrument; this indicates that at least part of the science could be achieved using the secondary instrument. In some cases, however, only one instrument is capable of making the necessary measurement, such as the detection of X-ray polarization by XPOL or timing of bursting neutron stars by HTRS.

**Table 1-1. Science Flowdown Matrix**

| Science Topic | Typical Target | # of Ptgs | Src Size (arcmin) | Typical Flux (erg/cm²/s) | Analysis | S/N required | Obs Time (Msec) | Abs Ast (arcsec) | FOV (arc-min) | Band-pass (keV) | PSF HPD (arcsec) | Mirror Effective Area Rqmt (sq m) 1.25 keV | 6 keV | 30 keV | Energy Res Rqmt FWHM(eV) | @ E (keV) | Rel Timing (μsec) | Instrum Primary | (Second) |
|---|---|---|---|---|---|---|---|---|---|---|---|---|---|---|---|---|---|---|---|
| Strong Gravity | MCG-6-30-15 | 20 | point | $5 \times 10^{-11}$ | spectra | 10 | 8 | N/A | N/A | 1-40 | N/A | 1.5 | 0.65 | 0.015 | 2.5 | 6 | N/A | XMS | (WFI/HXI) |
| SMBH Spin Survey | NGC 4051 | 200 | point | $10^{12}$ | spectra | 5-10/bin | 10 | N/A | N/A | 1-40 | N/A | 1 | 0.65 | 0.015 | 1000 | 30 | N/A | WFI/HXI | (XMS) |
| | MCG-6-30-15 | 10 | point | $5 \times 10^{-11}$ | polarization | 1% MDP | 10 | N/A | N/A | 2-10 | N/A | 2.5 | 0.5 | N/A | 1200 | 6 | N/A | XPOL | |
| Neutron Star EoS | 4U1636-536 | 15 | point | $10^{8}$ | spectra | 20/bin | 5.5 | N/A | N/A | 0.3-10 | N/A | 3 | 0.6 | N/A | 150 | 0.3-6 | 10 | HTRS | |
| Growth of SMBH | CDF-S | 38 | point | $3 \times 10^{-17}$ | imaging spectra | 5 at flux limit | 10 | 1 | 18 dia | 0.3-2 | 5 | 3 | 0.65 | 0.015 | 150 | 1 | N/A | WFI/HXI | (XMS) |
| Clusters / Feedback | z=0.1-2 cluster | 250 | 2-18 | $10^{13}$ | imaging spectra | 50 at flux limit | 14 | N/A | 2 × 2 | 0.3-40 | 5 | 3 | 0.65 | 0.015 | 2.5 | 6 | N/A | XMS | (WFI/HXI) |
| Cosmology | z=1-2 cluster | 1000 | 3 | $5 \times 10^{-14}$ | image, spectra | 2000 cts/obj | 15 | 10 | 5 × 5 | 0.3-7 | 10 | 1 | 0.1 | N/A | 10 | 6 | N/A | XMS | (WFI/HXI) |
| Cosmic Web of Baryons | QSO B1426 +428 | 30 | point | $10^{11}$ | spectra | 12/bin | 15 | N/A | N/A | 0.3-1 | 5 | N/A | N/A | N/A | 0.1 | 0.3 | N/A | XGS | (XMS) |

# 2. Technical Implementation

## 2.1 Payload Instrumentation

### 2.1.1 Instrument Descriptions

*Q1. Describe the proposed science instrumentation, and briefly state the rationale for its selection. Discuss the specifics of each instrument and how the instruments are used together.*

The International X-ray Observatory has a large diameter, grazing incidence mirror (the Flight Mirror Assembly, FMA); four instruments on a Moveable Instrument Platform (MIP) which are rotated into the mirror focus and operated one at a time for science data collection; and an X-ray Grating Spectrometer (XGS) that intercepts and disperses a fraction of the beam from the mirror onto a CCD (charge coupled device) camera, operating simultaneously with the observing MIP instrument (see Fig. 2-1). The four MIP instruments are the X-ray Microcalorimeter Spectrometer (XMS), the Wide Field Imager/Hard X-ray Imager (WFI/HXI), the High Time Resolution Spectrometer (HTRS), and the X-ray Polarimeter (XPOL). Table 2-1 summarizes the instrument capabilities and the science drivers for each.

The specifics of each instrument are described below, followed by the rationale for the selection of the instrument complement and the individual instruments.

#### 2.1.1.1. The Flight Mirror Assembly (FMA)

The FMA provides effective area of 3 m² at 1.25 keV, 0.65 m² at 6 keV, and 150 cm² at 30 keV. To meet the mission-level Point Spread Function (PSF) requirement of 5 arcsec half-power diameter (HPD), the FMA angular resolution must be 4 arcsec or better. Attaining the large effective area within the launch vehicle mass constraint requires a mirror with a high effective area-to-mass ratio of 20 cm²/kg, 50 times larger than Chandra and eight times larger than XMM-Newton.

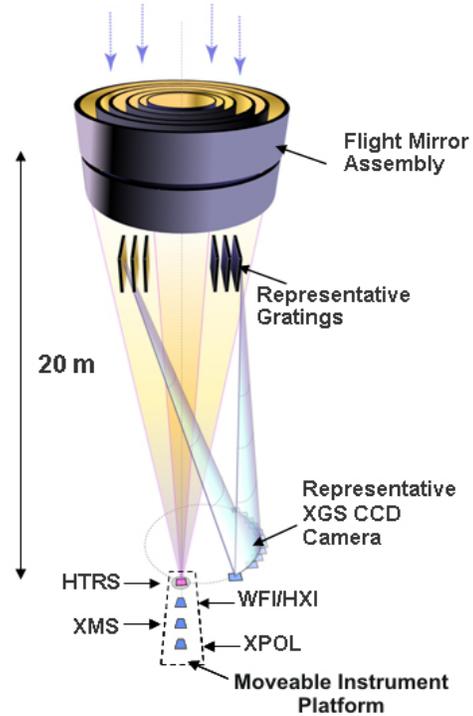

*Figure 2-1. IXO Payload Schematic.*

As the mirror is the major technical challenge for IXO, two technologies are being developed in parallel as a risk reduction strategy. These are thermally formed, segmented glass mirrors and silicon pore optics (SPO; see Fig. 2-2). Both approaches lead to a highly modular mirror design. The key technology hurdle for each is the con-struction of a module. The observatory can accommodate either mirror approach. Both technologies have demonstrated X-ray performance of ~15 arcsec HPD.

**Table 2-1. IXO Instrument Requirements**

| Instrument | | Bandpass | PSF (HPD) | FOV | Energy Resolution | Science Driver |
|---|---|---|---|---|---|---|
| | | keV | arcsec | arcmin | eV@keV | |
| **XMS** | Core | 0.3–12 | 5 | 2 × 2 | 2.5@6 | Galaxy Clusters |
| | Outer | | | 5 × 5 | 10@6 | |
| **WFI/** | WFI | 0.1–15 | 5 | 18 diameter | 150@6 | SMBH survey |
| **HXI** | HXI | 10–40 | 30 | 8 × 8 | 1000@30 | SMBH Spin |
| **XGS** | | 0.3–1.0 | 5 | N/A | E/ΔE = 3000 | Cosmic Web |
| **HTRS** | | 0.3–10 | N/A | N/A | 150@6 | NS EoS |
| **XPOL** | | 2.0–10.0 | 6 | 2.5 × 2.5 | 1200@6 | SMBH Spin |

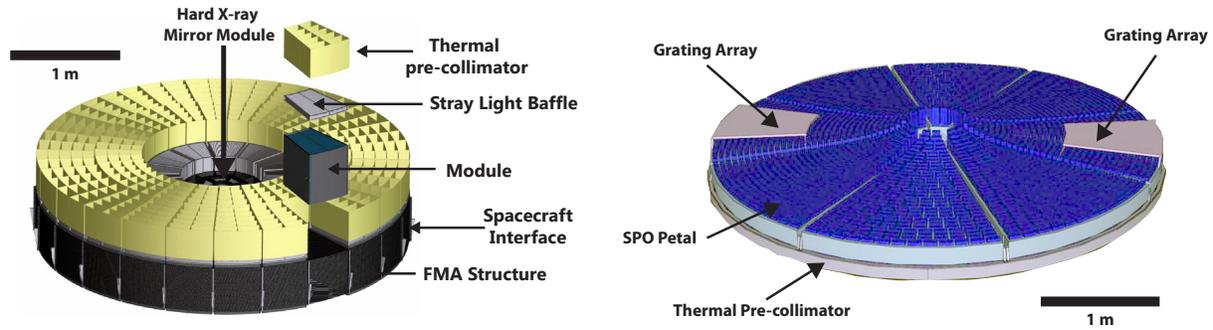

*Figure 2-2. The two FMA concepts: the segmented glass mirrors (left) and the SPO (right). Major elements of the segmented glass mirror are labeled. The segmented glass FMA is shown from the front (thermal pre-collimator on top) and the SPO is shown from the back (facing inside spacecraft).*

The iridium-coated segmented glass Wolter-1 FMA design incorporates 361 nested pairs of concentric shells separated into segments. The segments are grouped into 60 modules arranged in three concentric rings (Fig.2-2, left); the largest mirror has a diameter of 3.3 m. Each module comprises approximately 120 pairs of mirror segments, each 20 cm in axial length and 20–40 cm in azimuthal span, accurately aligned in a mounting structure. Segments are produced by thermally slumping 0.4 mm thick glass (the same as that manufactured for flat panel displays) onto figured fused quartz mandrels (Zhang et al. 2008). IXO requires ~14,000 segments. Segment mass production is being demonstrated by NuSTAR for which ~8,000 segments are required and the current weekly production rate is ~450 with nearly 100% yield. The response above 10 keV is provided by a 30 arcsec HPD hard X-ray mirror module mounted in the center of the FMA with segments coated with multilayers to enhance the 10–40 keV reflectivity and an aluminum pre-filter to restrict the bandpass to energies above 10 keV.

The SPO approach (Fig. 2-2, right) uses commercial, high-quality 1 mm thick silicon wafers as its base material. One side of a 6-cm-wide rectangular segment of a wafer is structured via etching or micromachining with accurately wedged ribs approximately 1 mm apart. The other side is coated with an X-ray reflecting metallic layer. Segments are then stacked atop an azimuthally curved mandrel and bonded together. This pro-cess utilizes techniques and assembly equipment adopted directly from the microelectronics industry. Two stacks are coaligned axially into a module, forming an approximation of a paraboloid-hyperboloid mirror (Collon et al. 2008). A total of 236 modules form a "petal," an azimuthal segment of the full mirror. Eight such petals form the complete mirror. The entire production chain—wafer to petal—has been demonstrated. Hard X ray sensitivity is provided by coating reflecting surfaces at the innermost radii with multilayers.

### 2.1.1.2. The X-ray Microcalorimeter Spectrometer (XMS)

The XMS provides high spectral resolution, non-dispersive imaging spectroscopy over a broad energy range. The driving performance requirements are to provide spectral resolution of 2.5 eV for the central 2 × 2 arcmin in the 0.3–7.0 keV band, and 10 eV to the edge of the 5.4 × 5.4 arcmin field of view. The XMS is composed of an array of microcalorimeters, devices that convert individual incident X-ray photons into heat pulses and measure their energy via precise thermometry. The microcalorimeters are based on Transition-Edge Sensor (TES) thermometers. The rapid change in electrical resistance in the narrow transition (<1 mK) of the superconducting-to-normal transition of the TES allows for extremely accurate thermometry, thereby enabling the determination of the energy of individual X-ray photons to an accuracy of <2.5 eV (Kilbourne et al. 2007).

The focal plane layout is depicted in Fig. 2-3. It consists of a core 40 × 40 array of 300 × 300 μm pixels with spectral resolution of 2.5 eV, filling the 2 × 2 arcmin field of view with 3 arcsec pixels

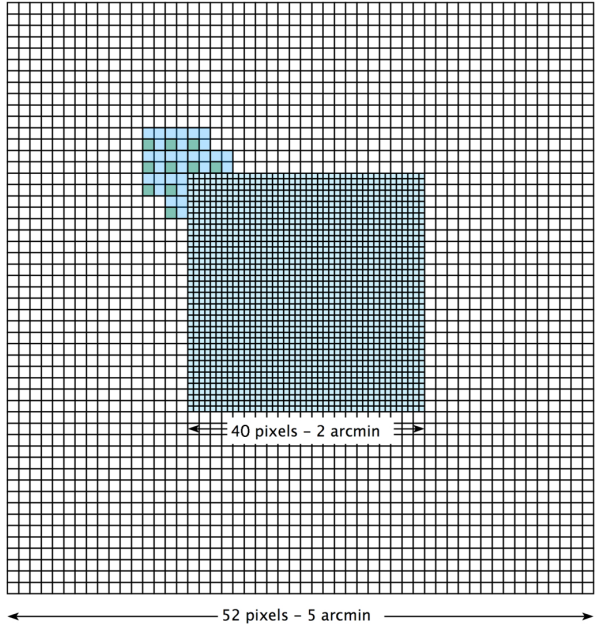

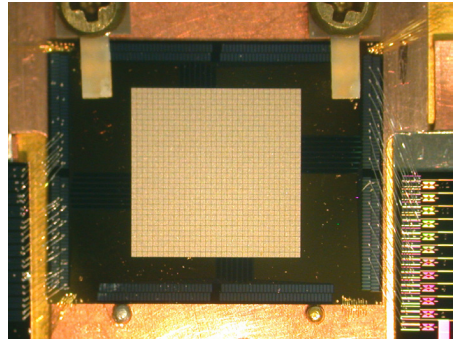

*Figure 2-4. A 32 × 32 microcalorimeter array with 300 micron pixels, high filling factor (95%) and high quantum efficiency (98% at 6 keV).*

*Figure 2-3. Schematic of a composite calorimeter array at the XMS focal plane. The inner array covers 2 arcmin and consists of 3 arcsec pixels with one absorber per TES. The full array covers 5.4 arcmin by use of 6-arcsec outer pixels which are read out by one TES per four absorbers as indicated by the groupings of one blue-green and 3 blue pixels ("blue splash" in upper left corner).*

(Fig. 2-4). Surrounding the core is a 52 × 52 array of 600 × 600 μm pixels (6 arcsec, 2,304 pixels total in the outer array) that extends the FOV to greater than 5 × 5 arcmin (5.4 × 5.4 arcmin) with better than 10 eV resolution [the change in spectral resolution is due to the increased pixel size which increases the heat capacity (Smith et al. 2008)].

The arrays are fabricated using standard microelectronics techniques. The pixels use Mo/Au bilayer superconducting films deposited on silicon-nitride membranes in a Si wafer. The X-ray absorbing elements are formed by electroplating Au/Bi films patterned so they provide a high array filling factor (95%), but only contact a small area of each TES to prevent electrical and chemical interaction with the sensitive thermometers.

Currently, 2.3 eV spectral resolution has been demonstrated in a non-multiplexed TES and an average of 2.9 eV has been achieved in a 2×8 array using a state-of-the-art, time-division SQUID multiplexer system (Kilbourne et al. 2008).

A Continuous Adiabatic Demagnetization Refrigerator (CADR) and a mechanical cryocooler provide cooling to 50 mK without expendable cryogens. Figure 2-5 shows a CAD drawing of the XMS dewar assembly. The cryocooler, being developed for the JWST/MIRI instrument that is already at TRL 6, is a baseline for the XMS. It will provide cooling at 5 K for precooling the CADR. The cryocooler for XMS consists of a three-stage pulse tube cooler and a $^4$He Joule-Thomson (JT) cooler. The pulse tube stages cool the radiation shields in the XMS cooler while the JT cooler provides the thermal interface to the CADR. The MIRI cooler performance would be improved with the use of a two-stage JT compressor. It provides 50 mW of cooling at 4.4 K for a total input power of 367 W, which matches the XMS requirements. Alternative cooling systems are under development by JAXA and ESA, and these are also based on strong flight heritage (SMILES, Herschel, Planck).

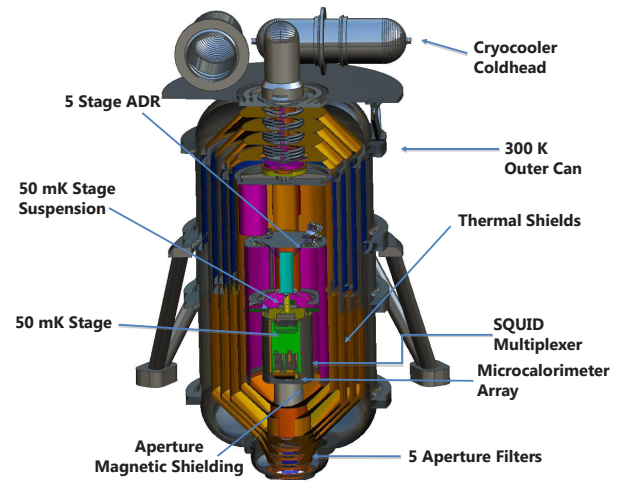

*Figure 2-5. Cutaway view of the XMS dewar assembly (CAD model) including details down to the detector assembly level.*

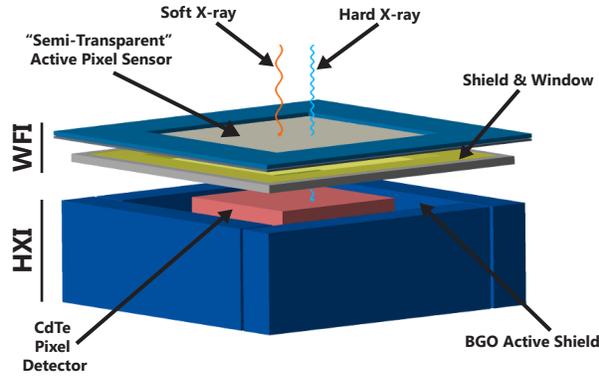

*Figure 2-6. WFI/HXI Schematic: The WFI soft X-ray Active Pixel Sensor (APS) is in front of the HXI CdTe detector.*

### 2.1.1.3. The Wide Field Imager/Hard X-ray Imager (WFI/HXI)

The WFI and HXI are two detectors incorporated into one instrument, with the HXI mounted directly behind the WFI (Fig. 2-6). The WFI is an imaging X-ray spectrometer with an 18 arcmin diameter FOV. It obtains images and spectra in the 0.1–15 keV band, with nearly Fano-limited energy resolution (50 eV at 300 eV; < 130 eV at 5.9 keV). A 100 × 100 µm pixel size, corresponding to 1 arcsec, oversamples the beam, minimizing pile up (de Korte et al. 2008).

The WFI's key component is the DEPFET (Depleted P-channel Field Effect Transistor) Active Pixel Sensor (APS). Each APS pixel also acts as an amplifier allowing the charge produced by an incident X-ray photon to be read directly from the pixel. This allows on-demand pixel readout, reduces readout noise, and offers radiation hardness against charge transfer inefficiency (Treis et al. 2008).

The HXI is a 5 × 5 cm Double-sided Strip Cadmium Telluride (DS-CdTe) detector located behind the WFI and observing simultaneously with it. It has nearly 100% detection efficiency up to 40 keV. The HXI will have energy resolution better than 1 keV (FWHM) at 30 keV and a FOV of 8 × 8 arcmin. To suppress background, five sides of the imager are surrounded by an active anticoincidence shield consisting of Bismuth Germanate (BGO) crystals viewed by Avalanche

Photodiodes (APDs). In addition, two layers of Double-sided Silicon Strip Detector (DSSD) are mounted above the CdTe to serve as particle background detectors and detectors of 7-30 keV X-rays (Takahashi et al. 2005).

### 2.1.1.4. The X-ray Grating Spectrometer (XGS)

The XGS is a wavelength-dispersive high-resolution spectrometer, offering spectral resolving power ($\lambda/\Delta\lambda$) of 3000 (FWHM) and effective area of 1000 cm² from 0.3 to 1.0 keV. The reference concept incorporates arrays of gratings that intercept a portion of the converging FMA beam and disperse the X-rays onto a CCD array. The existence of two viable grating technologies reduces risk. One implementation uses Critical Angle Transmission (CAT, Fig. 2-7, top) (Heilmann et al. 2008) gratings while the other approach uses Off-Plane Reflection Gratings (OPG, Fig. 2-7, bottom) (McEntaffer et al. 2008).

The CAT grating approach has been demonstrated on small prototypes, with measured diffraction efficiencies of 80–100% of theoretical values. CAT gratings with a grating-bar geometry meeting the IXO requirements have been fabricated. Prototype OPG are currently undergoing efficiency measurements, with an expectation of obtaining > 40% dispersion efficiency (sum of orders) from 0.3 keV to 1.0 keV.

An array of CCDs (9–32, depending on the grating technology used) used in photon-counting mode is used to image and read out the dispersed spectra. The CCD detectors provide better than the 80 eV resolution required for separation of the multiple diffraction orders produced by both kinds of grating.

### 2.1.1.5. The High Time Resolution Spectrometer (HTRS)

The HTRS performs precise timing measurements of bright X-ray sources (Barret et al. 2008). It can observe sources with fluxes of $10^6$ counts per second in the 0.3–10 keV band (about five times the intensity of the Crab) without performance degradation, while providing moderate spectral resolution (150 eV FWHM at 6 keV). The HTRS (Fig. 2-8) is an array of 37 hexagonal Silicon Drift Diodes (SDD), placed out of focus

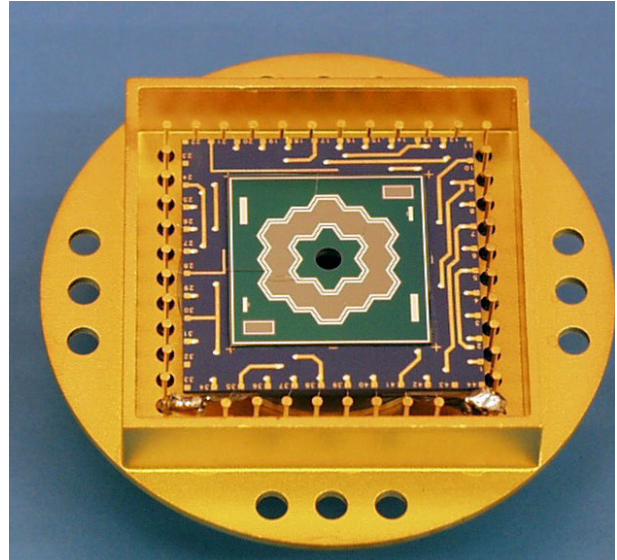

*Figure 2-8. Photograph of 12 hexagonal Silicon Drift Detectors (the 7 central SDDs are removed). The layout/design is the same as for the HTRS detector.*

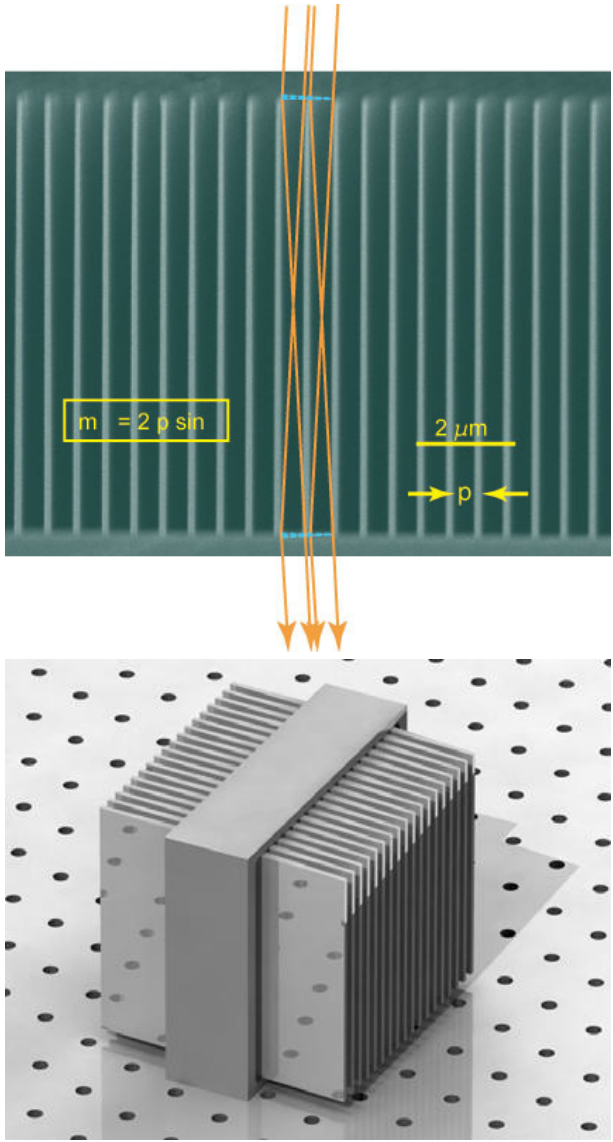

*Figure 2-7. Top: Scanning electron micrograph of a cross section through a 574 nm-period CAT grating prototype. Bottom: 18 off-plane reflection gratings mounted into a module.*

so that the converging beam from the FMA is distributed over the whole array.

### 2.1.1.6.  The X-ray Polarimeter (XPOL)

XPOL is an imaging polarimeter, with Minimum Detectable Polarization (MDP) of 1% for a source with a 2–6 keV flux of $1.5 \times 10^{-11}$ ergs cm$^{-2}$ s$^{-1}$ (1mCrab) in a $10^5$ s exposure. XPOL utilizes a fine grid Gas Pixel Detector (Fig. 2-9) to image the tracks of photoelectrons produced by incident X-rays and determine the direction of the primary photoelectron, which conveys in-formation about the polarization of the incoming radiation (Muleri et al. 2008). XPOL also has spectrographic capabilities with a resolving power of E/$\Delta$E of ~5 at 6 keV and timing resolution of a few $\mu$s. The field of view is $2.5 \times 2.5$ arcmin. The spatial resolution is a good match to the mirror PSF, giving a total angular resolution of 6 arcsec.

### 2.1.1.7.  Rationale for Selection

The ensemble of mirror plus instrumentation is baselined for IXO because it incorporates the most promising current technology for efficiently meeting the driving science requirements (Table 2-1). During Phase A an AO process will be used to select the final instrument complement.

The parallel FMA approaches offer high area-to-mass ratios plus modularity conducive to mass production. The XMS provides the highest available spectral resolution in an array providing the required spatial resolution and field of view. The DEPFET active pixel sensor array utilized by the WFI/HXI provides nearly Fano-limited energy resolution, great flexibility arising from the ability to control individual pixels, high readout speed, large sizes for monolithic detectors, and high intrinsic radiation hardness. The HXI detector provides the highest sensitivity for hard X-rays. The XGS meets the science requirements of resolving power R > 3000 and effective area > 1,000 cm$^2$ in the soft X-ray band between 0.3 and 1.0 keV. The HTRS provides the required capability to observe sources with extremely high count rates with sub-millisecond time resolution and CCD-like energy resolution (150 eV at 6 keV), over the ~0.3–10 keV band. The XPOL is the only polarimeter that combines high polarization sensitivity over a broad bandpass with imaging.

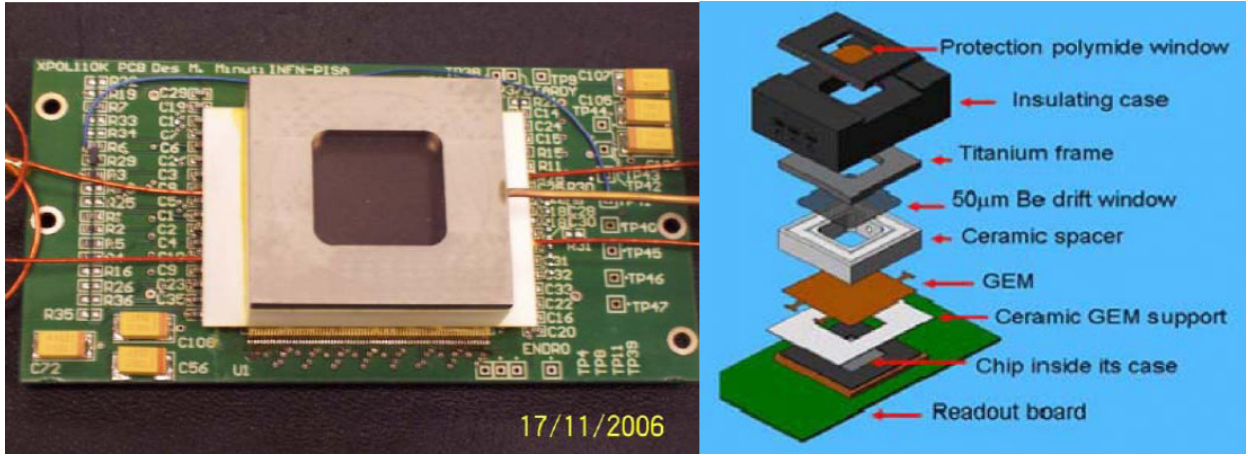

*Figure 2-9. XPOL photograph and schematic.*

### 2.1.1.8. How the Instruments Are Used Together

IXO is designed such that astrophysical data are collected using one of the four MIP-mounted instruments (XMS, WFI/HXI, HTRS or XPOL) at a time, in parallel with the XGS. In general, most observations will use one instrument only.

XGS observations will generally also include a simultaneous MIP-mounted instrument observation to extend the X-ray bandpass of the observations to E > 1 keV. The way the MIP-mounted instruments are used together is by performing distinct observations of the same object with alternate instruments in a way specified by an observer, with the intention of achieving a specific science goal.

## 2.1.2 Technical Maturity Levels

*Q2. Indicate the technical maturity level of the major elements and the specific instrument TRL of the proposed instrumentation, along with the rationale for the assessment. For any instrument rated at a Technology Readiness Level (TRL) of 5 or less, please describe the rationale for the TRL rating, including the de-*

*scription of analysis or hardware development activities to date, and its associated technology maturation plan.*

A careful evaluation has been performed to determine the technical readiness of each instrument and its major components to support the NASA-directed Independent Cost Estimate. The results of this evaluation, including recent progress, are summarized below.

The plans for reaching TRL 6 are described in Section 3 under Enabling Technology. Please also refer to the instrument technology roadmaps contained in the Supplemental Documents (see description in **Appendix D**) .

### 2.1.2.1. The Flight Mirror Assembly (FMA)

The components of the segmented glass mirror are at TRL 3 and 4. The mirror segments are at TRL 4. Current mirror segments are medium fidelity, meeting all but the figure quality requirement. They meet the effective area-to-mass ratio requirement (>20 cm²/kg); flight thickness surface coating with iridium has been demonstrated; they meet the surface smoothness requirement (microroughness of 4 Å); they meet the reflectivity requirement and coating density requirement; full-scale segments of the flight thickness have been fabricated and tested. The segments have been tested to environmental requirements: three nested mirror segments (~1 mm spacing) passed vibration and acoustic testing at qualification levels for EELV and Ariane 5 launch vehicles; segments have been performance-tested via X-ray imaging, producing results in good agreement with predictions. Mass production has been demonstrated through the NuSTAR program. Segment mounting and alignment is at TRL 3. Low fidelity breadboards have thus far been used. X-ray tests of aligned and permanently bonded segment pairs yielded HPD of 15 arcsec. The measured performance agrees with the results of analytical modeling.

The elements of the SPO mirror range from TRL 4 to 6. The silicon substrates are at TRL 6. They have adequate smoothness, flatness, and thickness uniformity to meet the IXO requirements. Deposition of an iridium surface with sufficient thickness, required surface smoothness and uniformity, including patterning for assembly, has been demonstrated. The production of a stack (stacking and bonding) is at TRL 4. Plates cut from wafers have been grooved, bent into shape, and robotically stacked. The existence of residual stack-up errors means that these stacks are low fidelity models of the IXO stacks.

The production of an SPO focusing module is at TRL 4. Angular resolution in X-rays of 17 arcsec HPD has been demonstrated from the first four plate pairs of a module, mounted inside a flight-representative petal. Technology development is required to reduce the influence of the particulate contamination causing the stack up error. Module mounting is at TRL 4. An isostatic mount has been developed, but requires refinement so that the loss of aperture due to the mount can be minimized.

### 2.1.2.2. The X-ray Microcalorimeter Spectrometer (XMS)

The XMS detector and readout technology is at TRL 4. XMS components (8 × 8 TES array and two-column SQUID multiplexer) have been integrated in a laboratory test environment and their performance validated. 32 × 32 arrays of the required pitch for XMS have been fabricated and are presently being tested. A separate demonstration of the outer array concept has been performed. The performance of all components agrees with analytical predictions; characteristics needed for scaling to XMS requirements are understood through models. Devices thus far demonstrated represent a low fidelity breadboard in terms of focal-plane layout and heat sinking of components. Testing has occurred in a relevant temperature, vacuum, and magnetic shielding environment; vibration and radiation tests have yet to be performed.

A JWST/MIRI demonstration model cooler has been operated for thousands of hours and both thermal vacuum tests and launch vibration qualification tests have been successfully completed. The MIRI cooler was certified in 2007 by a technical non-advocate review panel at TRL 6. The CADR is at TRL 4. A high-fidelity, four-stage brassboard CADR coupled to a commercial mechanical cryocooler has been built and its basic functionality has been demonstrated in critical environments (thermal and cryocooler-induced vibration) with a simulated cryocooler interface. The demonstrated performance agrees with analytical models; the performance has acceptable margin; scaling parameters are known; performance can be accurately modeled in the flight environment.

### 2.1.2.3. The Wide Field Imager/Hard X-ray Imager (WFI/HXI)

The WFI detector technology is at TRL 4. Detector prototypes of various formats (most relevant: 64 × 64 pixels of 75 × 75μm size, 64 × 64 pixels of 500 × 500 μm size) have been spectroscopically characterized with great success. These are low fidelity breadboards in terms of array size. Testing has been carried out in relevant thermal and vacuum environments. Radiation studies have been performed up to a multiple of the dose expected during a 10-year mission lifetime. Vibration tests have been successfully performed with mechanical prototypes similar to the WFI focal plane array at a multiple of the load expected during launch. Performance agrees with analytical predictions; the characteristics needed for scaling to full wafer-scale devices are understood.

The VELA/ASTEROID readout electronics are being operated continuously in critical test environments and can be considered TRL 4.

The HXI DSSD technology is at TRL 4. Component strips (2.5 cm and 4 cm wide DSSDs) have been integrated in a laboratory test environment and their performance validated. This demonstration represents a low fidelity breadboard in terms of focal-plane layout, together with read-out components. Tests were performed in a relevant temperature environment. Vacuum, vibration, and radiation tests have been performed at the component level. Performance agrees with analytical predictions; the characteristics needed for scaling to IXO-HXI requirements is understood through models.

The HXI CdTe imager technology is at TRL 4. Components (2.5 cm wide DS-CdTe's) have also been integrated into a low fidelity breadboard (focal-plane layout together with read-out components) in a laboratory test environment and their performance validated. Relevant environmental tests (temperature, vacuum, vibration, and radiation) have been done at the component level (e.g., 1-inch CdTe planar detector). The performance agrees with analytical predictions. Characteristics needed for scaling to the IXO-HXI requirements are understood through models.

### 2.1.2.4. The X-ray Grating Spectrometer (XGS)

The CAT and off-plane gratings are both at TRL 3. Eight CAT gratings/facets of various geometries have been measured with X-rays (in a relevant environment). IXO design parameters (aspect ratio, blaze angle) have been achieved in grating fabrication. The measured X-ray efficiency of a $3 \times 3$ mm prototype grating is 80–100% of theoretical. For the off-plane gratings, analytical predictions and laboratory demonstrations show that they can meet IXO's performance requirements. Off-plane gratings with properties similar to those needed for IXO have been performance-tested in a relevant temperature and vacuum environment, but no vibration testing has been performed. A higher-fidelity prototype IXO grating has been fabricated and is ready for X-ray testing.

The XGS CCD readout detectors are at TRL 5. Flight CCDs that meet all XGS requirements except readout rate have been demonstrated on-orbit on Chandra and Suzaku. These detectors represent a medium fidelity model of the XGS CCD array. The detector modifications needed for higher-speed readout (clock electrode strapping, higher responsivity output transistors) have been demonstrated successfully on optical CCD detectors in the laboratory.

### 2.1.2.5. The High Time Resolution Spectrometer (HTRS)

The HTRS has a TRL level of 6. Individual Silicon Drift Detectors identical to those consti-

tuting the IXO array are operational and their performance meets IXO requirements. A Detector Electronics Unit (DEU) similar to that for IXO flew on INTEGRAL; an upgraded version which can be considered a high fidelity analog of that needed for HTRS is under development for the ECLAIRs detector array on the Space multi-band Variable Object Monitor (SVOM) mission (2014).

### 2.1.2.6. The X-ray Polarimeter (XPOL)

The XPOL instrument is at TRL 5. A high-fidelity prototype detector has been demonstrated in a laboratory environment, meeting all performance requirements for IXO except dead time, and performing consistently with theoretical predictions. The detector array size and pitch meet IXO requirements. The detector underwent thermal vacuum testing between 15° C and 45° C, and was successfully vibrated. Breadboard detectors have undergone over 100 hours of laboratory testing, and over 15 prototypes have been built. A similar ASIC chip used in the Large Hadron Collider at CERN uses the same fabrication process as that needed for IXO. It is subjected to high dosages of radiation, needing only heavy ion testing for space qualification. All other components have been flown and are space qualified.

## 2.1.3 Instrument Risks

*Q3. In the area of instrumentation, what are the three primary technical issues or risks?*

The three primary technical risks are listed in Table 2-2. These risks have been identified and rated using the methodology discussed in the response to Programmatics and Schedule Q3, Section 5.3.

## 2.1.4 Instrument Tables

*Q4. Fill in entries in the instrument table. Provide a separate table for each instrument.*

See Table 2-5 to Table 2-10 at the end of Section 2.

## 2.1.5 Instrument Contingency

*Q5. If you have allocated contingency please include as indicated along with the rationale for the number chosen.*

**Table 2-2. Top Three Instrument Technical Risks and Mitigation Plans**

| Rank | Likelihood | Consequence | Risk Type | System | Risk Statement | Mitigation | Impact |
|---|---|---|---|---|---|---|---|
| 1 | 2 | 4 | Technical | FMA | If required angular resolution is not achieved with either mirror technology, then the SMBH at high redshift science will be significantly compromised. | Use parallel technology development through TRL 6 using segmented glass and Si pore optic approaches, prior to start of Phase B. Build and test an additional engineering unit prior to CDR. Thoroughly test the mirror through all stages of assembly. | Science - Compromise investigation of SMBH evolution |
| 2 | 3 | 2 | Technical | Inst - XMS | If the XMS cryogenic chain doesn't have sufficient reliability, then XMS cooling lifetime may not be achieved. | Provide thorough reliability analyses based on existing hardware test and on heritage orbit data. Use life testing as appropriate. Add redundancy to the design. | Mass & Power - Increased mass ( 60 kg) and power (300 W) for redundant system |
| 3 | 3 | 2 | Technical | Inst - XGS | If grating array throughput efficiency doesn't meet requirements, additional grating area coverage will be required. | Use parallel technology development. If necessary, add moveable (flip-up) deployment grating to remove grating area during observations where XGS is not required. | Cost - Increased grating size and/or additional mechanism |

For mass, rather than applying a uniform 30% contingency, IXO uses a variable "mass growth allowance" based on hardware type (structural, thermal, etc.) and design maturity level, as defined in *AIAA S-120-2006 Mass Properties Control for Space Systems*. Therefore, instrument mass contingencies range from 16% to 30%. For power and data rate, a 30% contingency is currently used which is consistent with the GSFC GOLD rules (GSFC-STD-1000D) for a project at this stage of maturity.

## 2.1.6 Payload Table

*Q6. Fill in the Payload Table.*

See Table 2-11, Payload Mass Table, at the end of Section 2.

## 2.1.7 Organizational Responsibilities

*Q7. Provide for each instrument what organization is responsible for the instrument and details of their past experience with similar instruments.*

Because all focal plane instruments are to be selected via open competition through an Announcement of Opportunity (AO), the organizations responsible for the various instruments have not yet been determined. The FMA will be procured via competition through a Request for Proposals (RFP). During the competition, past experience with similar instruments will be a selection factor.

Experienced instrument working groups are helping to develop the IXO reference mission and the critical technology. For the FMA, a team from GSFC/MSFC/SAO is developing the segmented glass technology, making use of their extensive experience with successful missions (Chandra, BBXRT, ASCA, Suzaku), and ESA is developing the SPO technology through contracting organizations. The XMS technology is being developed by a team lead by GSFC, NIST, SRON, and ISAS. The design builds on GSFC's successful work with suborbital instruments and the Suzaku XRS, GSFC/NIST work with readout electronics, SRON and ISAS and their European/Japanese partners working on the ground-based platform EURECA, as well as JAXA's successful work with cryocoolers. The WFI DEPFET technology is being developed at MPE, building upon their very successful EPIC-pn instrument on the XMM-Newton observatory. The HXI technology is based on JAXA development activity for ASTRO-H (2014 launch). Other technologies applicable to this device are being developed for NuSTAR. There are two concepts under development for the XGS: the OPGs are under study at Colorado/U. Iowa and CAT gratings are under study at MIT. The XPOL technology has been developed by an Italian consortium led by INFN (Pisa) and IASF (Rome). A related detector is under development at GSFC for the GEMS mission. The HTRS technology is being developed at CESR in France and at MPE where detector technology (Silicon Drift Diodes) is already mastered.

## 2.1.8    Instrumentation Studies

*Q8. For the science instrumentation, describe any concept, feasibility, or definition studies already performed.*

Numerous studies have been carried out for the instrumentation, in addition to substantial technology development. The most comprehensive documents arising from these studies are appended to this response (see description of supplemental documents in **Appendix D**). They include: 1) IXO Segmented Glass FMA Concept Study 2) IXO Payload Definition Document (PDD) with Corrigendum, 3) Mirror Technology Development Roadmap for the International X-ray Observatory, 4) IXO Silicon Pore Mirror Technology Development Plan (TDP), and 5) Instrument Technology Development Plans. Additionally, numerous papers about IXO instruments have been published in technical journals. Selected references are available at: **http://ixo.gsfc.nasa.gov/decadal_references/**.

## 2.1.9    Instrument Operations, Calibration, and Data Volume

*Q9. For instrument operations, provide a functional description of operational modes and ground and on-orbit calibration schemes. Describe the types of data and provide an estimate of the total data volume returned.*

### 2.1.9.1.    Instrument Modes

All instruments share three basic operational modes: (i) off; (ii) standby/engineering; and (iii) data collection. Examples of engineering modes are cool down of cryogenic instruments and diagnostics. Data collection modes depend on the specific instrument, for example, electronic window selection for the imaging detectors (XMS, WFI/HXI) and continuous clocking and frame transfer modes for the CCDs (XGS). Instrument calibration is performed using data collection modes. Nearly all mode switching is performed via software command. The IXO instruments have a limited number of mechanisms, all external to the instruments, operated with low duty cycle (one-time removal of protective covers, and occasional focus adjustment and filter wheel rotation).

### 2.1.9.2.    Ground and On-orbit Calibration

The IXO calibration philosophy is to calibrate systems and subsystems starting at the lowest possible level, and to utilize those calibrations as baselines in next-higher-assembly calibration up to and including the fully assembled observatory. The "test as you fly" (i.e., test in flight configuration) philosophy will be incorporated into the calibration approach. The strategy is to develop accurate theoretical and semi-empirical performance models for each subsystem's response to X-ray flux distributions: spectral, spatial, temporal and polarization, and to constrain and verify these models during ground calibration. Each calibration dataset will be used to constrain theoretical and semi-empirical models for the performance of the system or subsystem, and the full complement of these models represents the complete pre-flight calibration. This modeling approach carries over into the in-flight calibration where the instrument parameters will be verified and monitored. The science-based calibration re-quirements will be finalized during Phase A and reviewed during SRR.

### 2.1.9.3. Ground Calibration

For the overall ground calibration efforts, the IXO Science and Operations Center (ISOC; see also Section 4, Mission Operations) is responsible for generating the model requirements, data product requirements and data formats with support from the various instrument teams, mirror vendors, and the spacecraft vendor to ensure compatibility with the system-level calibration products and models. The ISOC is responsible for the Calibration Peer Reviews of the mirror and instruments. The individual instrument and mirror providers are responsible for pre-delivery calibration (including calibration requirements flowdowns and error budgets), and delivery of the calibration data products to the ISOC.

Ground calibration priorities include (i) alignment and co-alignment of mirror and detectors, (ii) effective area for the mirror as a function of energy and off-axis angle, (iii) PSF as a function of energy and off-axis angle, (iv) detector effects on effective area (e.g., QE, mode, gain, rate, rate linearity), and internal instrumental backgrounds, (v) detectors' energy/wavelength response and linearity, and (vi) parametric studies of the gain sensitivity to bias and temperature.

While there is significant commonality among the detectors, instrument-specific aspects of the calibrations exist. For the **FMA**, ground calibration will be performed on individual modules, integrated groups of modules, and the completed assembly. Calibration will be carried out using a combination of a vertical X-ray pencil beam and a horizontal long beam with large aperture (e.g., the MSFC X-Ray Calibration Facility); five months of time are scheduled for FMA calibration efforts. For the **XMS**, the temperature dependence of the gain will be calibrated. For the **WFI/HXI**, the detectors will be calibrated separately as well as fully assembled, including determination of the reduction in QE for the HXI due to obscuration by the WFI. For the **XGS**, the grating assemblies and CCD detectors will be calibrated both separately and as a system to determine the QE of the combined instrument (plus FMA) and the dispersion as a function of wavelength and spectral order. The **HTRS** calibration will focus on relative and absolute timing accuracy, including calibration of the spacecraft timing chain. For the **XPOL**, the modulation factor will be measured as a function of position.

### 2.1.9.4. Ground-to-Orbit Calibration Transfer

Much of the calibration data taken prior to launch is obtained under conditions different from on-orbit, including gravitational distortion, and finite distance effects at the calibration beamlines. The ISOC will oversee development of calibration models that provide transformations from ground-based to on-orbit performance predictions.

### 2.1.9.5. On-Orbit Calibration

The ISOC is responsible for planning, executing, and archiving on-orbit verification and calibration activities, with support from the instrument teams, mirror vendors, and the spacecraft vendor. Spacecraft systems and subsystems that impact IXO's scientific performance will be calibrated, including the aspect system boresights and the Inertial Reference Units (IRUs), both of which impact the aspect solutions, and the clocks which impact the absolute timing requirement. On-orbit verification of the FMA and instrument calibrations include on- and off-axis PSF (core and wings), vignetting functions, plate scales, effective areas, cross-calibrations between the IXO instruments (as well as with previous missions), and routine checks of the stability of the detector wavelength/energy scales, QE and QE uniformity, gains, astrometry, contamination, dark currents, system noise, etc. In addition, timing calibration is performed for the HTRS, and polarization calibration for the XPOL. Based on past experience with XMM-Newton and Chandra, routine calibration operations will take about 5% of the time available for science operations.

On-orbit calibration will be performed with celestial and on-board sources. Celestial sources will be chosen for flux stability and based on observations by previous missions. Continuum sources will be used for effective area calibration (e.g., hot white dwarfs), line sources for wavelength scale calibration and Line Response Functions (LRF) and dispersions relationships (e.g., Capella). X-ray and optical stray light calibration will be determined with off-axis observations of bright sources. Contamination monitoring sources will be continuum sources (e.g., BL Lacs), while open star clusters, which allow the simultaneous detection of many X-ray sources with well-determined optical positions, will be used to calibrate the relative pointing offsets and plate scale. Sources with known ephemerides will be used for timing calibration. IXO will also have on-board radioactive sources ($^{109}$Cd, $^{41}$Ca, $^{55}$Fe) as well as electron-impact sources and a polarized source for XPOL.

### 2.1.9.6. Data Analysis

IXO data analysis is straightforward and similar to three decades of common X-ray astronomy analysis (image, spectral, and photon event timing). Data analysis for XMS, WFI/HXI, XGS, and HTRS, data utilizes standard packages, such as CIAO and FTOOLS. Extending these packages to IXO data analysis is straightforward as they already support multi-mission processing. Analysis tools and techniques for polarization data will be developed in FTOOLS by the GEMS mission and can be extended to IXO.

### 2.1.9.7. Data Volume and Characteristics

All IXO detectors are photon-counting devices. The basic data are event based, and include time, position, energy, and associated event information. See Section 4.4 for details about data volume.

## 2.1.10 Flight Software

*Q10. Describe the instrument flight software, including an estimate of the number of lines of code.*

For each instrument, flight software carries out similar functions: spacecraft command and data interface, housekeeping gathering, instrument control and configuration, data acquisition and compression, as well as instrument-specific data processing. For the instruments with flight heritage, a considerable fraction of the software will be reused or easily modified; thus, not all of the lines of code listed below will be newly developed.

**FMA** – The FMA is passive. No instrument flight software is needed.

**XMS** – Based on the Suzaku XRS flight software, it is estimated that 20,000 lines of code are required. Instrument-specific XMS flight software is for data processing, thermometer readout, and CADR and cyrocooler control.

**WFI/HXI** – The lines of FPGA firmware code is estimated to be 100,000 while the number of lines of WFI flight software is estimated to be 80,000. Based on ASTRO-H, the HXI will require 50,000 lines of code with 100,000 FPGA lines of code.

**XGS** – The software for the XGS CCDs will be essentially the same as the software used for the Chandra ACIS, which comprised roughly 130,000 lines of high level language code (C++).

**HTRS** – For the HTRS, the lines of FPGA firmware code is estimated to be 60,000 while the number of lines of flight software is estimated to be 20,000.

**XPOL** – For each photon, instrument-specific software determines the impact point, photon energy and emission angle, by calculating the first, second and third moments of the charge distribution and identifying the side of the track from where the photoelectron is ejected. The anticipated number of onboard lines of code is roughly 10,000 in a high level language (C++). An FPGA will perform all the other tasks with firmware consisting of roughly 30,000 lines.

## 2.1.11 Non-US Participation

*Q11. Describe any instrumentation or science implementation that requires non-US participation for mission success.*

See discussion in Section 3.2.

## 2.1.12 Master Equipment List

*Q12. Please provide a detailed Master Equipment List (MEL).*

The abridged Payload MEL is listed in Table 2-12 at the end of Section 2. The Expanded Payload MEL is in Appendix A as Table A-1.

## 2.1.13 Instrument Flight Heritage and Space Qualification

*Q13. Describe the flight heritage of the instruments and their subsystems. Indicate items that are to be developed, as well as any existing hardware or design/flight heritage. Discuss the steps needed for space qualifications.*

The instrumentation required to fulfill the scientific requirements of IXO has significant flight heritage. In addition, a number of upcoming missions incorporate instruments similar to those anticipated for IXO. Instrument-related items needing development are discussed in Section 3, Enabling Technology.

Summarized below for each instrument are the flight heritage and the similarities to instruments under development.

### 2.1.13.1. FMA

The FMA has its heritage in several X-ray astronomical missions. The Wolter-I optical design has been used on all imaging telescopes used for non-solar X-ray astronomy, most notably Einstein, ROSAT, Chandra, and XMM-Newton. The modular approach has a heritage in the Japan-US ASCA and Suzaku missions, requiring similar numbers of mirror segments: ~14,000 for IXO vs. 6,800 for Suzaku. Iridium coatings have flight heritage from Chandra. The methodology of using normal incidence visible light metrology to accurately predict grazing incidence X-ray performance has been repeatedly demonstrated on all the aforementioned flight programs. For the segmented glass mirror approach, the segment fabrication technology has its heritage in the High-Energy Focusing Telescope (HEFT) balloon instrument, and is being demonstrated in the ongoing production of 8,000 glass segments for NuSTAR. The FMA alignment and metrology method has heritage in the extremely precise and successful Chandra mirror assembly. The SPO approach shares the same heritage as the segmented glass for the Wolter optics, coatings, and metrology. In addition, for the SPO approach, all fabrication and assembly steps are derived from mass production processes developed by the microelectronics industry.

### 2.1.13.2. XMS

The XMS instrument is based on X-ray microcalorimeters for high resolution, high throughput X-ray spectroscopy that have been developed over the last 20 years for astrophysics and laboratory spectroscopy. The first implementation of a microcalorimeter for astrophysics was on a sounding rocket payload (the X-ray Quantum Calorimeter, XQC) for measuring the spectrum of the diffuse X-ray background. The payload has been launched four times and is being prepared for another flight (McCammon et al. 2002). A major milestone for the XMS technology readiness is the XRS instrument on the Suzaku Observatory. This instrument featured a 32-channel microcalorimeter array operating at 60 mK, a single-stage CADR, and digital processing electronics capable of on-board optimal pulse height analysis. Both the Suzaku and XQC implementations have used ion-implanted Si for the thermometer with separately attached X-ray absorbers. A 6 × 6 calorimeter array with improved features (larger absorbers with higher uniformity and better energy resolution) is currently being developed for the Japan/US ASTRO-H mission, with a planned 2014 launch.

Cryocoolers demonstrating technology capable of meeting the IXO requirements have flown in space since the early 1990s with near-perfect performance. The most relevant cooler is the 20 K Stirling cooler flown on the JAXA Akari mission. The JAXA SMILES instrument, awaiting launch to the ISS (and thus at TRL 8), has a 4.5K JT cooler that meets all IXO requirements. The Suzaku XRS cooler operated successfully on-orbit, incorporating a single-stage ADR that cooled the XRS to 65 mK. Other relevant coolers include Stirling coolers on HIRDLS, ISAMS and MOPITT, and pulse tube coolers on Hyperion, SABER, AIRS and TES, all of which operate between 50 K and 80 K, and are still operational after tens of thousands of hours. A CADR has successfully cooled a calorimeter array during three suborbital flights of the XQC sounding rocket instrument.

### 2.1.13.3. WFI/HXI

The WFI detector draws from the heritage of an extremely successful X-ray CCD detector flown on XMM-Newton, the EPIC pn-CCD. The DEPFET technology is well tested with detectors developed for several different missions, including MIXS that will fly on Beppi-Columbo (~2013).

The HXI is an advanced version of the HXI to fly in 2014 on JAXA's ASTRO-H. Silicon strip detectors similar to the HXI DSSD, but with strips on one side, form the heart of the Fermi Gamma-Ray Observatory. DSSDs were successfully flown on a balloon experiment in 2003. An active BGO shield similar to that of the HXI is used in the Suzaku Hard X-ray Detector. BGO is also the planned shield material for the ASTRO-H HXI. A similar APD to those planned for the HXI is operating in orbit on the Cute-1.7 micro satellite. A similar analog electronics chain is working on-orbit on Suzaku (HXD-PIN detectors), and is being adapted for ASTRO-H.

### 2.1.13.4. XGS

Each of the two grating concepts builds on strong flight heritage.

Transmission gratings have been flown on the Objective Grating Spectrometer (OGS) on Einstein, and the Transmission Grating Spectrometer (TGS) on EXOSAT. Freestanding transmission gratings with hierarchical support structures have flown on IMAGE (2000) and TWINS A&B (2004, 2006). Transmission gratings are currently in use on the High- and Low-Energy Transmission Grating Spectrometers (HETGS and LETGS) onboard Chandra. The CAT grating optical design, including the grating array support structure, is based on the Chandra HETGS.

For the reflection gratings, the XMM-Newton Reflection Grating Spectrometer (RGS) provides fabrication, alignment, and metrology heritage. Off-plane reflection gratings have been flown on several UV and X-ray sounding rocket missions, including the recent flight of the University of Colorado's Cygnus X-ray Emission Spectroscopic Survey (CyXESS).

X-ray CCD detectors have a rich heritage in a variety of flight instruments, and have been in nearly continuous use in X-ray astronomy since the launch of ASCA in 1993. CCD-based instruments include Chandra's Advanced CCD Imaging Spectrometer (ACIS), XMM-Newton's RGS and European Photon Imaging Camera (EPIC), Swift's X-ray Telescope (XRT), and Suzaku's X-ray Imaging Spectrometer (XIS). Three of these instruments (ACIS, RGS, and EPIC) have all been operating successfully in space since 1999.

### 2.1.13.5. HTRS

SDDs were operated on the Mars Rover missions. Similar analog electronics are used on INTEGRAL/SPI spectrometer. The SDDs are used routinely in fast photon counting ground applica-tions, and have undergone extensive ground testing and qualification. One SDD discrete analog electronic chain has been running at CESR for several years.

### 2.1.13.6. XPOL

Gas proportional counters are an established space technology, utilized in virtually every X-ray observatory from Uhuru (1970) through BeppoSAX (1996). Gas Pixel Detectors (GPD) were flown on BeppoSAX. Polarization sensitive GPDs have undergone extensive ground testing and qualification over the past several years. A GPD variant, the Time Projection Chamber polarimeter, is the heart of the GEMS SMEX mission, recently selected by NASA for a 2014 launch. The XPOL Control Electronics (CE) are based on those used in AGILE.

### 2.1.13.7. Steps needed for space qualifications

Qualification and Engineering Test Units of new designs of the IXO payload will be space qualified, to protoflight levels, in accordance with the GSFC Standard Mission Assurance Requirements (MAR) document, CM Version (06-01-2009). The purpose of the space qualification program is to uncover deficiencies in design and method of manufacture. Examples of new designs include mirror modules, grating arrays, instrument focal plane assemblies, and detector readout subsystems. Test levels and durations will initially be set by GEVS-STD-7000 (2005) and refined during mission formulation.

## 2.2 Mission Design

### 2.2.1 Mission Design Overview

*Q1. Provide a brief descriptive overview of the mission design and how it achieves the science requirements.*

IXO is a facility-class observatory placed via direct insertion (no lunar swingby) into an 800,000 km semi-major axis halo orbit around the Sun-Earth L2 libration point using either an Evolved Expendable Launch Vehicle (EELV) or an Ariane V with a minimum throw mass of 6425 kg. The orbit progression for five years is shown in Fig. 2-12. IXO is built around a large area grazing incidence mirror assembly with a 20 m focal length and five science instruments. Flight-proven extensible masts allow the observatory to fit into either launch vehicle fairing. The mission design life is five years, with consumables sized for 10 years.

IXO meets all of the science requirements outlined in Section 1. All mission requirements were flowed down from the science objectives, the measurement requirements, and the payload accommodation and performance requirements. The power-rich and thermally stable quiescent L2 environment is ideal for IXO's observations, allowing undisturbed pointing at celestial objects for durations of $10^3$–$10^6$ sec with arcsecond-level pointing accuracy. All detectors are photon counting, thus longer integrations can be achieved by multiple exposures. IXO's field of regard is a 360° × 40° annulus, shown in Fig. 2-13, which over six months allows access to any location on the celestial sphere for a minimum of 1.5 months. With an average of two to three daily 60° (typical) repointings, each completed in a half hour, the overall observing efficiency is 85%. The allowed attitude relative to the sun line, shown in Fig. 2-13, is 70°–110° (pitch), ±180° (yaw); and ±10° (roll). These ranges provide steady thermal conditions, shade the detectors and radiators from the sun, and keep the unoccluded solar illumination on the solar arrays uninterrupted throughout the entire mission. Stationkeeping operations are performed every 21 days, while solar torque offloading is accomplished with frequent micro-impulses from small ACS thrusters, to improve observatory performance by keeping the reaction wheels unsaturated.

### 2.2.2 Mission Software Development

*Q2. Describe all mission software development, ground station development, and any science development required during Phases B and C/D.*

IXO mission software development includes the spacecraft flight software; flight software for the instruments; the science instrument EGSE software; and software for the IXO Science and Operations Center (ISOC), consisting of the Mission Data System (MDS) and the Science Data System (SDS) (the MDS and SDS provide the functions typically associated with the Mission Operations Center and the Science Operations Center, respectively; see also Section 4, Mission Operations). All ISOC activities are conducted in a single facility to reduce cost and increase synergy. There is no ground station development required for the IXO space/ground link since the Deep Space Network (DSN) will provide these services.

The ISOC will establish the interfaces to the GSE and to the software development environment at inception for use throughout the entire IXO development effort. A spacecraft simulator and a common software environment will be provided to the instrument teams, ensuring use of a single system from instrument development through operations. The spacecraft and science instrument flight software will be developed, integrated, validated and tested by the spacecraft contractor and instrument teams respectively, utilizing NASA and industry standards for soft-ware engineering and quality assurance, and providing for independent Verification & Validation (V&V). The instrument teams will develop any algorithms that may be required for calibration and/or data analysis; these algorithms will be implemented within the Science Data System by the ISOC to ensure functionality within the Science Data System and for use during the instrument and observatory calibrations. The science instrument software and associated EGSE software will be transferred to the IXO Science and Operation Center (ISOC) for use in ground calibration efforts after instrument delivery. This integration of ground data systems, software development and flat-sat development and test environments into the ISOC ensure a robust I&T environment.

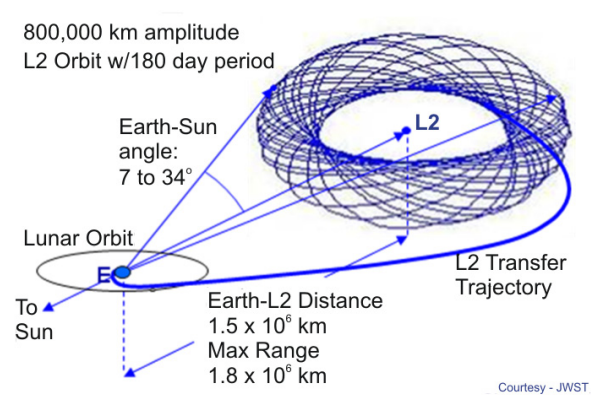

*Figure 2-12. IXO transfer and nominal mission operational L2 orbit.*

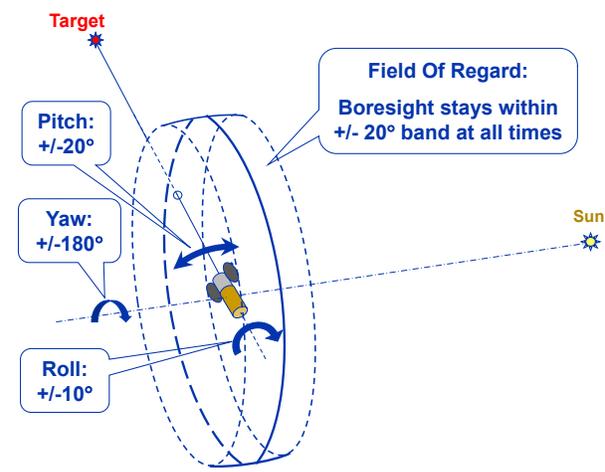

*Figure 2-13. IXO mission attitude and field of regard.*

The Mission Data System (MDS) and Science Data System (SDS) are described in Section 4 (Mission Operations Development). Science development during Phases B/C/D include supporting calibration requirements development and flow-downs to the instruments, supporting instrument and observatory level I&T activities, science mission planning system development, developing the documentation to support guest observers and the peer review, and extending the current set of X-ray data analysis tools that are available in standard analysis packages (e.g., CIAO, FTOOLs) to account for the higher spectral resolution data that will be available IXO's instruments.

The IXO spacecraft flight software (FSW) development effort is comparable to LRO. The software architecture draws from heritage missions such as LRO, which uses GSFC's core flight executive design, and will incorporate new development only for the mission specific components and as needed to mitigate obsolescence. Examples of software components that draw heavily from prior heritage include the real-time multi-tasking executive, pointing control, power management, command and data handling, memory loads and dumps, and fault protection. Examples of mission-specific functions include attitude determination, processing of sensor and actuator data, mechanism and deployable controls, and interfaces with the science instruments.

## 2.2.3 Mission Design Table

*Q3. Provide entries in the mission design table. For mass and power, provide contingency if it has been allocated.*
See Table 2-13.

## 2.2.4 Observatory Diagrams/Drawings

*Q4. Provide diagrams or drawings showing the observatory (payload and s/c) with the instruments and other components labeled and a descriptive caption. Provide a diagram of the observatory in the launch vehicle fairing indicating clearance.*

The drawings of the observatory in the deployed and launch configurations are shown in the **Observatory Quick Reference Guide**, and in Figs. 2-14, 2-15, and 2-16. IXO fits inside an EELV 5 m medium fairing static envelope with ample room. Any foreseeable future growth can easily be accommodated by making small adjustments.

## 2.2.5 Mission Risks

*Q5. For the mission, what are the three primary risks?*
The three primary IXO mission risks are described in Table 2-3.

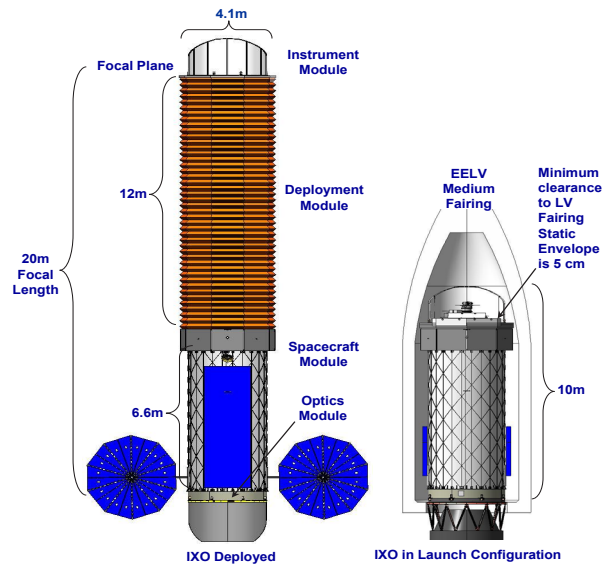

*Figure 2-14. IXO observatory in deployed and launch configurations*

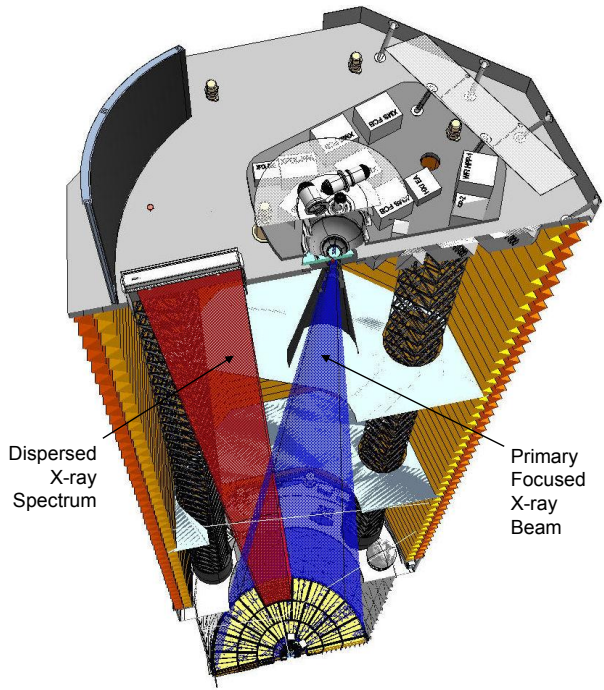

*Figure 2-15. IXO observatory cutaway view*

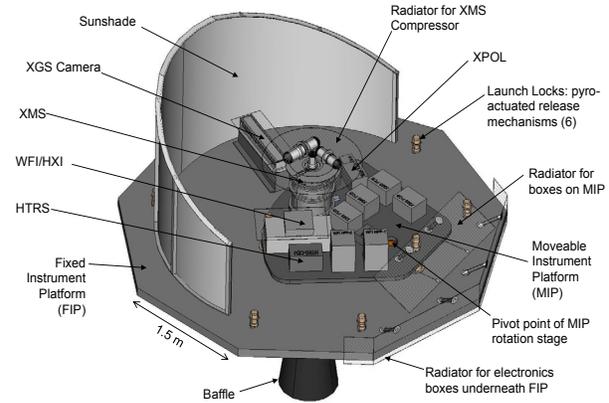

*Figure 2-16. IXO Instrument Module*

## 2.3 Spacecraft Implementation

IXO's maturity exceeds Pre-Phase-A expectations, having evolved over a decade at GSFC with recent contributions from ESA and JAXA. The systematic exploration and narrowing of IXO's trade-space produced a baseline design, which was further optimized through convergent iterations. The design has been reviewed by independent systems engineers at GSFC and external organizations. Performance estimates are tracked in error budgets, and supported by integrated Structural–Thermal–Attitude Control System modeling. IXO meets with margin all of its mission level requirements. The IXO spacecraft can be built with existing technology. All Spacecraft entries in the Spacecraft Master Equipment List (Table A-2) have a TRL 6 or higher.

### 2.3.1 Spacecraft Characteristics and Requirements

*Q1. Describe the spacecraft characteristics and requirements. Include a preliminary description of the spacecraft design and a summary of the estimated performance of the key spacecraft subsystems. Please fill out the Spacecraft Mass Table.*

The Spacecraft Mass Table is presented in Table 2-14. The IXO Systems Definition Document, submitted as a supplemental document (see description in **Appendix D**), describes the IXO mission in detail, presenting the operations concept, launch and flight dynamics parameters, the baseline configuration of the observatory, main functions, key performance metrics including pointing error budgets and resource budgets, and an overview of all of the subsystems.

IXO consists of four major modules to facilitate parallel development and integration and test: Instrument, Deployment, Spacecraft, and Optics Modules (see Fig. 2-14). Note that in the discussion below, spacecraft refers to the observatory (including all four modules), excluding the payload (FMA and instruments).

#### 2.3.1.1. Instrument Module (IM)

The IM (Fig. 2-16) accommodates the detector systems behind a fixed sunshade. All except the XGS camera mount to the MIP, which is comparable to moving platforms on Chandra and ROSAT. The XGS camera mounts on the Fixed Instrument Platform (FIP). The XMS, WFI/HXI, and the XGS camera have focus mechanisms. A Chandra heritage Telescope Aspect Determination System (TADS) assures the centering of the detectors in the converging X-ray beam, and provides the knowledge required for accurate aspect reconstruction.

**Table 2-3.   Top Three Mission Risks and Mitigation Plans**

| Rank | Likelihood | Consequence | Risk Type | System | Risk Statement | Mitigation | Impact |
|------|-----------|-------------|-----------|--------|----------------|------------|--------|
| 1 | 3 | 3 | Programmatics | FMA | If mirror build and test experiences significant delays, mission schedule margin will be eroded, resulting in launch delay. | Employ multiple sources and parallel development of mandrels, parallel lines for module assembly. Schedule margin, i.e., 10 months funded schedule slack on critical path. Modular nature of observatory minimizes impacts of FMA delays on the rest of the observatory. | Schedule - launch delay |
| 2 | 3 | 3 | Technical | Mission Systems | Given that the observatory is developed by an international consortium, there may be system level issues, such as interface incompatibility. | Provide full participation in reviews of all interfacing systems. International systems engineering team and IXO Management Council will resolve issues. Coordinate configuration management across partner organizations. ITAR agreements to allow information flow. Supply a thorough test and verification program with 10 months schedule slack on critical path. | Science - Possible degraded performance Schedule - launch delay |
| 3 | 2 | 4 | Technical | FMA | If required angular resolution is not achieved with either mirror technology, then the SMBH at high redshift science will be significantly compromised | Use parallel technology development through TRL 6 using segmented glass and Si pore optic approaches, prior to start of Phase B. Build and test an additional engineering unit prior to CDR. Thoroughly test the mirror through all stages of assembly. | Science - Compromise investigation of SMBH evolution |

#### 2.3.1.2.   Deployment Module (DM)

The DM is the portion of the metering structure which is extended on-orbit. It consists of three identical ADAM masts, similar to the one on NuSTAR. High deployment accuracy and repeatability was proven with the 60 m ADAM masts used in space on the Shuttle Radar Topography Mapper (SRTM). As the masts deploy, they pull up wire harnesses, X-ray baffles, and an accordion-like shroud to shield the instruments from stray light. The shroud is structured as a Whipple Shield (MLI thin foil layers spaced at specific distances) to minimize the number of micrometeroid penetrations.

#### 2.3.1.3.   Spacecraft Module (SM)

The SM is the central hub of the spacecraft and accommodates most of the Guidance, Navigation, & Control (GN&C), propulsion, power, avionics, and RF communications subsystem hardware. The reaction wheels, propulsion tanks, and electronics boxes mount to a nine-sided honeycomb spacecraft bus deck. The 6.6 m × 3.3 m diameter cylindrical composite isogrid Fixed Metering Structure (FMS) accommodates the thrusters, solar arrays, and the High Gain Antenna (HGA).

#### 2.3.1.4.   Optics Module (OM)

The OM includes the FMA and its covers, the XGS gratings, the star tracker/TADS periscope assembly, and the deployable sunshade. The Optics Module interfaces to the FMS at one end, and to the Launch Vehicle through the Separation System at the other.

A star-tracker, combined with the TADS's metering structure flex-body deflection sensing, supports sub-arcsec level end-to-end pointing performance, as follows (all 3σ numbers): image aspect knowledge is 1 arcsec required, 0.88 arcsec expected; image position control is 12 arcsec required, 1.13 arcsec expected; jitter is 200 milliarcsec required, < 20 milliarcsec expected. Metering structure deflection knowledge is 0.75 arcsec required, 0.6 arcsec expected. The expected pointing performance was verified by integrated Thermal-FEM-Control System modeling. The spacecraft subsystems are conventional, and are described in the answer to Q6. All observatory resource margins meet GSFC GOLD rules, (GSFC-STD-1000D), and are carefully managed. In later phases of the project, trading mass reserves to save cost will be considered.

### 2.3.2 Technical Maturity Levels

*Q2. Provide a brief description and an overall assessment of the technical maturity of the spacecraft subsystems and critical components. Provide TRL levels of key units. In particular, identify any required new technologies or developments or open implementation issues.*

All critical technology required for the IXO spacecraft is mature; no new technology development is needed and there are no implementation issues. The IXO Spacecraft MEL (see Table A-2 in Appendix A) lists every spacecraft component's TRL, with none lower than TRL 6. Many of the spacecraft requirements can be met with commercial off-the-shelf (COTS) components. EDUs, ETUs, and qualification units are called out in the MEL, and will be used during the development process as applicable. Integrated modeling is the principal verification tool used before the actual hardware exists, and great care shall be taken to incrementally build up the correlation between the models and the actual hardware as it becomes available. Verification will follow the "test as you fly" approach as much as feasible.

The structure is at TRL 8, and uses standard materials and design. The isogrid composite technology required to manufacture the Fixed Metering Structure is fully developed for the Minotaur fairing and the fuselage of the Boeing 787 Dreamliner. Mechanisms and actuators are at TRL 6 or 7. These include the FMA Cover Mechanisms at TRL 6 and the MIP Motor Assembly at TRL 7. The ADAM Masts are at TRL 7. The MLI shroud, shaped into a Whipple shield, is at TRL 6. The LV Separation System is also at TRL 6.

ACS components are at TRL 7–8, except for the TADS, which is at TRL 6, and uses the same concept and optical design as Chandra, with comparable requirements, parts, and technology. The AST 301 Star Tracker is at TRL 8; presently performing well on Spitzer. The Propulsion subsystem is at TRL 8. The Thermal subsystem is at TRL 8, and uses only standard off the shelf satellite thermal control technology: radiators, heaters, and Variable Conductance Heat Pipes (VCHPs). The Power subsystem is at TRL 8, using a simple design with existing flight proven technology. The Ultraflex arrays are at TRL 7, as they have flown only once, on Mars Phoenix Lander. The Avionics are at TRL 7, with a straightforward design, moderate data rates, and no real time processing requirements. The RF Comm subsystem is at TRL 8, except for the integrated S/Ka Transponder, which is at TRL 7. Note that separate S and Ka band transponders exist at TRL 9 and could be used with a 7 kg mass penalty, but a combined unit must be developed. Flight Software is based on LRO and is at TRL 7.

### 2.3.3 Lowest Technology Readiness Level (TRL) Units

*Q3. Identify and describe the three lowest TRL units, state the TRL level, and explain how and when these units will reach TRL 6.*

All components and technologies required for the IXO Spacecraft are at TRL 6 or above.

### 2.3.4 Risks

*Q4. What are the three greatest risks with the S/C?*

The three greatest IXO spacecraft risks are described in Table 2-4.

### 2.3.5 New Spacecraft Technologies

*Q5. If you have required new S/C technologies, developments, or open issues, describe the plans to address them (to answer you may provide technology implementation plan reports or concept study reports).*

No new technology development is required for the IXO Spacecraft. All technology required is at TRL 6 or above. The IXO spacecraft has no significant system or subsystem level open issues.

**Table 2-4. Top Three Spacecraft Risks and Mitigation Plans**

| Rank | Likelihood | Consequence | Risk Type | System | Risk Statement | Mitigation | Impact |
|---|---|---|---|---|---|---|---|
| 1 | 3 | 3 | Technical | Mission | Given that the observatory is developed by an international consortium, there may be system level issues, such as interface incompatibility. | Provide full participation in reviews of all interfacing systems. International systems engineering team and IXO Management Council will resolve issues. Coordinate configuration management across partner organizations. ITAR agreements to allow information flow. Supply a thorough test and verification program with 10 months schedule slack on critical path. | Science - Possible degraded performance Schedule - launch delay |
| 2 | 1 | 5 | Technical | Spacecraft | If the mirror covers fail to deploy then no X-rays will pass through the mirror resulting in loss of mission. If the Deployment Module does not fully deploy the focal length will not be achieved. | Mirror cover deployment has a single fault tolerant heritage design with proven industry standard redundant actuators, and an extensive ground qualification program. The ADAM Mast based deployment mechanism includes redundant actuators that are retractable and fault tolerant, based on a heritage design that flew successfully. The mast will have an extensive qualification program and end-to-end testing. | Science - Loss of mission |
| 3 | 1 | 4 | Technical | Spacecraft | If operation of the MIP is impeded, then it may not be possible to switch between the focal plane instruments. | MIP design includes redundant mechanism that is single fault tolerant. The MIP will have an extensive qualification program and end-to-end testing in test-as-you-fly configuration | Science - Loss of ability to switch instruments |

## 2.3.6 Subsystem Characteristics and Requirements

*Q6. Describe subsystem characteristics and requirements to the extent possible. Describe in more detail those subsystems that are less mature or have driving requirements for mission success.*

Note: All resource numbers in this section are CBE.

### 2.3.6.1. Structure

The structure is 24 m tall × 4.1 m (max) diameter (see Fig 2-14). The 20 m Metering Structure comprises a 6.6 m fixed composite cylinder, a spacecraft bus structure, and a 12.2 m deployable portion. IXO's structural properties were predicted using an observatory level Finite Element Model containing 42,978 nodes, shown in Fig. 2-17, and an FMA model containing 17,249 nodes for the primary structure and 125,738 nodes for each mirror module. Deployed end-to-end bending and torsion modes are all higher than 1 Hz, and in the launch configuration are over 12 Hz.

### 2.3.6.2. Mechanisms

Mechanisms include launch locks, the MIP, focus mechanisms for some of the instruments, three ADAM masts, a two-axis HGA gimbal, exterior and interior FMA covers, non-articulated Ultraflex solar array wings, and a deployable fore sunshield. The Structure/Mechanisms subsystem's mass is 1108 kg, but it only uses 1.9 W.

### 2.3.6.3. ACS

The ACS (107 kg, 54 W–333 W, depending on mode) provides the pointing performance critical to the mission, as described under Q1 of this section. Components include the AST-301 0.45 arcsec (3σ) Star Tracker, five Honeywell HR 16-150 0.2 N-m 150 N-m-s reaction wheels, an internally redundant SIRU, and 12 Adcole coarse

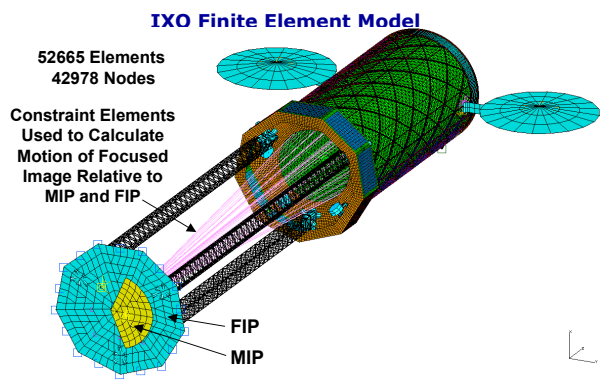

**IXO Finite Element Model**

52665 Elements
42978 Nodes

Constraint Elements
Used to Calculate
Motion of Focused
Image Relative to
MIP and FIP

FIP
MIP

*Figure 2-17. Observatory Level Finite Element Model.*

sun sensors for $4\pi$ steradian coverage. The control system accommodates the migration of the observatory center of mass by 5.6 cm caused by movements of the MIP.

### 2.3.6.4. Propulsion

The propulsion subsystem is pressure regulated (56 kg dry mass, 5 W) with twelve AMPAC DST-11H 22N biprop main thrusters, and four Aerojet MR-103 0.9N monoprop ACS thrusters, and is loaded with ~200 kg propellant in four spherical tanks. To mitigate pressure regulator lifetime concerns, a trade will be performed on switching after L2 orbit injection from pressure regulated mode to blowdown mode, a technique used on commercial communication satellites. Analysis predicts significant solar disturbance torques, the result of a 1.8 m offset between the observatory's center of mass and center of solar radiation pressure. The solar torque is offloaded continuously by imparting 16 μm/s impulses every 18 minutes by 0.11 s firings of a redundant set of 0.9 N ACS thrusters. This technique was demonstrated on Voyager. Integrated analysis has shown this approach to be the most efficient means of solar torque compensation: it keeps the reaction wheels unsaturated, and introduces minimal attitude deviations (0.17 arcsec), while using only 25 kg propellant over 10 years. The ACS thrusters' lifetime is addressed by using the same parts and qualification as Voyager, where thrusters have fired 500,000 times (vs. IXO's 300,000 for 10 years). As an extra contingency measure, IXO has the fallback option of using its reaction wheels to absorb solar torque as JWST does.

### 2.3.6.5. Thermal Subsystem

The thermal subsystem (158 kg, 460–1024 W) is conventional, with resistive heaters, passive radiators, and variable conductance heat pipes. A 3D observatory level thermal model with 1345 nodes, and one for the FMA with 9000 nodes for each of the 60 modules, analyzing a total of one million radiation couplings, has verified the thermal design. The 3D thermal models were also integrated with the structural model to confirm that the optical components' alignment requirements are met over the full range of mission temperatures.

### 2.3.6.6. Electrical Power System

The Electrical Power System (EPS) (119 kg, 167 W load) is a 28 VDC system with solar arrays on body-mounted and non-articulated, deployable panels (26 m$^2$ total area), generating 6600W BOL and 5200W EOL. An onboard 100 Ah Li-Ion battery, sufficient for approximately one hour with the observatory in safe mode, is used only at launch and in unforeseen contingencies, as the observatory is in sunlight continuously for 10 years.

### 2.3.6.7. Harness

The mass of the harness for the entire observatory is 273 kg, and it dissipates 21 W.

### 2.3.6.8. Avionics

The Spacewire-based avionics system (62 kg, 148 W) comprises four major units: C&DH with a 300-Gbit Solid State Recorder, Integrated Avionics (based on the Rad750 SBC in the present baseline), and fore and aft Remote Interface Units.

### 2.3.6.9. RF Comm Subsystem

The RF comm subsystem (30 kg, 44–80 W) uses CCSDS and Reed-Solomon encoding, operates in the Ka-band at 26 Mbps and in the S-band at 8 kbps and 2 kbps, and links to the DSN 34-m antenna thru a 0.7 m two-axis gimbaled HGA during a single daily 30-minute pass, or through the S-band omnidirectional antennas, if needed.

### 2.3.6.10. Flight Software

Flight software is based on LRO, and is described in Section 2.2 under Q2.

### 2.3.6.11. Mass, Power, and Data Storage

IXO's resource growth allowances and margins are comfortable, all in full compliance with the GSFC GOLD rules (GSFC-STD-1000D). Mass growth contingency percentages were assigned per AIAA S-120-2006, "Mass Properties Control for Space." The wet mass including growth allowance is 5121 kg compared to a launch vehicle throw mass 6425 kg. The mass growth margin over the Current Best Estimate (CBE) is 40.3%; over the contingent growth mass it is 20%. Power generated at BOL is 6600 W, with a minimum at EOL of 5200 W, contrasted to the maximum science mode power load (including 30 % growth contingency) of 3648 W, yielding an EOL power margin of 81%. The nominal 72-hr data volume is 52 Gbit (CBE); the data storage margin in the 300 Gbit Solid State Recorders is 600%. The margin is 50% for the 72-hr data volume at the maximum data rate of 192 Gbit (CBE).

As the baseline meets the IXO requirements with favorable margins, no lightweighting or power reduction is required.

### 2.3.7 Flight Heritage

*Q7. Describe the flight heritage of the spacecraft and its subsystems. Indicate items that are to be developed, as well as any existing hardware or design/flight heritage. Discuss the steps needed for space qualification.*

Most IXO components have substantial flight heritage. The IXO Spacecraft MEL (see Table A-2 in Appendix A) maintains a line-item level heritage database, listing heritage from a wide array of space missions, including COBE, Swift, RXTE, TRMM, EO-1, TOPEX, SMEX, Spitzer, Cassini, TRACE, HESSI, and Mars Probes.

The most significant flight heritage contributors are successful recent NASA missions, such as WMAP, LRO, and Chandra. Important test heritage is leveraged off confirmed future missions like SDO (2009) and JWST (2014). As all spacecraft components are at TRL 6 or higher, space qualification is not an issue.

Subsystem heritage highlights include:
- Orbit and mission: WMAP, JWST
- Fixed Metering Structure: Minotaur Composite Fairing, Boeing 787 Isogrid Fuselage
- Deployable Metering Structure: SRTMM

- Mechanisms: LRO, Chandra, Messenger, SOHO, Voyager, Shuttle Phoenix Lander
- ACS: Spitzer, NPOESS, TDRSS, GOES-R, Cassini, WMAP, Swift
- Propulsion: Chandra, Cassini, Mars Observer, Voyager
- Thermal: WMAP, LRO, SDO, COBE, TRMM
- Power: WMAP, LRO, Fermi, SDO
- Avionics: LRO, JWST, GOES-R, and MMS
- RF Comm: LRO, TRMM, Fermi, Terra (S band); LRO, Kepler, DS-1, SDO (Ka band)
- Flight Software: LRO
- Operations: Chandra, XMM-Newton

### 2.3.8 Science Instrument Accommodation

*Q8. Address to the extent possible the accommodation of the science instruments by the spacecraft. In particular, identify any challenging or non-standard requirements (i.e., jitter/ momentum considerations, thermal environment/temperature limits etc.) accommodation by the spacecraft.*

The IXO focal plane detectors are accommodated in the Instrument Module (IM), and the FMA and XGS Grating Arrays are in the Optics Module (OM). The observatory metering structures (deployed and fixed) provide the required 20 m focal length. Linear actuators at the mounts of the XMS, WFI/HXI, and XGS provide initial fine on-orbit focus adjustment, if necessary.

All four on-axis instruments mount side by side along with their proximity electronics on the MIP. The XGS Camera, the only off-axis instrument, mounts to the Fixed Instrument Platform (FIP). Supporting instrument electronics mount to the underside of the FIP. The instrument thermal requirements are met using a traditional cold-biased heater-controlled system. Radiators mount to the MIP and the FIP. These are connected to all of the instrument electonics boxes by variable conductance heat pipes to maintain temperatures as the instruments are turned on and off.

The IM Remote Interface Unit (RIU) provides power and data interfaces between the spacecraft and the instruments. It distributes commands and redundant regulated 28 VDC power to the instruments, and collects their science data and telemetry. Packetized instrument data are then transmitted on a Spacewire network from the RIU to the central avionics for storage and downlink.

The OM spacecraft adapter ring provides the mechanical interface between the FMA and the observatory. The XGS Grating Arrays mount to four mirror modules on the interior side of the FMA. An RIU in the OM, also on the Spacewire network, provides power and command and telemetry interfaces to the spacecraft. Significant power is provided to maintain the FMA temperature at 20C. Heaters mounted on the FMA collimators are controlled through heater control boxes mounted to each FMA mirror module; additional heaters on the fixed metering structure help maintain the mirror temperature. Interior and exterior covers protect the FMA from contamination and attenuate acoustic loads during launch, and are permanently deployed on-orbit after the observatory has completed outgassing. The particulate and molecular contamination-control requirements for the mirror and the instruments are on the order of Level 100A. Purge until T-0 is required for the FMA and the instruments.

The Spacecraft Module and Deployment Module provide a clear field of view for the focused beam from the FMA to the instruments at the focal plane. The Deployment Module shroud protects the focal plane instruments from stray light. Two X-ray baffles are located within the DM, with cut-outs lined with high-Z material, sized for the X-ray beams directed to the on-axis instruments and the XGS camera. An additional conical baffle is located on the FIP for the on-axis instruments, and the XGS CCD array has a separate baffle. Energetic particle background control is provided by a magnetic broom which directs particles away from instrument apertures.

The GN&C subsystem provides subarcsecond level pointing, as described under Q1. Integrated modeling predicts the jitter is at < 0.02 arcsec levels.

## 2.3.9   Schedule and Organization

*Q9. Provide a schedule for the spacecraft, indicate the organization responsible, and describe briefly past experience with similar spacecraft buses.*

A summary spacecraft development schedule is provided in Fig. 2-18. The schedule provides top-level development activities for each of the four modules of IXO. The schedule assumes implementation by industry partners; thus it starts at award and ends at delivery to observatory level integration and test.  The planned design milestones, allocated flight build and test periods, and durations for integration and test are consistent with the IXO mission development activities. Each module has its own schedule reserve to ensure timely delivery to I&T and mitigate mission level impacts.

The Spacecraft Module provider will be the prime contractor of observatory development and integration. The Instrument Module and the Deployment Module may be developed by different agencies and delivered to the prime contractor for observatory integration. The decision on allocation of mission responsibilities will take place in Phase A as discussed in Section 5 (Programmatics & Schedule).

## 2.3.10   Non-US Participation

*Q10. Describe any instrumentation or spacecraft hardware that requires non-US participation for mission success.*

For discussion of IXO instrumentation, see Section 3.2. All spacecraft hardware required for mission success is available from US sources.

## 2.3.11   Spacecraft Characteristics Table

*Q11. Fill out the Spacecraft Characteristics Table.*

See Table 2-17.

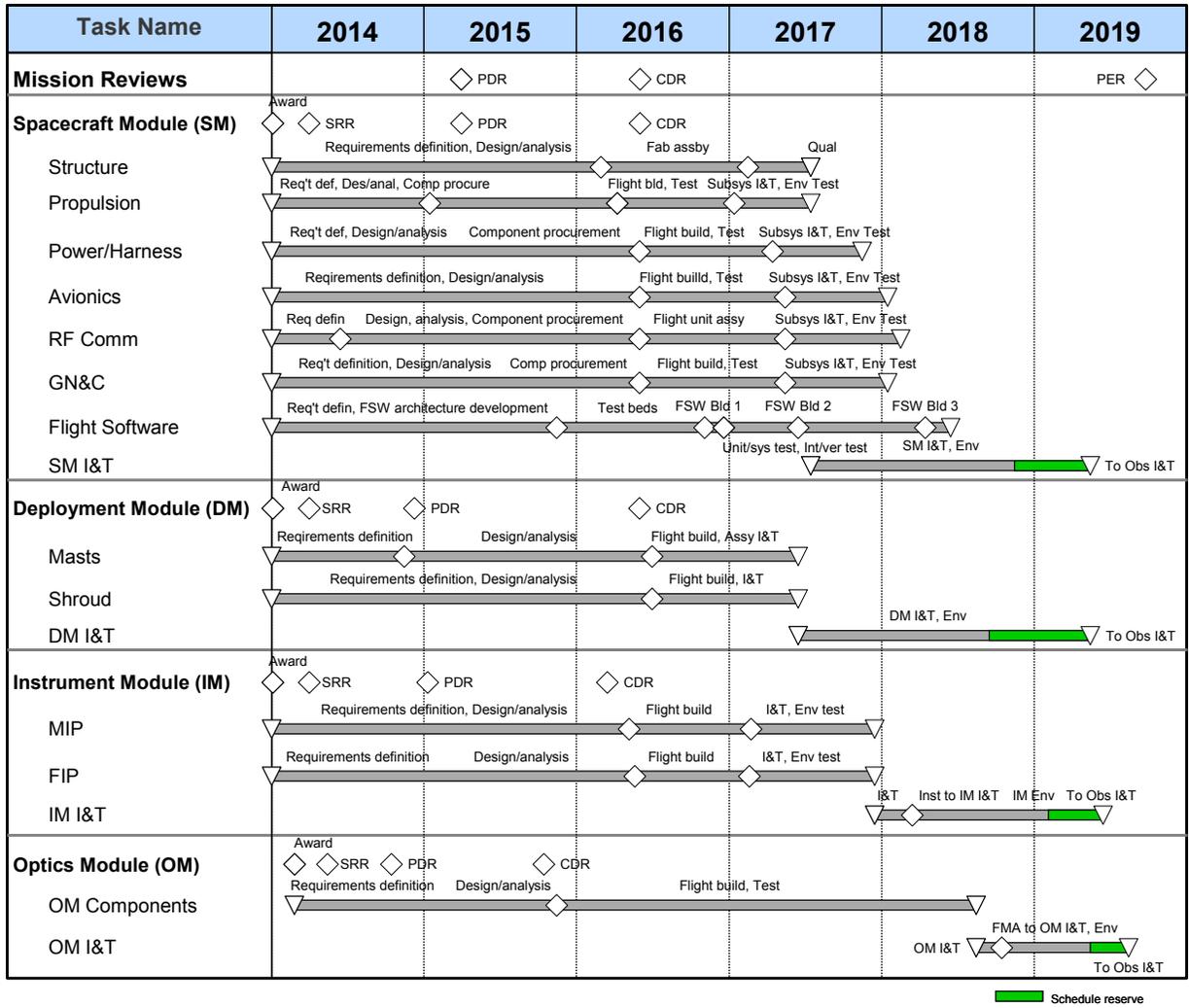

*Figure 2-18. Spacecraft development schedule.*

**Table 2-5. Instrument Table, Flight Mirror Assembly (FMA)\*\* (See discussion of this table in the text, page 16)**

| Item | Value | Units |
|---|---|---|
| Type of instrument | Wolter Type 1 Mirror Assembly | |
| Number of channels | N/A | |
| Size/dimensions | 3.3 × 3.3 × 1.2 | m × m × m |
| Instrument mass without contingency (CBE*) | 1731 | kg |
| Instrument mass contingency | 16 | % |
| Instrument mass with contingency (CBE+Reserve) | 2009 | kg |
| Instrument average payload power without contingency[1] | 1540 | W |
| Instrument average payload power contingency | 30 | % |
| Instrument average payload power with contingency | 2002 | W |
| Instrument average science data rate^ without contingency | N/A | kbps |
| Instrument average science data^ rate contingency | N/A | % |
| Instrument average science data^ rate with contingency | N/A | kbps |
| Instrument Fields of View (if appropriate) | 18 | arcmin |
| Pointing requirements (knowledge) | N/A | degrees |
| Pointing requirements (control) | N/A | degrees |
| Pointing requirements (stability) | N/A | degrees |

\*\* Segmented glass mirrors
\* CBE = Current Best Estimate.
[1] Power for FMA thermal control heaters includes: 1110 W for mirror + 420 W for fixed metering structure + 10 W for HXMM
^Instrument data rate defined as science data rate prior to on-board processing

**Table 2-6. Instrument Table, X-ray Microcalorimeter Spectrometer (XMS) (See discussion of this table in the text, page 16)**

| Item | Value | Units |
|------|-------|-------|
| Type of instrument - Imaging X-ray spectrometer | | |
| Number of channels | 68 amp. chains | |
| Size/dimensions (for 13 components) | | m × m × m |
| Dewar Assembly | 1.00 × 0.75 Dia | m × m × m |
| Filter Wheel | 0.21 × 0.64 × 0.40 | m × m × m |
| Pre-Amplifier/BiasBox (PBB) | 0.15 × 0.23 × 0.20 | m × m × m |
| Feedback/Controller Box's (FCB) -total of 4 boxes each | 0.23 × .28 × 0.20 | m × m × m |
| Pulse Processing Electronics (PPE) | 0.28 × 0.28 × 0.20 | m × m × m |
| ADR Controller (ADRC) | 0.13 × 0.25 × 0.38 | m × m × m |
| Cryocooler Control Electronics (CCE) | 0.20 × 0.20 × 0.20 | m × m × m |
| Filter Wheel Control Electronics (FWC) | 0.25 × 0.20 × 0.5 | m × m × m |
| Power Distribution Units (PDU) -total of 2 boxes each | 0.25 × .38 × 0.20 | m × m × m |
| Instrument mass without contingency (CBE*) | 263 | kg |
| Instrument mass contingency | 24 | % |
| Instrument mass with contingency (CBE+Reserve) | 327 | kg |
| Instrument average payload power without contingency | 649 | W |
| Instrument average payload power contingency | 30 | % |
| Instrument average payload power with contingency | 844 | W |
| Instrument average science data rate^ without contingency | 25.6 | kbps |
| Instrument average science data^ rate contingency | 30 | % |
| Instrument average science data^ rate with contingency | 33.3 | kbps |
| Instrument Fields of View @ 2.5 eV | 2 | arcmin |
| Instrument Fields of View @ 10 eV | 5.4 | arcmin |
| Pointing requirements (knowledge - diameter) (Pitch&Yaw) - see note 1& 2 | 3.4 | arcsec |
| Pointing requirements (control - diameter) (Pitch&Yaw) - see note 2 | 12 | arcsec |
| Pointing requirements (stability - diameter) (Pitch&Yaw) - see notes 1& 2 | 3.4 | arcsec/sec |

*CBE = Current Best Estimate.
^Instrument data rate defined as science data rate prior to on-board processing
Note 1: Pointing values given for knowledge and stability are the values required after ground processing of data using the Telescope Aspect Determination System (TADS). The use of the TADS eliminates the need to place stringent knowledge and stability requirements on either the instruments or the spacecraft, thereby simplifying instrument and spacecraft implementation and reducing cost.
Note 2 : The instrument does not place a controlling requirement on roll. Spacecraft roll pointing requirements are derived from other system considerations.

**Table 2-7.   Instrument Table, Wide Field and Hard X-ray Imager (WFI/HXI) (See discussion of this table in the text, page 16)**

| Item | Value | Units |
|------|-------|-------|
| Type of instrument - Imaging X-ray Spectrometer | | |
| Number of channels | 32 ADC channels for WFI 18 ASIC Channels for HXI | |
| Size/dimensions -  10 components | | |
| WFI Focal Plane Array (FPA) | 0.08 × 0.32 × 0.32 | m × m × m |
| HXI Sensor Head (HXI - S) - co-aligned with WFI FPA | 0.085 × 0.3 × 0.3 | m × m × m |
| WFI Hemisphere Preprocessor Boxes (HPP) - 2 each | 0.35 × 0.25 × 0.20 | m × m × m |
| WFI Frame Builder / Brain Box (FBB) | 0.35 × 0.40 × 0.25 | m × m × m |
| WFI Power Conditioner Units (PCU) - 2 each | 0.35 × 0.25 × 0.20 | m × m × m |
| WFI Filter Sled | 0.015 × 0.20 × 0.60 | m × m × m |
| HXI Digital Electronics (HXI-D) | 0.20 ×  0.20 × 0.10 | m × m × m |
| HXI PSU (HXI-E) | 0.20 × 0.20 × 0.10 | m × m × m |
| Instrument mass without contingency (CBE*) | 89 | kg |
| Instrument mass contingency | 26 | % |
| Instrument mass with contingency (CBE+Reserve) | 111 | kg |
| Instrument average payload power without contingency | 268 | W |
| Instrument average payload power contingency | 30 | % |
| Instrument average payload power with contingency | 348 | W |
| Instrument average science data rate^ without contingency | 55 | kbps |
| Instrument average science data^ rate contingency | 30 | % |
| Instrument average science data^ rate with contingency | 72 | kbps |
| Instrument Fields of View - WFI | 18 | arcmin |
| Instrument Fields of View - HXI | 8 | arcmin |
| Pointing requirements (knowledge - diameter) (Pitch&Yaw) - see notes 1& 2 | 1.7 | arcsec |
| Pointing requirements (control - diameter) (Pitch&Yaw) - see note 2 | 12 | arcsec |
| Pointing requirements (stability - diameter) (Pitch&Yaw) - see notes 1& 2 | 3.4 | arcsec/sec |

*CBE = Current Best Estimate.
^Instrument data rate defined as science data rate prior to on-board processing
Note 1: Pointing values given for knowledge and stability are the values required after ground processing of data using the Telescope Aspect Determination System (TADS). The use of the TADS eliminates the need to place stringent knowledge and stability requiremetns on either the instruments or the spacecraft, thereby simplifying instrument and spacecraft implementation and reducing cost.
Note 2 : The instrument does not place a controlling requirement on roll. Spacecraft roll pointing requirements are derived from other system considerations.

**Table 2-8.   Instrument Table, X-ray Grating Spectrometer (XGS)** (See discussion of this table in the text, page 16)

| Item | Value | Units |
|---|---|---|
| Type of instrument - Spectrometer | | |
| Number of channels | 4 nodes × 32 CCDs = 128 channels | |
| Size/dimensions (for 4 components) | | m × m × m |
| Focal Plane Assembly | 0.88 × 0.24 × 0.13 | m × m × m |
| Detector Electronics Assembly Box (DEA) | 0.11 × 0.11 × 0.27 | m × m × m |
| Digital Processing Assy Box (DPA) | 0.11 × 0.11 × 0.15 | m × m × m |
| CAT Grating Assembly - approximate - 4 each | 0.49 × 0.33 × 0.03 | m × m × m |
| Instrument mass without contingency (CBE*) | 50 | kg |
| Instrument mass contingency | 21 | % |
| Instrument mass with contingency (CBE+Reserve) | 61 | kg |
| Instrument average payload power without contingency | 77 | W |
| Instrument average payload power contingency | 30 | % |
| Instrument average payload power with contingency | 100 | W |
| Instrument average science data rate^ without contingency | 128 | kbps |
| Instrument average science data^ rate contingency | 30 | % |
| Instrument average science data^ rate with contingency | 166 | kbps |
| Instrument Fields of View (if appropriate) | N/A | arcmin |
| Pointing requirements (knowledge - diameter) (Pitch&Yaw) - see notes 1& 2 | 3.4 | arcsec |
| Pointing requirements (control - diameter) (Pitch&Yaw) - see note 2 | 16 | arcsec |
| Pointing requirements (stability - diameter) (Pitch&Yaw) - see notes 1& 2 | 3.4 | arcsec/sec |

*CBE = Current Best Estimate.
** The CAT XGS shown was used in the payload accommodation study. The Off-Plane Grating XGS should have similar parameters.
^Instrument data rate defined as science data rate prior to on-board processing
Note 1: Pointing values given for knowledge and stability are the values required after ground processing of data using the Telescope Aspect Determination System (TADS). The use of the TADS eliminates the need to place stringent knowledge and stability requirements on either the instruments or the spacecraft, thereby simplifying instrument and spacecraft implementation and reducing cost.
Note 2 : The instrument does not place a controlling requirement on roll. Spacecraft roll pointing requirements are derived from other system considerations.

**Table 2-9. Instrument Table, High Time Resolution Spectrometer (HTRS) (See discussion of this table in the text, page 16)**

| Item | Value | Units |
|---|---|---|
| Type of instrument - Spectrometer | | |
| Number of channels | 37 | |
| Size/dimensions (for 2 components) | | |
| Focal Plane Assembly (Detector Unit+Filter Wheel+DEU) | 0.3 × 0.20 × 0.39 | m × m × m |
| Central Electronic Unit (HTRS-CEU) | 0.20 × 0.35 × 0.20 | m × m × m |
| Instrument mass without contingency (CBE*) | 23 | kg |
| Instrument mass contingency | 22 | % |
| Instrument mass with contingency (CBE+Reserve) | 27 | kg |
| Instrument average payload power without contingency | 109 | W |
| Instrument average payload power contingency | 30 | % |
| Instrument average payload power with contingency | 142 | W |
| Instrument average science data rate^ without contingency | 50 | kbps |
| Instrument average science data^ rate contingency | 30 | % |
| Instrument average science data^ rate with contingency | 65 | kbps |
| Instrument Fields of View (if appropriate) | 15 | arcmin |
| Pointing requirements (knowledge - diameter) (Pitch&Yaw) - see notes 1& 2 | 60 | arcsec |
| Pointing requirements (control - diameter) (Pitch&Yaw) - see note 2 | 60 | arcsec |
| Pointing requirements (stability - diameter) (Pitch&Yaw) - see notes 1& 2 | 14.3 | arcsec/sec |

*CBE = Current Best Estimate.
^Instrument data rate defined as science data rate prior to on-board processing
Note 1: Pointing values given for knowledge and stability are the values required after ground processing of data using the Telescope Aspect Determination System (TADS). The use of the TADS eliminates the need to place stringent knowledge and stability requirements on either the instruments or the spacecraft, thereby simplifying instrument and spacecraft implementation and reducing cost.
Note 2 : The instrument does not place a controlling requirement on roll. Spacecraft roll pointing requirements are derived from other system considerations.

**Table 2-10. Instrument Table, X-ray Polarimeter (XPOL) (See discussion of this table in the text, page 16)**

| Item | Value | Units |
|---|---|---|
| Type of instrument - Polarimeter | | |
| Number of channels | 1 | |
| Size/dimensions (for 3 components) | | |
| Focal Plane Assembly | 0.17 × 0.19 × 0.27 | m × m × m |
| Back End Electronics (BEE) | 0.19 × 0.14 × 0.11 | m × m × m |
| Control Electronics (CE) | 0.29 × 0.20 × 0.11 | m × m × m |
| Instrument mass without contingency (CBE*) | 8.8 | kg |
| Instrument mass contingency | 20 | % |
| Instrument mass with contingency (CBE+Reserve) | 10.6 | kg |
| Instrument average payload power without contingency | 46 | W |
| Instrument average payload power contingency | 30 | % |
| Instrument average payload power with contingency | 60 | W |
| Instrument average science data rate^ without contingency | 300 | kbps |
| Instrument average science data^ rate contingency | 30 | % |
| Instrument average science data^ rate with contingency | 390 | kbps |
| Instrument Fields of View (if appropriate) | 2.5 × 2.5 | arcmin |
| Pointing requirements (knowledge - diameter) (Pitch&Yaw) - see notes 1& 2 | 2.5 × 2.5 arcmin | arcsec |
| Pointing requirements (control - diameter) (Pitch&Yaw) - see note 2 | 14.4 | arcsec |
| Pointing requirements (stability - diameter) (Pitch&Yaw) - see notes 1& 2 | 4 | arcsec/sec |

*CBE = Current Best Estimate.
^Instrument data rate defined as science data rate prior to on-board processing
Note 1: Pointing values given for knowledge and stability are the values required after ground processing of data using the Telescope Aspect Determination System (TADS). The use of the TADS eliminates the need to place stringent knowledge and stability requirements on either the instruments or the spacecraft, thereby simplifying instrument and spacecraft implementation and reducing cost.
Note 2 : The instrument does not place a controlling requirement on roll. Spacecraft roll pointing requirements are derived from other system considerations.

**Table 2-11. Payload Mass Table (See discussion of this table in the text, page 17.)**

| Payload | Current Best Estimate (CBE) (kg) | Mass Contingency | CBE Plus Contingency (kg)* |
|---------|----------------------------------|------------------|----------------------------|
| FMA | 1731 | 16% | 2009 |
| XMS | 263 | 24% | 327 |
| WFI/HXI | 89 | 26% | 111 |
| XGS | 50 | 21% | 61 |
| HTRS | 23 | 22% | 27 |
| XPOL | 9 | 20% | 11 |
| TOTAL | 2164** | 18% | 2546 |

*Data is rounded from MEL payload totals
** Rounding differences account for discrepancies between individual items and totals shown.

**Table 2-12. Payload Master Equipment List (Abridged\*) (See discussion of this table in the text, page 20.)**

| Item | Quantity | Mass CBE Total Flight Mass (kg) | Power (W) Science Average | Safehold | Peak |
|---|---|---|---|---|---|
| **FMA** | **1** | **1731** | **1120** | **973** | **1120** |
| **FMA Primary Structure Assy** | **1** | **340** | **-** | **-** | **-** |
| Ring 1 Modules | 12 | 287 | - | - | - |
| **Ring 2 Modules** | **24** | **393** | **-** | **-** | **-** |
| Ring 3 Modules | 24 | 536 | - | - | - |
| **HXMM Assembly** | **1** | **51** | **10** | **8** | **10** |
| Thermal Hardware | 1 | 119 | 1110 | 965 | 1110 |
| **Fasteners** | **1** | **5** | **-** | **-** | **-** |
| **XMS** | **1** | **263** | **649** | **0** | **703** |
| Filter Wheel Mechanism | 1 | 7 | - | - | - |
| **Gate Valve** | **1** | **0.03** | **-** | **-** | **-** |
| Pyro Devices for Gate Valve | 1 | 0.2 | - | - | - |
| **XMS Cryostat Assembly** | **1** | **87** | **-** | **-** | **-** |
| Dewar Bipod Assembly | 3 | 9 | - | - | - |
| **XMS Electronics Boxes** | **1** | **109** | **434** | **0** | **488** |
| XMS Thermal Subsystem | 1 | 19 | - | - | - |
| **Cryocooler** | **1** | **32** | **215** | **0** | **215** |
| **WFI/HXI** | **1** | **89** | **268** | **0** | **308** |
| **WFI** | **1** | **65** | **222\*\*** | **0** | **262\*\*** |
| **Focal Plane Array (FPA)** | **1** | **18** | **25** | **0** | **43** |
| Hemisphere Preprocessor Boxes (HPP) | 2 | 13 | 98 | 0 | 98 |
| **Frame Builder / Brain Box (FBB)** | **1** | **9** | **24** | **0** | **24** |
| Power Conditioner Units (PCU) | 2 | 13 | 6 | 0 | 6 |
| **Filter Sled (4 positions: open, closed, calibration, filter)** | **1** | **11** | **3** | **0** | **12** |
| **HXI** | **1** | **24** | **46** | **0** | **46** |
| **HXI Sensor (HXI-S)** | **1** | **15** | **20** | **0** | **20** |

**\*See Appendix A for Expanded Payload MEL (Table A-1)**

**\*\* Power total includes 70% dc-dc converter efficiency**

**Table 2-12. Payload Master Equipment List (Abridged\*) (Cont.)  (See discussion of this table in the text, page 20.)**

| Item | Quantity | Mass | Power (W) | | |
| --- | --- | --- | --- | --- | --- |
| | | CBE Total Flight Mass (kg) | Science Average | Safehold | Peak |
| HXI Analog Electronics (HXI-E) | 1 | 5 | 20 | 0 | 20 |
| HXI Digital Electronics (HXI-D) | 1 | 4 | 6 | 0 | 6 |
| XGS | 1 | 50 | 77 | 0 | 83 |
| Readout Camera Assembly | 1 | 41 | 77 | 0 | 83 |
| Focal Plane Assembly | 1 | 28 | 7 | 0 | 9 |
| Camera Structure | 1 | 1.6 | - | - | - |
| Detector Electronics Assembly Box (DEA) | 1 | 6 | 50 | 0 | 50 |
| Digital Processing Assy Box (DPA) | 1 | 5 | 20 | 0 | 24 |
| Thermal Subsystem | 1 | 0.4 | - | - | - |
| CAT Grating Assembly | 1 | 9 | - | - | - |
| Grating Assembly A | 1 | 2.3 | - | - | - |
| Grating Assembly B | 1 | 2.3 | - | - | - |
| Grating Assembly C | 1 | 2.3 | - | - | - |
| Grating Assembly D | 1 | 2.3 | - | - | - |
| HTRS | 1 | 23 | 109\*\* | 0 | 109\*\* |
| Detector Unit | 1 | 2.5 | 22 | 0 | 22 |
| Filter wheel | 1 | 2.1 | 20 | 0 | 20 |
| Detector Electronic Unit (DEU) | 1 | 6.4 | 20 | 0 | 20 |
| Central Electronic Unit (CEU) | 1 | 12 | 34 | 0 | 34 |
| XPOL | 1 | 8.8 | 46 | 0 | 46 |
| Focal Plane Assembly | 1 | 4.8 | 12 | 0 | 12 |
| GPD+FW | 1 | 3.3 | 2 | 0 | 2 |
| Back End Electronics (BEE) | 1 | 1.6 | 10 | 0 | 10 |
| Control Electronics (CE) | 1 | 4 | 34 | 0 | 34 |

**\*See Appendix A for Expanded Payload MEL (Table A-1)**
**\*\* Power total includes 70% dc-dc converter efficiency**

**Table 2-13. Mission Design Table (See discussion of this table in the text, page 24.)**

| Parameter | Value | Units |
|---|---|---|
| Orbit Parameters (apogee, perigee, inclination, etc.) | L2 Orbit: 800,000 km Y Amplitude (in ecliptic normal to Earth-Sun line), ≤ 500,000 km Z Amplitude (normal to ecliptic) | |
| Mission Lifetime | 60 months required 120 months goal | mos |
| Maximum Eclipse Period | 0 (continuous uneclipsed full sun during entire 10 year mission) | min |
| Launch Site | KSC or Kourou | |
| Spacecraft* Dry Bus Mass without contingency | 4211 | kg |
| Spacecraft Dry Bus Mass contingency** | 17.1 | % |
| Spacecraft Dry Bus Mass with contingency | 4930 | kg |
| Spacecraft Propellant Mass without contingency | 163 | kg |
| Spacecraft Propellant contingency*** | 47 | % |
| Spacecraft Propellant Mass with contingency**** | 191 | kg |
| Launch Vehicle | EELV / 5 m Medium Fairing or Ariane 5 | Type |
| Launch Vehicle Mass Margin | 1014 | kg |
| Launch Vehicle Mass Margin (%) | 19.8 | % |
| Spacecraft Bus Power without contingency | 3675 (max) | W |
| Spacecraft Bus Power contingency | 30 | % |
| Spacecraft Bus Power with contingency | 4777 (max) | W |

* In the above Table, the term "Spacecraft" refers to the entire observatory, and "Spacecraft Bus Power" refers to the Power Load of the entire observatory

** Mass contingency percentages assigned per AIAA standard "AIAA_S-120-2006, Mass Properties Control for Space Systems".

*** Propellant contingency calculated with full propellant tanks (281 kg propellant) thrusting an observatory at its present CBE dry mass of 4211 kg.

****The Spacecraft propellant mass with contingency is based on the contingent growth mass of the observatory, and calculated per 3 sigma delta-v numbers, burdened by ACS taxes, varying small delta-v contingency, and ullage.

**Table 2-14. Spacecraft Mass Table (See discussion of this table in the text, page 25.)**

| Spacecraft bus | Current Best Estimate (kg) | Percent Mass Contingency | CBE Plus Contingency (kg) |
|---|---|---|---|
| Structures & Mechanisms | 1108 | 15 | 1269 |
| Thermal Control | 158 | 19 | 189 |
| Propulsion (Dry Mass) | 56 | 3 | 57 |
| Attitude Control | 107 | 6 | 114 |
| Command & Data Handling | 62 | 22 | 75 |
| Telecommunications | 30 | 4 | 32 |
| Power* | 392 | 23 | 483 |
| Total Spacecraft Dry Bus Mass | 1912** | 16 | 2219 |

Contingency percentages assigned per AIAA standard "AIAA_S-120-2006, Mass Properties Control for Space Systems". The Mass Growth contingency specified in the standard for existing hardware from another program, based on measured mass of qualification hardware, is in the 3 to 5% range.

* Includes 128 kg (CBE plus contingency) for the Power Subsystem and 355 kg (CBE plus contingency) for the observatory harness

** Rounding differences account for discrepancies between individual items and totals shown.

**Table 2-15. Spacecraft Characteristics (See discussion of this table in the text, page 32.)**

| Spacecraft bus | Value/Summary, units |
|---|---|
| **STRUCTURE** | |
| **Structures material** | **Aluminum composite honeycomb panel Advanced grid stiffened CFRP** |
| **Number of articulated structures** | **2 (HGA, MIP)** |
| **Number of deployed structures** | **7 (Deployable Metering Structure, 2 FMA Covers, HGA, 2 Ultraflex Solar Arrays, Deployable OM Sunshade)** |
| **THERMAL CONTROL** | |
| **Type of thermal control used** | **Conventional cold biased thermal control with heaters, passive radiators and variable conductance heat pipes** |
| **PROPULSION** | |
| **Estimated delta-V budget, m/s** | **98 m/s (10 years)** |
| **Propulsion type(s) and associated propellant(s)/ oxidizer(s)** | **Pressure regulated biprop (~200 kg MMH/NTO) with a mixture ratio of 0.86, also Monoprop manifolded to the MMH tank.** |
| **Number of thrusters and tanks** | **Twelve 22N biprop station keeping thrusters Four 0.9 N monoprop ACS thrusters for solar pressure offloading One COPV titanium He tank MEOP of 2,176 psia Two monolithic titanium MMH tanks MEOP of 400 psia Two monolithic titanium NTO tanks MEOP of 400 psia** |
| **Specific impulse of each propulsion mode, seconds** | **278 s biprop and 150 s monoprop while in pressure regulated mode, both bi- and mono- propellant thrusters will also work in end-of-life blowdown mode with slightly lower Isp** |
| **ATTITUDE CONTROL** | |
| **Control method (3-axis, spinner, grav-gradient, etc.).** | **3 axis stabilized** |
| **Control reference (solar, inertial, Earth-nadir, Earth-limb, etc.)** | **Stellar referenced inertial** |
| **Attitude control capability, degrees** | **Image positioning to 1.13 arcsec (1.13/3600 degree) ($3\sigma$)** |

**Table 2-15. Spacecraft Characteristics (Cont.) (See discussion of this table in the text, page 32.)**

| Spacecraft bus | Value/Summary, units |
|---|---|
| Attitude knowledge limit, degrees | 0.88 arcsec (0.88/3600 degree) (3$\sigma$) |
| **ATTITUDE CONTROL (cont.)** | |
| Agility requirements (maneuvers, scanning, etc.) | 60 deg yaw and 20 deg pitch completed in 1 hour required / 0.52 hrs expected |
| Articulation/#–axes (solar arrays, antennas, gimbals, etc.) | High Gain Antenna: 2 axes<br>Moving Instrument Platform: 1 axis |
| Sensor and actuator information (precision/errors, torque, momentum storage capabilities, etc.) | AST 301 Star Tracker: 0.15 as (1$\sigma$) after calibration<br>TADS flexbody deflection monitor: 0.23 arcsec (1$\sigma$)<br>Honeywell HR16 Reaction Wheels: 0.2 N-m, 150 N-m-s<br>SIRU: 0.00015 deg/rt-hr random walk<br>12 Adcole Coarse Sun Sensors for safe mode 4$\pi$ steradian coverage |
| **COMMAND AND DATA HANDLING** | |
| Spacecraft housekeeping data rate, kbps | Maximum 7.4 kbps |
| Data storage capacity, Mbits | 300,000 Mbits (300 * 10$^9$ bits) |
| Maximum storage record rate, kbps | 3,283 kbps |
| Maximum storage playback rate, kbps | 26,000 kbps (26 Mbps) |
| **POWER** | |
| Type of array structure (rigid, flexible, body mounted, deployed, articulated | One body mounted rigid panel<br>Two deployed nonarticulated Ultraflex arrays |
| Array size, meters x meters | 6.75 m $\times$ 2 m Body Mounted Array<br>Two 3.4 meter diameter deployed Ultraflex arrays |
| Solar cell type (Si, GaAs, Multi-junction GaAs, concentrators) | 29.3% efficiency triple junction GaInP2/GaAs/Ge |
| Expected power generation at Beginning of Life (BOL) and End of Life (EOL), watts | 6600 W BOL max<br>5200 W EOL min |
| On-orbit average power consumption, watts | 3648 W (incl. 30% contingency) |
| Battery type (NiCd, NiH, Li-ion) | Li-Ion |
| Battery storage capacity, amp-hours | 100 Ah at 80% DoD |

# 3. Enabling Technology

## 3.1 Introduction

The enabling technologies for the IXO mission that have not yet achieved TRL 6 are described below. Since all spacecraft technology is at TRL 6 and above, the only enabling technologies are for the mirror and instruments. Regarding the first part of Question 1, a full discussion of the current TRL rating and rationale for each enabling technology is available in Section 2.1 and so is not repeated here. The issue of non-US technology is discussed in Section 3.3. A technology development plan for each is included in the Supplemental Documents listed in **Appendix D**.

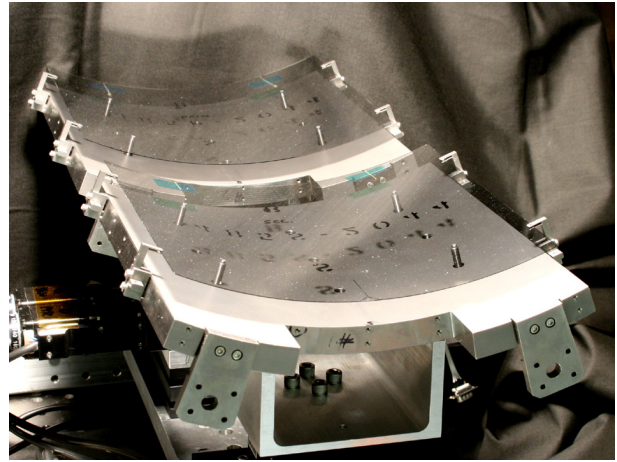

*Figure 3-1. Mirror segments aligned and mounted for X-ray testing.*

*Q1. For any technologies rated at a TRL of 5 or less, please describe the rationale for the TRL rating, including the description of analysis or hardware development activities to date, and its associated technology maturation plan.*

*Q2. Describe the critical aspect of the enabling technology to mission success and the sensitivity of mission performance if the technology is not realized.*

### 3.1.1 Flight Mirror Assembly

Technology development for the two parallel mirror concepts concentrates on two major areas: fabrication of the optical components, and mounting and alignment to form the mirror assembly (See Fig 3-1, 3-2).

#### 3.1.1.1. Segmented Glass Mirror

The Glass Mirror Technology Roadmap (see Supplementary Documents described in **Appendix D**) presents details of mirror technology development. Mirror segment fabrication, mounting and assembly technologies are developed separately to TRL 5, then merged to reach TRL 6. Table 3 of the Roadmap shows detailed schedules.

Mirror segment development focuses on three primary error contributors to performance: mandrel figure, mid-frequency figure error, and axial sag error. Mandrel figure is not a technology development issue. Two mandrels have been figured to the required ~ 2.3 arcsec HPD requirement. Mid-frequency error improvement is proceeding in three different approaches: improved application and smoothing of the currently used boron nitride (BN) release layer, an alternative release layer of sputtered platinum, and a second alternative of sputtered BN. Incorporation of an improved release layer reduces the mid-frequency error contribution from ~8 to 2 arcsec HPD. It has been demonstrated that coating stress is reduced using a chromium binder layer beneath the iridium (Ir) coating. Reduced coating stresses will reduce the sag contribution from ~5 to 1 arcsec HPD. Combined, these developments reduce the measured 15 arcsec HPD performance to <10 arcsec HPD. The remaining ~ 8 arcsec of error are due to mirror segment alignment and mounting.

Two different alignment and mounting methods are being developed: an active and a passive approach. In the active approach, prior to bonding, mirror pair focal length and low spatial frequency errors can be corrected with a set of actuators. In new results, *completed after the June 8th Astro2010 submissions*, mounting and alignment has been demonstrated for a pair of mirrors to the allocated alignment contribution of 1 arcsec HPD. The passive approach utilizes a temporary mount for the mirrors, making them more rigid for aligning and mounting. Recent results demonstrate repeatable mounting without figure degradation.

Alignment is a repetitive operation; thus, co-aligning a limited number (3) of multiple adjacent pairs of segments demonstrates the technology and procedures necessary to align all the mirror segments. In addition, both active and passive approaches are amenable to feedback controlled automated alignment using optical alignment metrology and piezo-electric mirror adjusters or positioners.

Development plans for the two approaches are similar: repeatable aligning and mounting of a mirror segment pair in a moderate fidelity FMA simulator will satisfy TRL 4 (2/10). Both approaches then progress to co-aligning two or three adjacent pairs to allocated alignment tolerances. X-ray testing will verify optical alignment metrology. Vibro-acoustic testing in a simulated flight environment will be performed to achieve TRL 5 (11/10). As part of meeting TRL 5, analyze the scalability of the approaches to the largest size mirror modules will be assessed. A selection between the two approaches will then be made, and a simulated FMA module will be constructed. Several segment pairs will be coaligned, and the remainder of the module will be populated by segment simulators. X-ray testing of the configuration will demonstrate 4 arcsec HPD angular resolution and effective area in agreement with predictions. This will be a high fidelity scalable prototype achieving critical performance under realistic conditions, in agreement with predictions, achieving TRL 6 in January 2012. As part of meeting TRL 6, the largest (3.2 m diameter), and thus the most challenging, mirror pair will slumped, aligned, and mounted.

### 3.1.1.2. Silicon Pore Optic

The SPO Technology Development Plan (see Supplementary Documents described in **Appendix D**), discusses the development required to achieve TRL 6. This plan focuses on assembly and alignment of the pore optics, mounting the stacks into a petal, developing mass production techniques, and "ruggedization" of the assembly to meet vibro-acoustic qualification loads.

Particulate contamination during stacking creates local deformations that propagate as plates are added. To address this, four solutions are being implemented: (1) the assembly environment is being kept cleaner and includes particle counting during assembly; (2) the Si plates are carefully cleaned prior to stacking; (3) an improved, cleaner, assembly robot is being used; and (4) in-situ particle detection and removal has been in-troduced into the stacking process. X-ray testing planned for later this year will test the efficacy of these improvements. Stacking and alignment development will also focus on building stacks for the inner part of a petal. This necessitates a smaller cylindrical radius of curvature, requiring more bending of the plates to the mandrel in building up a stack.

TRL 4 was achieved early 2009, at the conclusion of the high-resolution pore optics development activity, at which point a pair of aligned stacks was assembled that is capable of meeting X-ray performance requirements. TRL 5 is reached in mid 2011, when an SPO module will be compatible with environmental requirements has been constructed, taking advantage of materials studies and manufacturing improvements, marking the end of the first phase of the ruggedization activity. TRL 6 will be reached in early 2012, at the conclusion of the ruggedization and environmental testing activity. At that point a pair of aligned stacks meeting all performance and environmental requirements will have been demonstrated.

### 3.1.1.3. Critical technology and mission sensitivity (Q2)

The critical aspects of the FMA enabling technologies are the PSF and effective area. Mission performance sensitivity to not realizing the technology is generally a graceful degradation of performance. For example, a reduction in effective area will result in a longer observation being required to achieve the same signal to noise. The consequence of a degraded FMA PSF is mitigated by the influence of the rest of the telescope. This is shown in Fig. 3-3, where the total system PSF is plotted as a function of the FMA PSF. From the figure we see that the change in total PSF is slower than linear with respect to FMA performance. The confusion limit scales approximately as the cube of the PSF, and background scales as the square of the PSF. In addition, changes in the PSF result in changes in the XGS spectral resolving power. This is shown in Fig. 3-4, where spectral resolving power is plotted as a function of FMA PSF. Here, too, there is a gradual change in resolving power that is slower than linear with FMA PSF. Resolving power changes by only ~ 3% in response to a change in FMA PSF of ~ 32%.

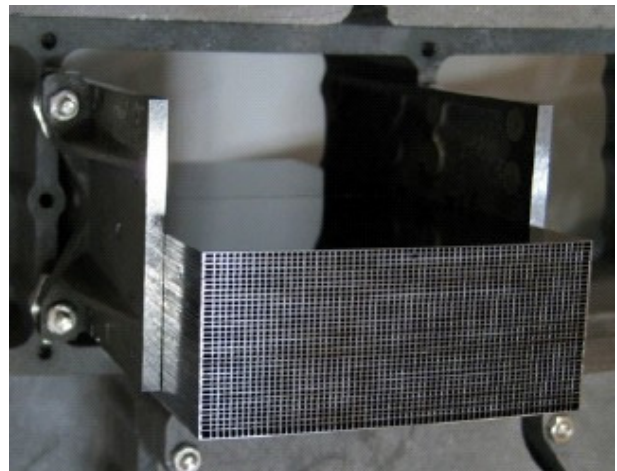

*Figure 3-2. Mounted SPO mirror*

Detailed science impacts of requirement sensitivity are covered in Section 1.

### 3.1.2  X-ray Microcalorimeter Spectrometer

The enabling XMS technology challenge lies in producing full arrays with readout electronics (multiplexing) to achieve the spectral resolution over the XMS FOV requirements of 5 × 5 arcmin.

The XMS uses a 4 kilopixel TES array with 1600 pixels in the core array, plus an additional 2,304 pixels in the surrounding outer array (see Fig. 2-3). 32 × 32 arrays and SQUID multiplexer readouts are being fabricated to assess energy resolution performance and quantify noise budgets. Work is underway to improve heat sinking so that the energy resolution requirement is met with high uniformity, and to implement readout of multiple absorbers by a single TES for the outer array (Smith et al. 2008).

Two approaches are under development for the readout. A prototype Time-Division-Multiplexing (TDM) SQUID readout system has been successfully tested. Two columns of 8 TESs were read out, and an average resolution of 2.9 eV was achieved with very high uniformity, ± 0.02 eV (Kilbourne et al. 2008). An alternate approach, Frequency Division Multiplexing with base-band feedback, will be demonstrated later in 2009.

Multiplexing 32 rows of TES pixels while maintaining the required energy resolution entails two improvements to the TDM system: lower SQUID noise and faster switching speed. Reduction in SQUID noise will be achieved by better heat sinking of the multiplexer chip and by implementation of a new generation of quieter series-array SQUIDs placed on the 50 mK stage. Increasing the row-switching speed by a factor of four will allow 32 rows to be read out, with each pixel being sampled at the same rate as currently used to sample eight rows. These advances will be incorporated into a readout system capable of multiplexing three columns of 32 pixels. Environmental tests will be performed by the end of 2009, bringing the core array technology to TRL 5. A 6 × 32 readout of both the core and extended arrays will be accomplished by the mid-2013 (TRL 6). The system will be verified at counting rates up to 200 counts/s per pixel.

The XMS requires a mechanical cryocooler to cool to below 5 K and a Continuous Adiabatic Demagnetization Refrigerator (CADR) to cool from below 5 K to ~50 mK. A five-stage CADR is baselined. Flight-qualified single stage CADRs have operated at temperatures as low as 35 mK. A full-scale four-stage CADR breadboard has demonstrated cooling to 50 mK with a 5 K heat sink, thereby achieving TRL 5. The additional fifth stage provides a stable stage at 1K for heat sinking wires and a thermal shield.

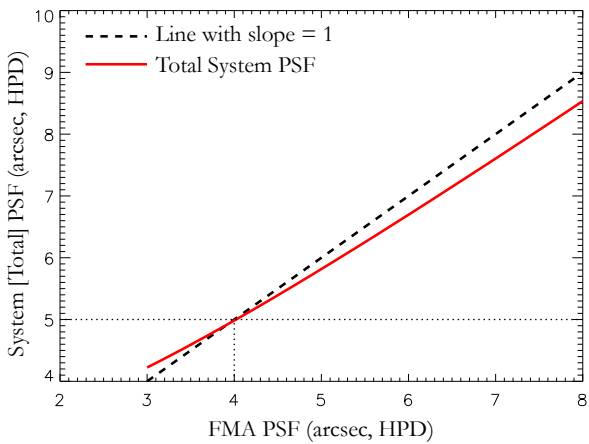

*Figure 3-3. The system PSF grows more slowly than linear as a function of FMA PSF.*

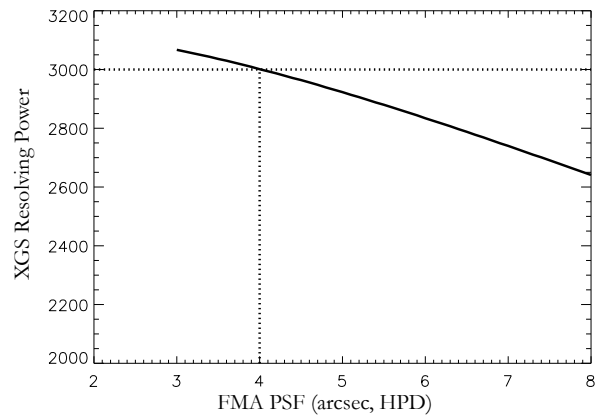

*Figure 3-4. The XGS resolving power decreases much slower than linear as a function of the FMA PSF.*

The next steps are to provide higher-resolution temperature readout electronics to improve overall thermal stability of the base temperature, and then fabricate and environmentally test the five-stage system (TRL 6).

Multi-stage mechanical cryocooler technology has reached a high level of maturity and is being implemented on several flight missions. The JWST/MIRI instrument uses a mechanical cryocooler that can meet the requirements for the XMS and has already achieved TRL 6.

### 3.1.2.1. Critical technology and mission sensitivity (Q2)

The critical aspects of the XMS technology are the detector FOV and the spectral resolution. Since spectral resolution scales inversely with pixel size and since multiplexer degradation scales with the number of detectors multiplexed, FOV can be traded for improved spectral resolution. Thus, any limitations in spectral resolution can ultimately be realized as a reduced FOV. The resulting consequence to mission performance would be a graceful degradation for extended sources. The consequence for mission science of reduced energy resolution and/or FOV are discussed further in Section 1.

### 3.1.3 Wide Field Imager/Hard X-ray Imager

Separate technology development plans for achieving TRL 6 exist for the WFI and the HXI.

**WFI:** For the DEPFET sensor, WFI prototypes (128 × 512 with 75 μm pixels) have been produced; operation is expected in late 2009. Operation and characterization of these high-fidelity breadboard systems demonstrates TRL 5 (early 2011). Large physical area prototype devices (512 × 512 pixels, 100 μm) are scheduled to go into production in September 2009, and are expected to be available by end 2010. This production run serves as a yield determination study for large area devices. This study and radiation hardness studies on large area detectors will identify and address critical scaling issues, leading to TRL 6 (2012).

Analog electronics will be developed along several lines. Two parallel technologies, VLSI Electronic for Astronomy (VELA) and Active current Switching Technique ReadOut In X-ray spectroscopy with DEPFET (ASTEROID), are currently at TRL 4 having undergone proof of concept testing in a laboratory environment. This development is being undertaken because their thinner gate structure makes them intrinsically radiation hard. In addition, they offer two to five times faster processing speeds than previous technology. TRL 5 is planned for 2010 with the production of ASTEROID for the MIXS flight. Finally, TRL 6 will be achieved in 2011 after a long term stability study, radiation hardness characterization, and speed improvements.

**HXI:** All major components are currently at TRL 4. All systems have been operated in a laboratory environment, with both the CdTe and DSSD detectors achieving an energy resolution of 1.2 keV (vs. the requirement of 1 keV) at 30 keV. Each of these will achieve TRL 5 (2011) based on enviromental performance testing of the 2.5 cm wide CdTe-imager on the ASTRO-H engineering model, and TRL 6 (2012) with a 3.2-cm-wide CdTe-imager onboard the ASTRO-H satellite.

### 3.1.3.1. Critical technology and mission sensitivity (Q2)

The primary critical aspect of the WFI/HXI technology is the wide 18 arcmin FOV. A reduction in the FOV can be compensated for by mosaicking observations, although this would result in a corresponding increase in observation times. Science impacts are discussed in Section 1.

### 3.1.4 X-ray Grating Spectrometer (XGS)

With the current level of development, both the CAT and OPG XGS gratings are TRL 3; each has a technology development plan to achieve TRL 6.

**CAT**: Grating facet sizes must be increased to achieve the required throughput efficiency. Accomplishing this size increase results in TRL 4. To maximize grating efficiency over a grating module area the ratio of the grating "open area" to support area must be increased. These efforts will al-low the effective area requirements to be met, and demonstrate TRL 5 (2012). Successful X-ray testing after environmental testing of the larger size, higher aspect ratio gratings in a realistic mounting structure will result in TRL 6 in 2013.

**OPG**: The ruling density and the ruling of radial patterns (the rulings must converge at the system focus), simultaneous with grating efficiency, remain to be demonstrated. Flatness, alignment, and mounting of the grating plates also require improvement. X-ray testing of the resolving power and efficiency of a grating with a flight-like groove profile will demonstrate TRL 4 in late 2009.

A flight-like grating on a Be substrate will be mounted into a flight-like module to provide a medium fidelity subsystem capable of demonstrating overall performance with realistic support elements, bringing the TRL to 5 in early 2011.

Two developments are required to achieve TRL 6: development of a method to efficiently reproduce high quality gratings onto flight substrates, and mounting of these gratings into a single flight-like module to verify the alignment strategy. Five of these replicated gratings spaced across a module will be adequate to ensure proper alignment. The remaining slots will be filled with mass simulators. The aligned module will undergo environmental tests, X-ray efficiency tests and spectral resolution tests, resulting in the demonstration of TRL 6 (2012).

### 3.1.4.1.  Critical technology and mission sensitivity (Q2)

Mission performance if the technology is not fully realized degrades gracefully. Reductions in either the grating effective area or resolution will reduce the signal-to-noise for the Cosmic Web observations, which can be compensated for with longer observing times. These impacts are discussed more fully in Section 1.

### 3.1.5    X-ray Polarimeter

The XPOL technology maturation plan is to: (a) modify ASIC to increase system speed to meet the deadtime requirement for bright sources, (b) complete radiation hardness testing, and (c) test the gas pixel detector with the ASIC before and after environmental testing. This development and testing will achieve TRL 6 in 2012.

### 3.1.5.1.  Critical technology and mission sensitivity (Q2)

The critical aspect of the enabling technology is the instrument deadtime for high count rates. The sensitivity of mission performance would be a limitation on the brightest sources that can be observed. However, this is mitigated by longer observations using the neutral density filter in the XPOL filterwheel.

### 3.1.6    High Time Resolution Spectrometer

Listed here for completeness, the High Time Resolution Spectrometer (HTRS) is at TRL 6.

## 3.2  Non-US Technology

*Please indicate any non US technology that is required for mission success and what back up plans would be required if only US participation occurred.*

The IXO mission as currently envisioned is predicated upon a close international collaboration between NASA, ESA, JAXA and European member states. Non-US technology required for the IXO mission success is limited to the WFI detector. In the highly unlikely circumstance that no foreign participation is forthcoming, then a US-led replacement consisting of an active pixel sensor or CCD array is a viable alternative.

## 3.3  Cost and Schedule for Enabling Technology

*Q3. Provide specific cost and schedule assumptions by year for Pre-Phase A and Phase A efforts that allow the technology to be ready when required.*

Cost estimates by year for pre-Phase A and Phase A technology development of enabling technologies are provided in Table 6-4 (See Section 6, "Costs"). These cost estimates were generated by technology developers using grassroots methodology. Estimates were generated in FY09 fixed year currency and inflated according to NASA New Start Inflation Index. Reserves, allocated on a technology by technology basis, are shown in total amount in Table 6-4. The phasing of the technology funding is consistent with the technology maturation plans.

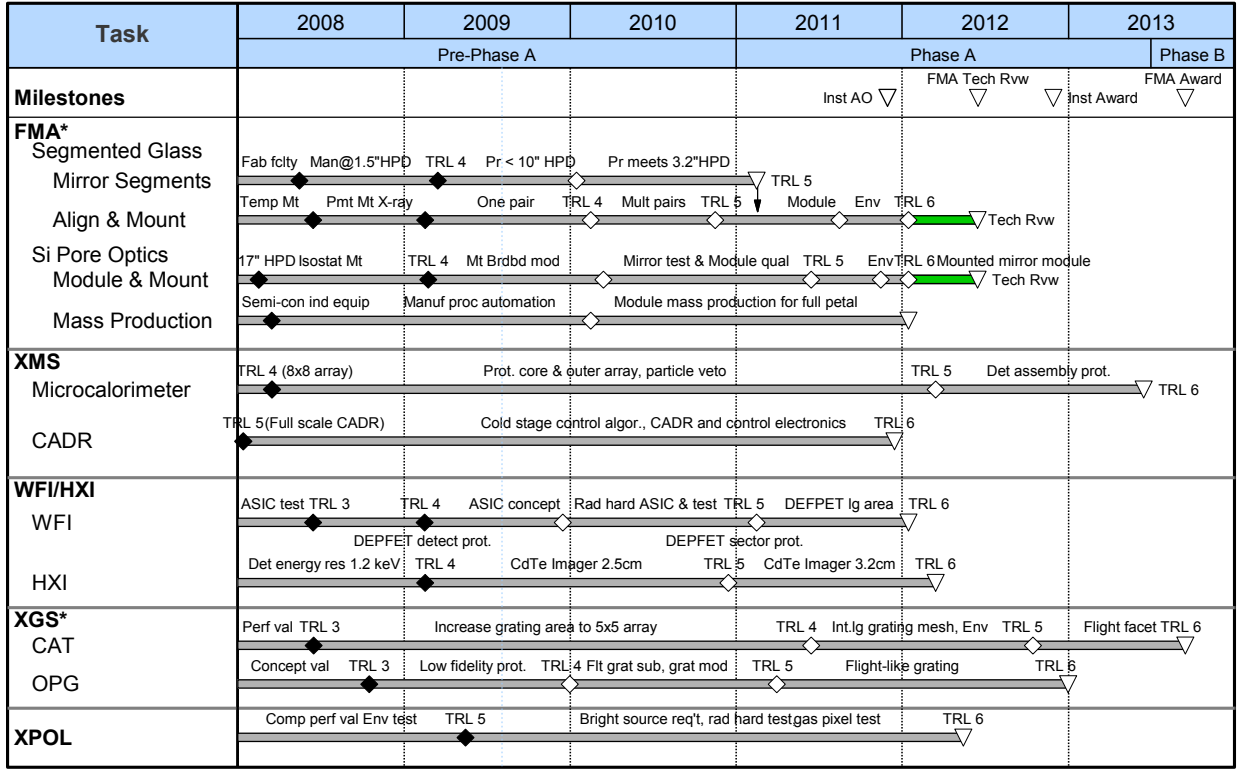

*Figure 3-5. Technology maturation summary schedule.*

The funding requirements shown assume that additional leveraged funding is applied toward the technology development milestones. For example, HXI technology development is leveraging CdTe and DSSD development from Astro-H. The WFI ASIC technology leverages development from the MIXS on Beppi Columbo.

The key technology activities and TRL milestones are summarized in the schedule in Fig. 3-5. Accomplishment of the technology activities is dependent on receipt of funds. An explicit funded schedule reserve of four months is included on the FMA technology development activities. The remaining technologies are not on the critical path; schedule reserve has not been called out for these. However, funding reserve is available to accommodate schedule growth.

# 4. Mission Operations Development

## 4.1 Mission Operations Overview

***Q1. Provide a brief description of mission operations, aimed at communicating the overall complexity of the ground operations. If the NASA DSN network will be used provide time required per week as well as the number of weeks (timeline) required for the mission.***

Mission operations are conducted from the IXO Science and Operations Center (ISOC) that will provide the command and control, mission planning, and the science processing, archive, and user support functions for the mission (Fig. 4-1 and Table 4-1). IXO is to be launched on either an EELV or Ariane into an L2 orbit, and the remainder of the five-year prime operations phase. The orbit at L2 is chosen for high target visibility and ensures no Earth and only minor lunar eclipses of the Sun, simplifying thermal and power management. The resulting observing efficiency (85%) is higher than any previous X-ray observatory (~70% Chandra, ~60% XMM).

During routine operations, the flight team at the ISOC contacts the spacecraft once per day for 30 minutes to perform a health and safety check and dump recorded telemetry. Command loads for future observations will be uplinked as needed (~once per week). On-board storage is sized to allow for two days of missed contacts, assuming worst case data volumes. Routine contact activity can be performed while observing, including S-band Telemetry Tracking & Control (TT&C), ranging, and Ka-band HGA downlink. The goal is to perform routine operations with a single shift (compared with Chandra's three), with staff on call for anomaly resolution, high solar radiation, and implementing target of opportunity observations. The IXO flight software is planned with a set of standard operating modes (includ-ing normal pointing, maneuver, sun hold, and safemode), four science observing modes, ~2,000 separate commands and a telemetry stream that will include ~2,500 telemetry points (with safety limits) for monitoring and trending. The requirements described above do not present significant drivers for the ground system and can be met with existing multi-mission or a standard Government off-the-shelf/commercial off-the-shelf (GOTS/COTS) command and telemetry systems.

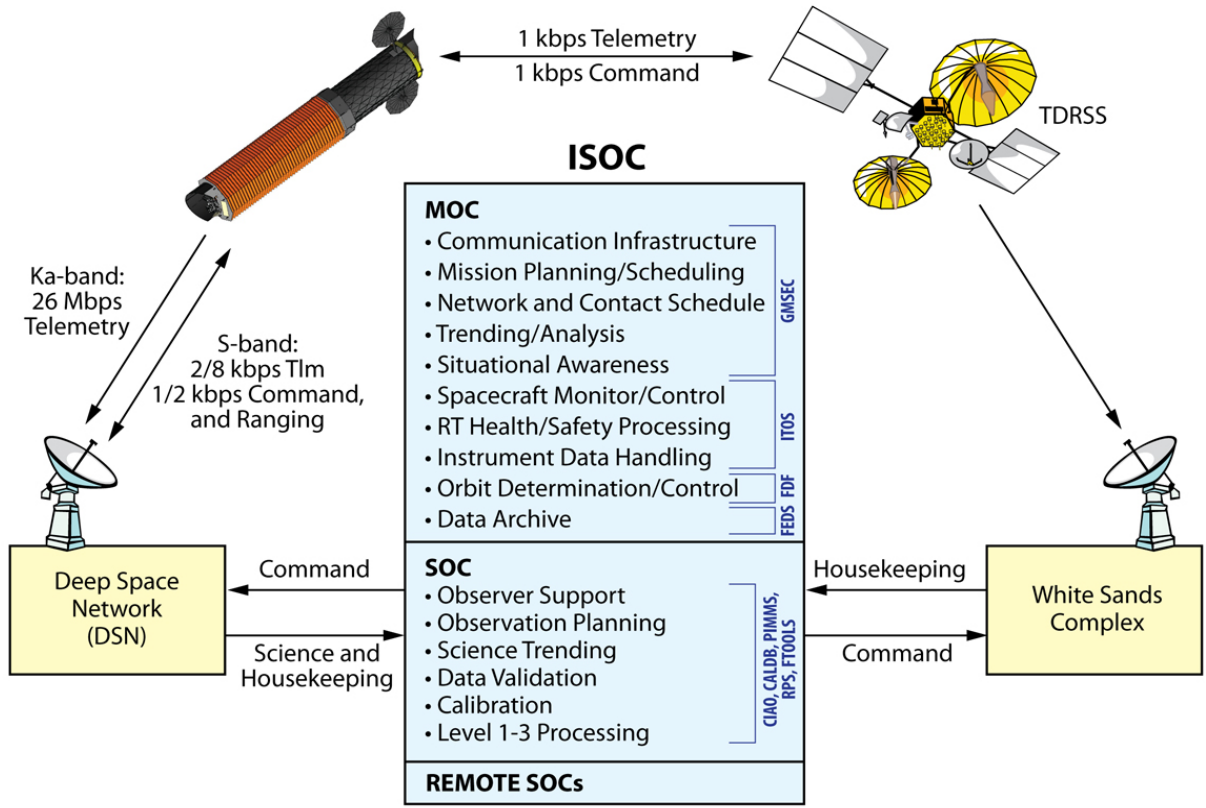

*Figure 4-1. IXO Operations Concept showing key functions of the IXO Science and Operations Center (ISOC). GOTS/COTS packages and their functionalities are shown for the MOC. Re-use of Chandra packages are shown for the SOC.*

The observing schedule and subsequent pointing profile for IXO is expected to be similar to Chandra with ~1,000 maneuvers per year (~3 times per day on average) to observe an average of ~800 science targets per year, reaction wheel unloading burns as required (~1–2 per week), and orbit station-keeping burns ~3 weeks apart (performed during slews). Wheel unloads and station-keeping burns are performed during real-time DSN contacts. Solar pressure offloading takes place via small impulses approximately every 18 minutes. The command sequences for the spacecraft will be automated where possible to incorporate momentum dumps and high gain antenna slewing activities based on the DSN contact schedule. Based on the field of regard (Fig. 2-13), the mission planning flexibility will be comparable with XMM-Newton and slightly less than Chandra. IXO will, however, have an increased efficiency due to the absence of radiation belt transitions.

The IXO DSN requirements call for use of the 34 m network with 2 × 12 hour tracks per day (24-hour coverage) for 28 days covering the launch, ascent, and orbital activation and checkout phase. This is followed by 2 × 30 minute tracks per day during the remainder of the 100 day cruise and injection to L2. During the balance of the five year operations phase, 1 × 30 minute track per day and 1 × 3 hour track approximately every two weeks is required (to downlink bright source data). In addition to the DSN requirements, short duration TDRSS coverage is planned during the launch vehicle separation event approximately 30 minutes after launch.

## 4.2 Special Ground Support Requirements

*Q2. Identify any unusual constraints or special communications, tracking, or near real-time ground support requirements.*

IXO has no unusual constraints for tracking or near real-time ground support. As described above, IXO requires standard DSN coverage during launch, activation, cruise, and injection to L2. The real-time TDRSS coverage during separation is also standard, and the DSN coverage during normal science operations is quite modest at ~25% of the Chandra requirement. The orbit station keeping and other L2 orbital characteristics do not require special coverage and the science instrument operation can be accommodated within the routine track requirements for the science operations phase. As with other DSN missions, additional contacts will be required for anomaly resolution and will be obtained through the standard scheduling process with JPL.

## 4.3 Operational Constraints

*Q3. Identify any unusual or especially challenging operational constraints (i.e., viewing or pointing requirements).*

IXO has no unusual or especially challenging operational constraints. IXO will be injected into a 800,000 km 180 day L2 orbit with an Earth distance of between 1.5 and 1.8 million km, and an Earth-Sun angle of between 7 and 30 deg (essentially identical to the JWST orbit). This orbit is free of earth shadows and has occasional low obscuration (<14%) lunar penumbras. The orbit imposes only modest station keeping and momentum management requirements for IXO.

## 4.4 Science and Data Products

*Q4. Describe science and data products in sufficient detail that Phase E costs can be understood compared to the level of effort described in this section.*

The IXO data processing system is composed of a suite of pipelines utilized to reduce IXO data, starting from raw spacecraft telemetry. The steps include deriving and applying the aspect solution to remove the spacecraft motion, applying stan-dard calibrations to the data, computing good data intervals and applying filters to remove poor quality data, producing the photon event lists, and performing data analysis to detect sources, extract source spectra, and other source properties. This approach has been used on many missions, starting with Einstein, continuing with ROSAT, ASCA, RXTE, etc. and is currently in use for Chandra, XMM-Newton, Suzaku, and Swift. Data processing for the five instruments proceeds through a set of standard levels that include:

**Level 0**: Raw spacecraft telemetry is decommutated and split into functionally independent parallel streams formatted as binary FITS files;

**Level 1**: Applies the aspect correction to locate photons on the sky, instrument-specific calibrations such as detector gain, and good time intervals, to produce photon lists;

**Level 1B:** Data from multiple observation intervals that constitute an observation are combined to create merged event files, images, and other observation-level data products. Images are generated for WFI and HXI observations; dispersed spectra are extracted from XGS observations, spatial/spectral data cubes are generated for XMS observations, photon lists for HTRS, and polarization angle and fraction for XPOL.

**Level 2**: Standard data analysis tools are applied to extract per-observation source and spectral properties such as source position and basic properties, source spectra and lines, temporal variability, and polarization parameters. Products include postage stamp images, spectral line lists, and tables of properties.

**Level 3:** Combine Level 2 properties to produce catalogs and mosaics.

The IXO raw data rate of 199 kbps daily average (195 kbps science data, 4 kbps spacecraft housekeeping) and ~3,300 kbps peak rate twice per month for 12 hours produces ~1 TB of telemetry per year. Based on experience with Chandra and XMM-Newton, we estimate an expansion from raw telemetry of a factor ~3 (compressed) in the processed data products, yielding ~3 TB/year of data or ~20 TB over five years (with 30% margin). With periodic reprocessing of all data increasing the volume by a factor of two, the IXO archive is sized to be ~40 TB after five years.

## 4.5 Science and Operations Center

*Q5. Describe the science and operations center for the activity: will an existing center be expected to operate this activity?; how many distinct investigations will use the facility?; will there be a guest observer program?; will investigators be funded directly by the activity?*

The ISOC provides the mission and science operations activities for the mission. All ISOC activities are conducted in a single facility to reduce cost and increase synergy. The present plan is to utilize the facilities of the Chandra X-ray Center for IXO. ESA and JAXA will operate mirror science centers for data analysis and distribution. We assume here an architecture where the operations will be conducted from the US, however, IXO could be operated by ESA as determined by agreement between the agencies.

The ISOC conducts the annual peer review on behalf of NASA HQ, performs mission planning, flight operations, data receipt and monitoring, science processing, archiving, data distribution, provides analysis tools to users, maintains the observatory and instrument calibration, and administers the grant and EPO programs. The ISOC consists of the Mission Data System (MDS), and the Science Data System (SDS).

The MDS provides the data capture, command and control, telemetry processing, monitoring and flight mission planning functions for IXO. These functions can be provided by a standard GOTS/COTS command and telemetry system (e.g., GMSEC, ITOS/ASIST and FEDS), or the Chandra ground system.[1] For costing, we baselined the Chandra system with customization as required for the data capture, telemetry processing, and command and telemetry database software. Prior to CDR, a trade study will be conducted between in-house customization and available commercial command and control systems to ensure the most cost effective solution.

The SDS provides the science observation planning, instrument operation, science data processing, archive, analysis tools, and calibration functions for IXO. The SDS benefits from significant reuse of the Chandra science data system which consists of a pipeline data processing sys-tem, archive, analysis tool suite, proposal submission system, and supporting infrastructure.

---

[1] Chandra is designed as a multi-mission ground system.

The ISOC will annually support ~200 peer reviewed guest observer science programs comprised of ~800 targets per year. A grant program will be administered by the ISOC that provides awards to US-based PIs of winning observing and archive proposals. Based on Chandra and XMM usage, we expect more than 1000 unique investigators each year. The program is planned to operate in a similar fashion to the Chandra grants program. ESA and JAXA will support their investigators separately.

## 4.6 Data Archive

*Q6. Will the activity need and support a data archive?*

The ISOC will need and support an active data archive with capabilities comparable to the Chandra and XMM-Newton archives. These include making data available through a convenient web interface, full search capabilities, handling of proprietary periods, provision of multiple levels of data products, calibration products (e.g., CALDB), handling of reprocessing and provision of Level 3 catalogs once available. All data will also be stored in the permanent archive available through NASA's High Energy Astrophysics Science Archive Research Center (HEASARC).

**Table 4-1.  Mission Operations and Ground Data Systems Table**

| Downlink Information | Value, units |
| --- | --- |
| Number of Contacts per Day | 1 |
| Downlink Frequency Band, GHz | S-band: 2.025 – 2.120 GHz<br>Ka-band: 31.8 – 32.3 GHz |
| Telemetry Data Rate(s), bps | S-band: 2000/8000 bps<br>Ka-band: 26 Mbps |
| S/C Transmitting Antenna Type(s) and Gain(s), DBi | LGA omni: 1 dBi<br>HGA: 44 dBi (Ka-band) |
| Spacecraft transmitter peak power, watts. | 30 W (Ka-band) |
| Downlink Receiving Antenna Gain, DBi | 78 dBi (Ka-Band) |
| Transmitting Power Amplifier Output, watts | 10 W |
| Uplink Information | Value, units |
| Number of Uplinks per Day | 1 |
| Uplink Frequency Band, GHz | 2.2 – 2.3 GHz |
| Telecommand Data Rate, bps | 1000/2000 bps |
| S/C Receiving Antenna Type(s) and Gain(s), DBi | LGA omni: 1 dBi<br>HGA: 20 dBi (S-band) |

# 5. Programmatics and Schedule

## 5.1 Organization

*Q1. Provide an organizational chart showing how key members and organizations will work together to implement the program.*

IXO mission will be implemented in partnership between the National Aeronautics and Space Administration (NASA), the European Space Agency (ESA), and the Japan Aerospace and Exploration Agency (JAXA). All three agencies have extensive expertise and experience in space science missions, spaceflight system development, and X-ray instrumentation. Successful international collaborations on past X-ray astrophysical missions have included XMM-Newton (ESA-NASA), ASCA and Suzaku (JAXA-NASA), and Chandra (NASA-Netherlands). The IXO response to the RFI#1 summarized the current, Pre-Phase A, international organization for the IXO international team.

The mission roles and responsibilities for IXO implementation will be finalized by NASA, ESA, and JAXA during Phase A based on the output of the mirror technology review, mission studies, and programmatic considerations. In the scenario presented in this RFI, NASA leads the mission including the Spacecraft and Deployment Modules, the Optics Module with the Flight Mirror Assembly, the XMS, the XGS, and science and mission operations. NASA will also lead mission system engineering and mission integration and test. ESA will provide the Instrument Module and launch services. The European member states, through ESA, will provide the WFI/HXI, HTRS, and XPOL instruments, and contribute to the XMS. JAXA will contribute the Hard X-ray Imaging detector system to the WFI/HXI, the Hard X-ray Mirror Module to the FMA, and portions of the Deployment Module. The agencies will share science operations and data analysis.

The modular nature of IXO provides clearly definable interfaces to facilitate sharing the development between international partners. Either NASA or ESA may be selected to lead the Flight Mirror Assembly (FMA), based on technical, cost, and schedule considerations at the time of the technology review in mid-2012. Selection of instrument providers will be through a coordinated multi-agency Announcement of Opportunity (AO).

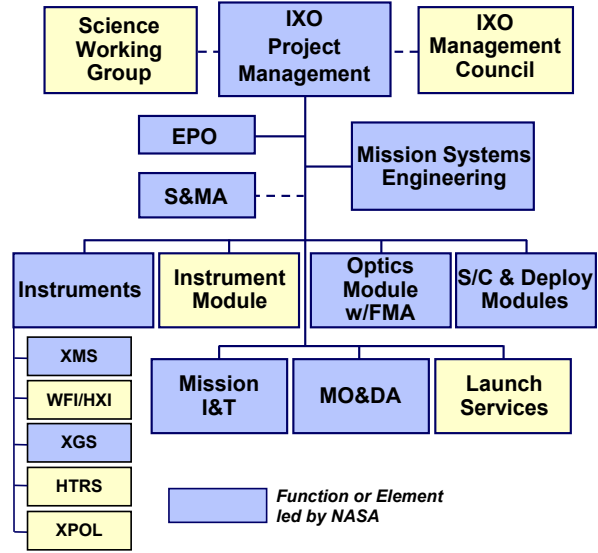

*Figure 5-1. IXO Project Organization for Mission Development (Phase B/C/D), consistent with NASA leadership of the mission, as presented in this submission for RFI#2.*

There have been many successful examples of international cooperation in developing X-ray instruments, e.g., the LETG on Chandra, the RGS on XMM-Newton, the XIS and XRT on Suzaku, and the XRT on Swift. Within ESA it is typical for instruments to be multi-national consortia and there is heritage for JAXA/NASA collaborations on missions.

A summary-level project organization chart for mission development (Phases B-D) is shown in Fig. 5-1. The IXO Project Management will be staffed by the lead agency to provide overall management and integration of the mission elements. The mission lead agency will competitively select an industry partner as prime contractor for development of the spacecraft and observatory integration. Formal communications interfaces will be through the international partner organizations. Lessons learned on effective management and oversight of systems across international partners will be applied. In particular, systems engineers from across the mission flight and ground segments will be members of the mission system engineering team, and will work together to define interfaces, perform system level analyses, etc. En-gineers and managers across the mission, including all partner organizations, will also participate in requirement, design, and other key reviews at both the element and mission system level. Resources will be applied early in the program to manage issues related to International Traffic in Arms Regulations (ITAR), Export Administration Regulations (EAR) and establishing Technical Assistance Agreements. The key stakeholders across the international project team will form the IXO Program Management Council that will expedite resolution of issues across institutional boundaries. NASA's role in IXO will be managed by the Goddard Space Flight Center.

### 5.2 Top Risks to the Program

*Q2. Provide a table and a 5 by 5 risk chart of the top 8 risks to the program. Briefly describe how each of these risks will be mitigated and the impact if they are not.*

All risks for the program were developed by instrument technology developers and the project engineers and subsequently reviewed by the Project Management team. The top eight mission risks are summarized with risk description, mitigation, and impact, if not mitigated, in Table 5-1. The 5 by 5 risk chart for these risks is given in Fig. 5-2. The risk standard scale for consequence and likelihood is consistent with GSFC-STD-0002 Risk Management Reporting.

### 5.3 Mission Schedule

*Q3. Provide an overall (Phase A through Phase F) schedule highlighting key design reviews, the critical path and the development time for delivery required for each instrument, the spacecraft, development of ground and mission/science operations, etc.*

The overall IXO mission schedule is provided in Fig. 5-3. Milestones and key decision points, consistent with NASA Procedural Requirements 7120.5D, were used in the overall planning and are indicated along the top row of the schedule. The critical path is highlighted on the schedule as are planned start and delivery dates for all major elements.

The schedule supports a May 2021 launch readiness, with eight years and one month for de-velopment through on-orbit checkout (Phases B, C, and D). This includes a total of 10 months of funded schedule reserve on the critical path. Five years of mission operations after launch are nominally planned, with an option to extend the science mission to 10 years. Additional schedule reserve is held on the key technology development efforts for the mirror during Phase A, and flight development activities not on the critical path.

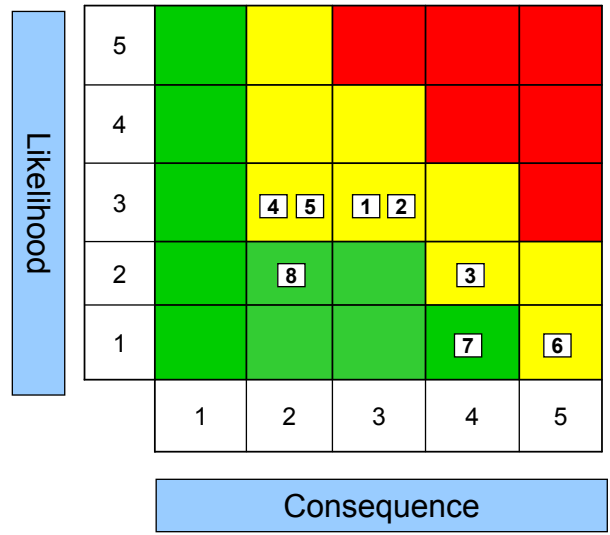

*Figure 5-2. Top 8 Risks to the Program on 5 by 5 matrix*

The schedule reflects the ability to capitalize on the modular nature of IXO. The four observatory modules (Optics Module, Instrument Module, Spacecraft Module, and Deployment Module) will be developed and qualified in parallel, and then delivered for final observatory Integration and Test (I&T). Time has been allocated as appropriate for each of the processes for solicitation, selection, and contract awards.

The critical path runs through the development of Flight Mirror Assembly (FMA) to Optics Module I&T to observatory I&T to the launch site operations and launch. A detailed FMA development schedule, based on the segmented glass mirror technology, is provided in **Appendix F.2**. Lessons learned from 1) establishing NuSTAR mirror segment facilities and mirror segment fabrication to date, 2) current IXO mirror technology development, and 3) Chandra and XMM-Newton mirror calibration and test effort, have been incorporated into the FMA schedule durations and flow. The overall schedule duration for the FMA using silicon pore optics is expected to be comparable to that for the segmented glass. Observatory I&T reflects activities, flows, and durations that have been developed based on experience from other space observatories of comparable size and type with an emphasis from the Chandra development.

**Table 5-1.   Top 8 Risks to Program and Mitigation Plans**

| Rank | Risk Type | System | Risk Statement | Mitigation | Impact |
|---|---|---|---|---|---|
| 1 | Programmatic | FMA | If mirror build and test experiences significant delays, mission schedule margin will be eroded, resulting in launch delay. | Employ multiple sources and parallel development of mandrels, parallel lines for module assembly. Schedule margin, i.e., 10 months funded schedule slack on critical path. Modular nature of observatory minimizes impacts of FMA delays on the rest of the observatory. | Schedule - launch delay |
| 2 | Technical | Mission Systems | Given that the observatory is developed by an international consortium, there may be system level issues, such as interface incompatibility. | Provide full participation in reviews of all interfacing systems. International systems engineering team and IXO Management Council will resolve issues. Coordinate configuration management across partner organizations. ITAR agreements to allow information flow. Supply a thorough test and verification program with 10 months schedule slack on critical path. | Science - Possible degraded performance Schedule - launch delay |
| 3 | Technical | FMA | If required angular resolution is not achieved with either mirror technology, then the SMBH at high redshift science will be significantly compromised. | Use parallel technology development through TRL 6 using segmented glass and Si pore optic approaches, prior to start of Phase B. Build and test an additional engineering unit prior to CDR. Thoroughly test the mirror through all stages of assembly. | Science - Compromise investigation of SMBH evolution |
| 4 | Technical | Inst - XMS | If the XMS cryogenic chain doesn't have sufficient reliability, then XMS cooling lifetime may not be achieved. | Provide thorough reliability analyses based on existing hardware test and on heritage orbit data. Use life testing as appropriate. Add redundancy to the design. | Mass & Power - Increased mass ( 60 kg) and power (300 W) for redundant system |
| 5 | Technical | Inst - XGS | If grating array throughput efficiency doesn't meet requirements, additional grating area coverage will be required. | Use parallel technology development. If necessary, add moveable (flip-up) deployment grating to remove grating area during observations where XGS is not required. | Cost - Increased grating size and/ or additional mechanism |
| 6 | Technical | Spacecraft | If the mirror covers fail to deploy then no X-rays will pass through the mirror resulting in loss of mission. If the Deployment Module does not fully deploy the focal length will not be achieved. | Mirror cover deployment has a single fault tolerant heritage design with proven industry standard redundant actuators, and an extensive ground qualification program. The ADAM Mast based deployment mechanism includes redundant actuators that are retractable and fault tolerant, based on a heritage design that flew successfully. The mast will have an extensive qualification program and end-to-end testing. | Science - Loss of mission |
| 7 | Technical | Spacecraft | If operation of the MIP is impeded, then it may not be possible to switch between the focal plane instruments. | MIP design includes redundant mechanism that is single fault tolerant. The MIP will have an extensive qualification program and end-to-end testing in test-as-you-fly configuration. | Science - Loss of ability to switch instruments |
| 8 | Technical | Inst - XMS | If the 32x XMS multiplexer does not achieve the speed required to meet the spectral resolution, then the heat load on the cooling system will increase due to additional multiplexer chains. | Use parallel technology development of time division multiplexing and frequency division multiplexing. Slow down multiplexer to maintain required spectral resolution, and accept a lower maximum XMS count rate at full resolution. Provide ample margin on heat load in cooling chain design. | Science - Reduced energy resolution at high count rates |

*Figure 5-3. Top Level Mission Schedule*

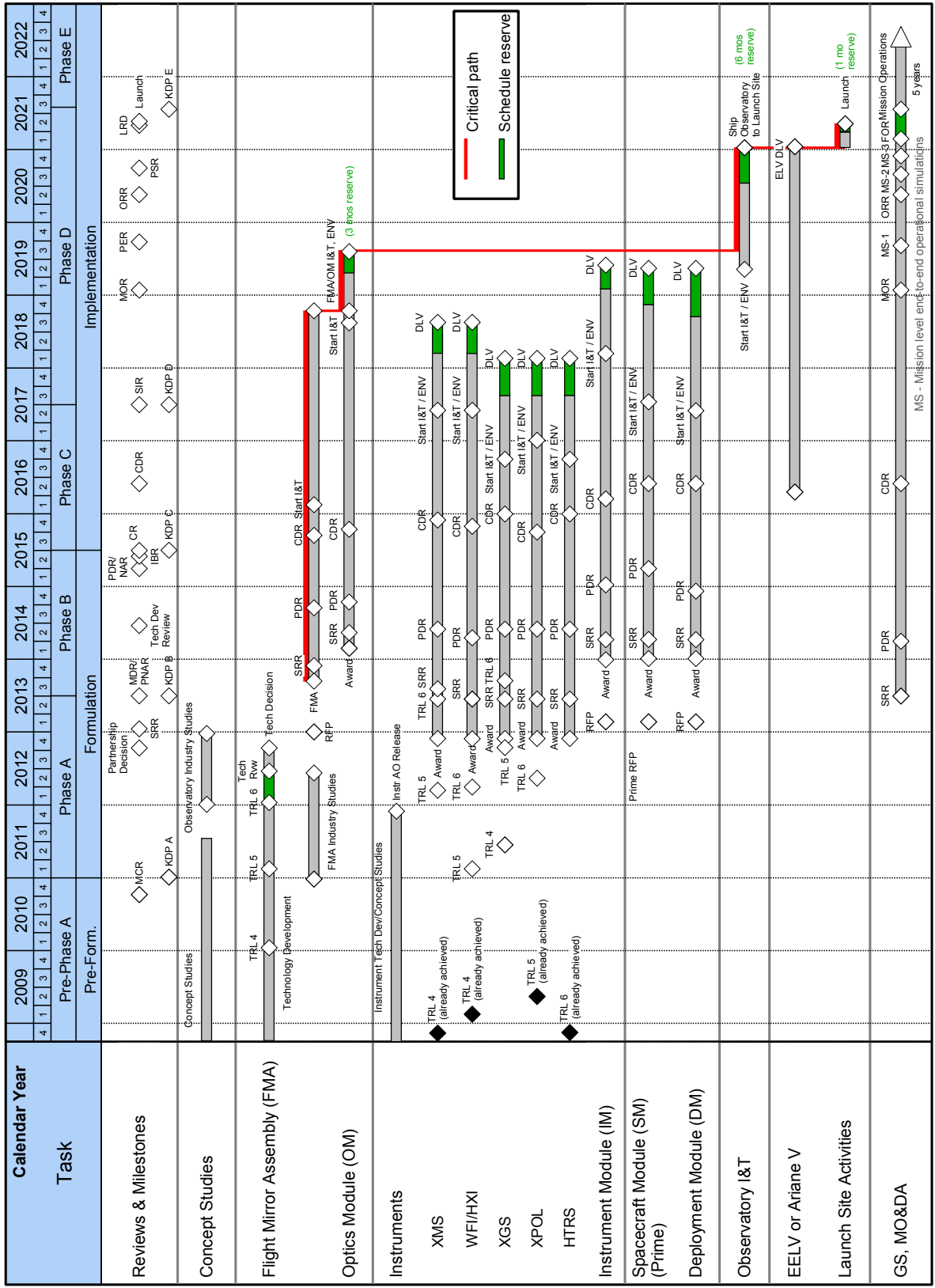

As mentioned above, a total of 10 months of schedule reserve has been allocated on the critical path in Phases B through D. Three months of schedule reserve is allocated to the FMA and OM development in Phases B/C/D. Six months reserve is held within the observatory I&T effort of 20 months. One additional month of reserve is held for launch site activities, which have a duration of almost four months. Schedule reserve is also allocated for activities that are not on the critical path, including the instruments, the Instrument Module, Spacecraft Module, Deployment Module, and Mission Operations and Ground System development activities as indicated in Fig. 5-1. The schedule reserve by phase meets the requirements in Goddard Procedural Requirement 7120.7.

In addition to the schedule reserve on the mission development discussed above, four months of reserve in Phase A are allocated between the mirror TRL 6 demonstrations and the mirror technology review, which takes place in mid-2012. This review initiates the technology selection process which concludes in late 2012 with NASA, ESA, and JAXA finalization of their responsibilities.

In 2012, parallel Phase A studies of the overall observatory implementation will be conducted by multiple industry contractors. Following these studies, after the agency roles and responsibilities have been defined, the lead agency will issue a Request for Proposals (RFP) to select the Observatory prime contractor, who will also provide the Spacecraft Module (SM).

The Announcement of Opportunity (AO) for science instrument teams will be released in late 2011 and instrument contracts awarded by the end of 2012. The instrument design reviews (Preliminary Design Review and Critical Design Review) will be timed to support instrument development and will occur prior to similar reviews at the mission level. All instruments will be fully tested and qualified prior to delivery to the In-strument Module (IM). After the instruments are integrated onto the IM, the IM will be tested and delivered for integration with the rest of the observatory in mid-2019. Completion of IM I&T with the instruments prior to final observatory I&T will provide early verification of the instruments in their flight assembly, reducing overall schedule risk.

Observatory I&T will commence with integration of the Deployment Module (DM) with the Spacecraft Module (SM). Integration of the IM and the Optics Module (OM) will be next, followed by observatory environmental and functional testing. A more detailed breakdown of the Spacecraft schedule is provided in Section 2.3

The development of the Ground System (GS), Mission Operations (MO), and science Data Analysis (DA) system will be tied to and will occur in parallel with the observatory development, with major reviews as indicated in Fig. 5-3. Mission level simulations, including the fully integrated observatory and ground systems, are also shown. Phase E duration is five years with the option to extend to 10 years.

***Q4. Fill out the Key Phase Duration table indicating the length of time required (months) for: each Phase (A through F), ATP to PDR, ATP to CDR, and other key metrics for schedule analysis***

The Key Phase Duration table is provided in Table 5-2.

***Q5. Fill out the Key Event Dates table indicating the dates (month/year) for the key development and operations milestones***

The Key Event Dates table is provided in Table 5-3.

**Table 5-2. Key Phase Duration Table**

| Project Phase | Duration (months) |
|---|---|
| Phase A - Conceptual Design | 30 |
| Phase B - Preliminary Design | 24 |
| Phase C - Detailed Design | 24 |
| Phase D - Integration and Test | 49 |
| Phase E - Primary Mission Operations | 60 |
| Phase E2 - Extended Mission Operations | 60 (Not Costed) |
| Start of Phase B to PDR | 21 |
| Start of Phase B to CDR | 35 |
| Start of Phase B to Delivery of FMA/OM | 73 |
| Start of Phase B to XMS | 61 |
| Start of Phase B to WFI/HXI | 61 |
| Start of Phase B to XGS | 55 |
| Start of Phase B to XPOL | 55 |
| Start of Phase B to HTRS | 55 |
| Start of Phase B to Delivery of Spacecraft | 72 |
| Start of Phase B to Delivery of Observatory | 90 |
| System Level Integration and Test | 20 |
| Project Total Funded Schedule Reserve | 10 |
| Total Development Time Phase B-D | 97 |

**Table 5-3. Key Event Dates**

| Project Phase | Milestone Date |
|---|---|
| Start of Phase A | January, 2011 |
| Start of Phase B | July, 2013 |
| Preliminary Design Review (PDR) | March, 2015 |
| Critical Design Review (CDR) | May, 2016 |
| Delivery of FMA/OM | August, 2019 |
| Delivery of XMS | August, 2018 |
| Delivery of WFI/HXI | August, 2018 |
| Delivery of XGS | February, 2018 |
| Delivery of XPOL | February, 2018 |
| Delivery of HTRS | February, 2018 |
| System Integration Review (SIR) | July, 2017 |
| Pre-Ship Review (PSR) | September, 2020 |
| Launch Readiness Date (LRD) | May, 2021 |
| End of Mission - Primary | August, 2026 |
| End of Mission - Extended (EOM-E) | August, 2031 (Not Costed) |

# 7. Changes Since the Previous NRC Recommendation

*Activities ranked in either the 2000 "Astronomy and Astrophysics in the New Millennium" survey or in the "Beyond Einstein Program Assessment Committee" should provide up to four (4) additional pages describing the changes in the activity science goals, technical implementation, and/or estimated cost since AANM and the most recent previous NRC report. We need to understand your explanation of changes that significantly affect the scientific return, the activity risk, and/or estimated cost of the activity, and the reasons for them.*

The science reach of IXO relative to other space and ground-based missions was assessed and prioritized in 2000 by the Astronomy and Astrophysics in the New Millennium (AANM) Survey. In this survey IXO—then known as Constellation-X or Con-X—was ranked as the second highest priority large space based facility (after JWST). The NRC Connecting Quarks to the Cosmos study in 2003 also assessed the capabilities of Con-X and called out the unique ability of the mission to address science at the intersection of astronomy and physics. The strategy laid out in both reports was reaffirmed by the mid-term review undertaken by the Committee on Astronomy and Astrophysics (reported in a letter to NASA HQ on 2005, Feb 11). The Beyond Einstein Program Assessment Committee (BEPAC) review in 2008 found that "the Constellation-X mission will make the broadest and most diverse contributions to astronomy of any of the candidate Beyond Einstein missions."

Over the past decade, the Con-X mission implementation has evolved as launch vehicle capabilities and costs changed, as technical progress brought new ideas and approaches into play, and, most importantly, by the decision to merge the NASA Con-X mission and the ESA/JAXA X-ray Evolving Universe Spectroscopy (XEUS) mission to form the International X-ray Observatory (IXO). The science goals and measurement requirements of IXO clearly reflect a merger of the two predecessor mission concepts, and thus are similar but not identical to the Con-X objectives and requirements presented to previous NRC panels. The basic science goals of Con-X were unchanged from AANM to BEPAC, but with the merger with XEUS, have expanded to include AGN at high-$z$, polarization studies, and high count rate science, all part of the initial XEUS science case.

The mission cost to NASA increased from AANM to BEPAC due to a decade of inflation, increases in launch vehicle cost, the inclusion of pre-Phase A, Phase A and B costs, and changes in NASA cost rules to include full cost accounting. Changes from BEPAC Con-X to Astro2010 IXO include increased science capabilities, restructuring of the mission configuration using a single optic, and a four-year launch delay. Since BEPAC, the project has performed higher-fidelity 70% CL cost analyses based on extensive mission and instrument concept studies (See **Appendix D**). These analyses are supported with both internal and independent validations. Cost-sharing with ESA, JAXA and ESA-member states reduces the overall cost to NASA from $3.3B to $1.8B.

The evolution of the implementation approach is greatly facilitated by the fact that the project adopted from the start a modular approach, where the basic parts can be assembled in different ways, in response to the launch vehicle capabilities available at any given time. The segmented design allows the mirror to be assembled in different ways, either as a single unit, or multiple smaller diameter mirrors with independent focal planes for which the data are combined after the fact on the ground. This modular approach has made the mission robust to the changing launcher landscape. Through the evolution of the mission concept from AANM to BEPAC the science drivers and measurement capabilities have been maintained. The recent merger with XEUS augments the capabilities and allows additional science.

The evolution from Constellation-X in AANM to IXO for Astro2010 is shown graphically in Fig. 7-1.

## 7.1 The AANM Constellation-X Mission

The mission concept submitted to the AANM was a multiple-telescope, multi-spacecraft approach. The required collecting area was achieved with a design that split the collecting area across four identical satellites to be carried in pairs on either two Atlas V or two Delta IV launchers. Each satellite carried a 1.6 m diameter spectroscopy X-ray telescope (SXT) with a 10 m focal length, each with a microcalorimeter array at the focus and a reflection grating spectrometer. The design was optimized to maintain spectral resolving power of at least 300 across the bandpass, and resolution of 2400 at the iron K line. The angular resolution requirement was 15 arcsec HPD, with a goal of 5 arcsec. Hard X-ray telescopes (HXTs) extended the bandpass up to 60 keV to provide constraints on the X-ray continuum. For this design there were a total of three HXTs per spacecraft, for a total of 12 individual HXTs.

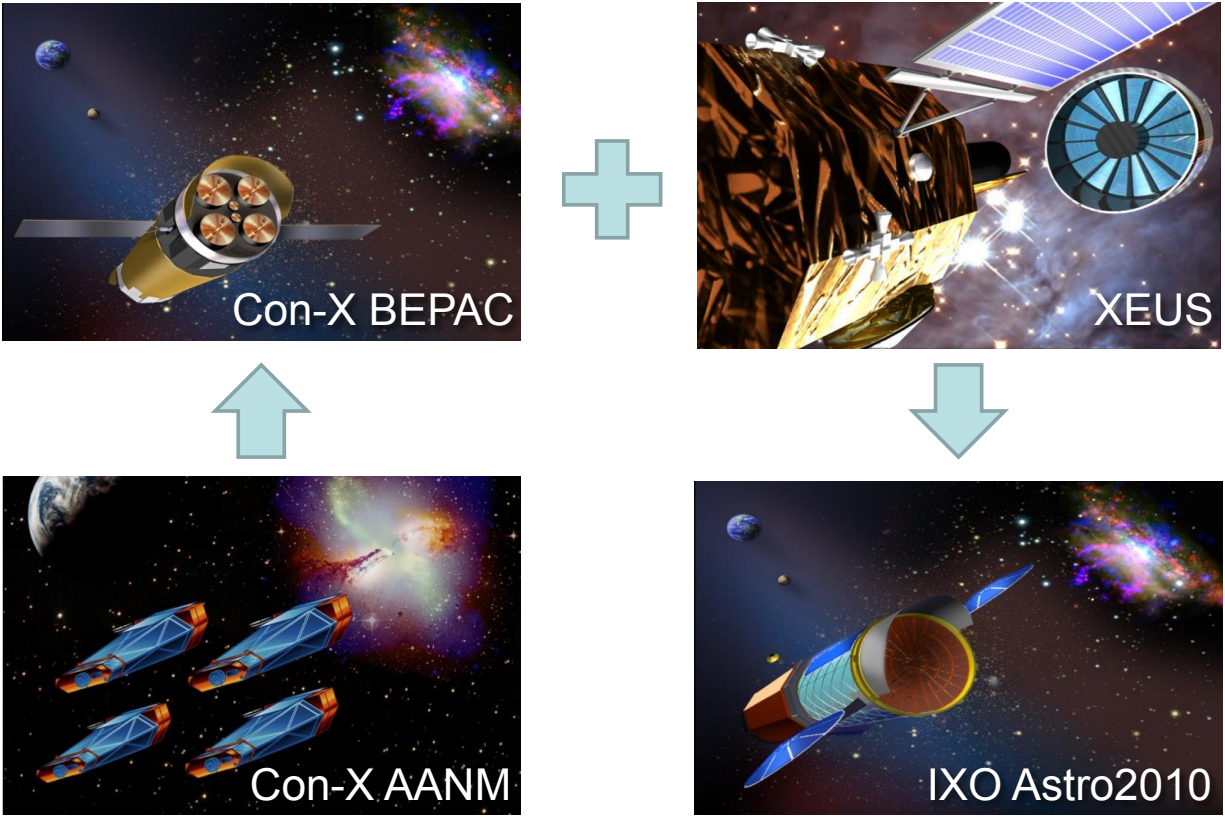

*Figure 7-1. Changes in the configuration of Con-X/IXO from AANM to Astro2010, clockwise from lower left, starting with the four satellite configuration for AANM, the single satellite/four telescope configuration for BEPAC, the XEUS formation flying configuration at the time of the merger, and the single telescope configuration for Astro2010.*

The mission costs presented to the AANM were based on a grass roots estimate assuming a launch in 2008. They were given in FY1999 fixed year dollars and included a four-year phase C/D, two launch vehicles and five years of operations, and came to $800M in total.

## 7.2 The BEPAC Constellation-X Mission

The cost of launch vehicles increased with time; with the advent of the EELV capability, a single spacecraft, single launch became the most cost effective mission implementation during the time of the BEPAC review in 2007. This implementation maintained 4 SXTs with the same 10 m focal length but slightly smaller 1.3 m diameter, each with identical microcalorimeter arrays at the focal plane. Effective area for the microcalorimeter was maintained by optimization of the mirror design, including use of higher reflectivity Ir coatings rather than Au and by placing grating spectrometers behind only two of the four mirrors. The grating effective area was maintained by increasing the fractional area covered by gratings for the two mirrors. The number of CCD arrays was also thus reduced from four to two. In the intervening time there had been great progress in the technology development of the HXTs and also a reassessment of the requirements. The number of HXTs went from twelve to two. A viable configuration with gratings on one mirror and a single HXT was also considered at the time of the BEPAC review.

The end-to-end mission cost (pre-phase A through five years of operations) was $1.7B in FY2007 fixed-year dollars and $2.2B in RY$ for a June 2017 launch. This project estimate was based on grassroots, with 30% contingency. The BEPAC independent cost assessment was $3.1B in RY$ with 70% confidence. There was limited opportunity to compare Project and BEPAC estimates and therefore no assessment or reconciliation of the differences.

### 7.3 The Astro2010 IXO mission

The science case for a large X-ray observatory had also been given priority in ESA's planning for their future programs. In the ESA Cosmic Visions planning, XEUS was selected in 2007 as one of three candidate Large Missions. The proposed XEUS mission was the result of about a 10-year study. There was considerable overlap in the science goals of XEUS and those of Con-X, but also some significant differences. The Con-X mission science emphasized spectroscopy with instrumentation dedicated to that purpose. The XEUS mission included many of the Con-X science goals, but also included additional science such as a survey of AGN in the $z > 7$ Universe, to constrain the growth of the first super-massive black holes. This survey required a larger field of view camera than planned for Con-X and an angular resolution requirement of 5 arcsec (the Con-X goal). The XEUS mission also included two other additional science objectives: 1) study bright X-ray sources with high count rates, and 2) study the X-ray polarization of sources, using a new technology not available when Con-X was first proposed.

Given the finite resources available and the upcoming key decisions at both agencies, it was recognized that merging the two missions would be highly desirable. In the spring of 2008, Constellation-X was merged with XEUS to create the IXO mission as presented to Astro2010. This agreed merger plan is detailed in a letter signed at an ESA-NASA HQ bilateral meeting on 2008 July 15 and 16.

The mirror collecting area requirements for the IXO mission at 1.25 keV ($3\,m^2$) and 6 keV ($0.65\,m^2$) are similar to those of Con-X as reviewed in AANM and Quarks to Cosmos. Driven by the XEUS imaging science objectives, the angular resolution requirement for IXO has been established as 5 arcsec HPD. We note that this improvement in angular resolution was already incorporated into the Con-X requirements by early 2008, to resolve structure in clusters of galaxies discovered by Chandra and to use galaxy clusters as cosmological probes. The additional science goals from the XEUS mission required new capabilities for wide field imaging, high time resolution studies of bright sources and X-ray polarization which were not included in the Con-X mission. These require dedicated instruments to address each goal (the WFI/HXI, HTRS and XPOL, respectively). The simplest mission design that can accommodate the suite of instruments is a single large telescope with a 20 m focal length and a translating focal plane to move each instrument into the focus— similar to the approach taken with Chandra. To accommodate the 20 m focal length within available EELVs our IXO design incorporates an Extendible Optical Bench which is based on designs already successfully flown.

It is notable that the total number of instruments decreased from AANM (to BEPAC) to Astro2010. This is because with the four large SXTs on Con-X there were four microcalorimeter arrays and four (or two) grating spectrometers. In the new IXO design, there is only one microcalorimeter array and one grating spectrometer. The merger with XEUS brought three new instruments (which we anticipate may come from ESA or JAXA) but still results in a net decrease in focal plane instruments (see Table 7-1). While there was a cost saving in having multiple instruments of identical design in the Con-X concept, the expectation that new instruments will come from ESA or JAXA reduces the overall cost to NASA while providing a more capable observatory. The hard X-ray capability is now integrated into the FMA, and the HXT detector is part of a hybrid low and high energy device (the WFI/HXI).

**Table 7-1. Mission Characteristics Comparison for NRC Studies**

| Mission Parameters | AANM | BEPAC | Astro2010 |
|---|---|---|---|
| Mirror Area (1.25 keV) | 3.4 m² | 3.0 m² | 3.0 m² |
| Number of Optics | 4 | 4 | 1 |
| Focal Length | 10 m | 10 m | 20 m |
| Number of LV | 2 | 1 | 1 |
| Number of Instruments[1] | 12 | 7 | 5 |
| Effective Area @ [2] | | | |
| 1.25 keV | 1.5 m² | 1.5 m² | 2.4 m² |
| 6 keV | 0.6 m² | 0.6 m² | 0.65 m² |
| 30 keV | 0.2 m² | 0.015 m² | 0.015 m² |
| Angular Resolution | | | |
| 0.1–7 keV | <15 arcsec | <15 arcsec | 5 arcsec |
| 7–40 keV | <60 arcsec | <30 arcsec | <30 arcsec |
| Count rate | | ¼ Crab at full capability | 1 Crab (~10⁶ c/s) with <10% dead time |
| Polarimetry | N/A | N/A | 1% MDP; 100 ksec, 5 × 10⁻¹² cgs (2–6 keV) |
| Spectral Resolution | | | |
| 0.3–7 keV | 2.5eV, 2.5 × 2.5 arcmin | 2.5eV, 2.5 × 2.5 arcmin | 2.5 eV, 2 × 2 arcmin |
| 0.3–10 keV | N/A | N/A | 10 eV, 5 × 5 arcmin |
| 0.1–15 keV | N/A | N/A | 150 eV, 18 × 18 arcmin |
| 0.2 –1 keV | >300 | >1250 | 3,000 |
| Launch Date | 2008 | 2017 | 2021 |
| Cost Assumptions | FY1999 for Phase C/D and 5 years Phase E | FY2008 Pre-phase A, A/B/C/D and 5 year phase E | Pre-phase A, A/B/C/D and 5 year phase E |
| Project Cost Estimate[3] | FY1999 $0.8B FY2009 $1.1B | FY2007 $1.7B FY2009 $1.8B | FY2009 $3.3B |
| FY09 Cost to NASA | $1.1B | $1.8B | $1.8B |

Notes:
[1] Instrument count conveys level of complexity: AANM = 12 in 4 identical telescopes each with 3 instruments (microcalorimeter, grating, and HXT array); BEPAC = 7 in 4 microcalorimeters, 2 gratings, 1 HXT array; Astro2010 = 5 in 1 microcalorimeter, 1 grating, 1 combined WFI/HXI, 1 XPOL, 1 HTRS.
[2] Effective areas are for the imaging instruments, and reflect a change to more efficient reflectors going from AANM to BEPAC, and a change to highly transparent (or removable) gratings going from BEPAC to Astro2010.
[3] Conversions from FY1999$, FY2007$ assume 3% inflation/year.

# Appendix A. Master Equipment Lists

Table A-1. Expanded Payload MEL.

| Item | Per Assy | Flight Set | QUAL | EDU | ETU | Spares | CBE Unit Mass (kg) | CBE Total Flight Mass (kg) | AIAA Maturity Code | Mass Growth Allow (%) | Max Exp. Mass (kg) | Science Average (W) | Safehold (W) | Peak (W) | Comments, Material, Heritage |
|---|---|---|---|---|---|---|---|---|---|---|---|---|---|---|---|
| **FMA** | 1 | 1 | 0 | 0 | 0 | 0 | 1731 | 1725 | - | 16% | 2009 | 1120 | 973 | 1120 | |
| FMA Primary Structure Assy | 1 | 1 | 1 | 1 | 0 | 0 | 340 | 340 | 2 | - | 391 | n/a | n/a | n/a | |
| Main Beam Assy | 12 | 12 | 12 | 1 | 1 | 1 | 12.0 | 143 | 2 | - | 165 | n/a | n/a | n/a | |
| Main Beam | 1 | 12 | 12 | 1 | 1 | 1 | 11.07 | 11.1 | 2 | 0.15 | 12.7 | n/a | n/a | n/a | Structural: CFRP_M55J954-3 |
| Mounting Foot | 1 | 12 | 12 | 1 | 1 | 4 | 0.64 | 0.64 | 2 | 0.15 | 0.7 | n/a | n/a | n/a | Structural: Titanium, Ti6A4V |
| Shear Interface Assy | 3 | 36 | 36 | 3 | 3 | 4 | 0.08 | 0.25 | 2 | - | 0.3 | n/a | n/a | n/a | |
| Shear Insert | 1 | 36 | 36 | 3 | 3 | 6 | 0.05 | 0.05 | 2 | 0.15 | 0.1 | n/a | n/a | n/a | Structural: Titanium, Ti6A4V |
| Shear Plate | 2 | 72 | 72 | 6 | 6 | 7 | 0.02 | 0.03 | 2 | 0.15 | 0.0 | n/a | n/a | n/a | Structural: Titanium, Ti6A4V |
| Shear Bushing | 1 | 72 | 72 | 6 | 6 | 7 | 0.001 | 0.002 | 2 | 0.15 | 0.0 | n/a | n/a | n/a | Structural: Aluminum Bronze, C63000 |
| Secondary Beam Assy | 12 | 12 | 12 | 1 | 1 | 2 | 7.0 | 84 | 2 | - | 97 | n/a | n/a | n/a | |
| Secondary Beam | 1 | 12 | 12 | 1 | 1 | 2 | 6.2 | 6.2 | 2 | 0.15 | 7.1 | n/a | n/a | n/a | Structural: CFRP_M55J954-3 |
| Mounting Foot | 1 | 12 | 12 | 1 | 1 | 2 | 0.64 | 0.64 | 2 | 0.15 | 0.7 | n/a | n/a | n/a | Structural: Titanium, Ti6A4V |
| Shear Interface Assy | 2 | 24 | 24 | 2 | 2 | 2 | 0.08 | 0.16 | 2 | - | 0.2 | n/a | n/a | n/a | |
| Shear Insert | 1 | 24 | 24 | 2 | 2 | 5 | 0.05 | 0.048 | 2 | 0.15 | 0.1 | n/a | n/a | n/a | Structural: Titanium, Ti6A4V |
| Shear Plate | 2 | 48 | 48 | 4 | 4 | 5 | 0.02 | 0.032 | 2 | 0.15 | 0.04 | n/a | n/a | n/a | Structural: Titanium, Ti6A4V |
| Shear Bushing | 1 | 48 | 48 | 4 | 4 | 5 | 0.001 | 0.002 | 2 | 0.15 | 0.002 | n/a | n/a | n/a | Structural: Aluminum Bronze, C63000 |
| Ring 0 Assy | 1 | 1 | 1 | 1 | 1 | 0 | 7.2 | 7.2 | 2 | - | 8.3 | n/a | n/a | n/a | |
| Ring 0 | 1 | 1 | 1 | 1 | 1 | 0 | 7.2 | 7.2 | 2 | 0.15 | 8.3 | n/a | n/a | n/a | Structural: CFRP_M55J954-3 |
| Ring 1 Assy | 1 | 1 | 1 | 1 | 1 | 0 | 10.8 | 10.8 | 2 | - | 12.4 | n/a | n/a | n/a | |
| Ring 1 | 12 | 12 | 12 | 1 | 1 | 0 | 9.81 | 9.8 | 2 | 0.15 | 11.3 | n/a | n/a | n/a | Structural: CFRP_M55J954-3 |
| Ring Shear Interface Assy | 12 | 12 | 12 | 1 | 1 | 2 | 0.08 | 0.98 | 2 | - | 1.1 | n/a | n/a | n/a | |
| Ring Shear Insert | 1 | 12 | 12 | 2 | 2 | 1 | 0.05 | 0.048 | 2 | 0.15 | 0.1 | n/a | n/a | n/a | Structural: Titanium, Ti6A4V |
| Ring Shear Plate | 2 | 24 | 24 | 4 | 4 | 2 | 0.02 | 0.032 | 2 | 0.15 | 0.04 | n/a | n/a | n/a | Structural: Titanium, Ti6A4V |
| Ring Shear Bushing | 1 | 24 | 24 | 4 | 4 | 2 | 0.001 | 0.002 | 2 | 0.15 | 0.002 | n/a | n/a | n/a | Structural: Aluminum Bronze, C63000 |
| Ring 2 Assy | 1 | 1 | 1 | 1 | 1 | 0 | 18.4 | 20.3 | 2 | - | 23 | n/a | n/a | n/a | |
| Ring 2 | 24 | 24 | 24 | 1 | 1 | 0 | 18.4 | 18.4 | 2 | 0.15 | 21 | n/a | n/a | n/a | Structural: CFRP_M55J954-3 |
| Ring Shear Interface Assy | 24 | 24 | 24 | 2 | 2 | 2 | 0.08 | 2.0 | 2 | - | 2.3 | n/a | n/a | n/a | |
| Ring Shear Insert | 1 | 24 | 24 | 4 | 4 | 5 | 0.05 | 0.048 | 2 | 0.15 | 0.1 | n/a | n/a | n/a | Structural: Titanium, Ti6A4V |
| Ring Shear Plate | 2 | 48 | 48 | 4 | 4 | 5 | 0.02 | 0.032 | 2 | 0.15 | 0.04 | n/a | n/a | n/a | Structural: Titanium, Ti6A4V |
| Ring Shear Bushing | 1 | 48 | 48 | 4 | 4 | 5 | 0.001 | 0.002 | 2 | 0.15 | 0.002 | n/a | n/a | n/a | Structural: Aluminum Bronze, C63000 |
| Ring 3 Assy | 1 | 1 | 1 | 1 | 1 | 0 | 30 | 30 | 2 | - | 34 | n/a | n/a | n/a | |
| Ring 3 | 1 | 1 | 1 | 1 | 1 | 0 | 28 | 28 | 2 | 0.15 | 32 | n/a | n/a | n/a | Structural: CFRP_M55J954-3 |
| Ring Shear Interface Assy | 24 | 24 | 24 | 2 | 2 | 2 | 0.082 | 2.0 | 2 | - | 2.3 | n/a | n/a | n/a | |
| Ring Shear Insert | 1 | 24 | 24 | 4 | 4 | 5 | 0.048 | 0.048 | 2 | 0.15 | 0.1 | n/a | n/a | n/a | Structural: Titanium, Ti6A4V |
| Ring Shear Plate | 2 | 48 | 48 | 4 | 4 | 5 | 0.016 | 0.032 | 2 | 0.15 | 0.04 | n/a | n/a | n/a | Structural: Titanium, Ti6A4V |
| Ring Shear Bushing | 1 | 48 | 48 | 4 | 4 | 5 | 0.001 | 0.002 | 2 | 0.15 | 0.002 | n/a | n/a | n/a | Structural: Aluminum Bronze, C63000 |
| Ring 4 Assy | 1 | 1 | 1 | 1 | 1 | 0 | 38.1 | 38 | 2 | - | 44 | n/a | n/a | n/a | |
| Ring 4 | 1 | 1 | 1 | 1 | 1 | 0 | 38 | 38 | 2 | 0.15 | 44 | n/a | n/a | n/a | Structural: CFRP_M55J954-3 |
| Bonding Brackets | 96 | 96 | 96 | 8 | 8 | 10 | 0.06 | 5.8 | 2 | 0.15 | 6.6 | n/a | n/a | n/a | Structural: CFRP_M55J954-3 |
| **Ring 1 Modules** | 12 | 12 | 1 | 1 | 1 | 1 | 23.4 | 281 | 2 | - | 332 | n/a | n/a | n/a | |
| Module Structure Assy, Ring 1 | 1 | 12 | 1 | 1 | 1 | 1 | 7.29 | 7.3 | 2 | - | 8.4 | n/a | n/a | n/a | |
| Module Front, Ring 1 | 1 | 12 | 1 | 2 | 2 | 1 | 1.08 | 1.08 | 2 | 0.15 | 1.2 | n/a | n/a | n/a | Structural: Titanium, T15Mo |
| Module Side, Ring 1 | 2 | 24 | 2 | 2 | 2 | 2 | 1.6 | 3.2 | 2 | 0.15 | 3.7 | n/a | n/a | n/a | Structural: Titanium, T15Mo |
| Module Back, Ring 1 | 1 | 12 | 1 | 1 | 1 | 1 | 0.72 | 0.7 | 2 | 0.15 | 0.8 | n/a | n/a | n/a | Structural: Titanium, T15Mo |
| Module Back Close-out, Ring 1 | 1 | 12 | 1 | 1 | 1 | 10 | 1.53 | 1.5 | 2 | 0.15 | 1.8 | n/a | n/a | n/a | Structural: Titanium, T15Mo |
| Radial Rail Supports, Ring 1 | 8 | 96 | 8 | 8 | 8 | 10 | 0.09 | 0.7 | 2 | 0.15 | 0.8 | n/a | n/a | n/a | Structural: Titanium, T15Mo |
| Kinematic Mount Assy | 1 | 1 | 1 | 1 | 1 | 0 | 0.09 | 0.1 | 2 | - | 0.1 | n/a | n/a | n/a | |
| Shear Pin | 1 | 36 | 3 | 3 | 3 | 4 | 0.008 | 0.01 | 2 | 0.15 | 0.01 | n/a | n/a | n/a | Structural: Titanium, Ti6A4V |
| Shear Pin Bushing | 1 | 36 | 3 | 3 | 3 | 4 | 0.0096 | 0.008 | 2 | 0.15 | 0.009 | n/a | n/a | n/a | Structural: Titanium, Ti6A4V |
| Deployment Screw | 2 | 72 | 6 | 6 | 6 | 6 | 0.0002 | 0.0004 | 2 | 0.2 | 0.0005 | n/a | n/a | n/a | Structural: Stainless Steel, Inconel 718 |
| Segments 744-1386 (P and S pair) | 143 | 1716 | 143 | 143 | 143 | 315 | Various | 14 | 2 | 15% | 16 | n/a | n/a | n/a | Optics: D263 Glass |
| Mirror Attachment Rails | 16 | 192 | 16 | 16 | 16 | 19 | 0.01 | 0.2 | 2 | 15% | 0.2 | n/a | n/a | n/a | Structural: Titanium, T15Mo |
| Bonding Tabs | 16 | 16 | 16 | 16 | 16 | 19 | 0.0002 | 0.0 | 2 | 15% | 0.5 | n/a | n/a | n/a | Structural: Titanium, T15Mo |
| | 143 | 27456 | 2288 | 2288 | 2288 | 2746 | | | | | | | | | |

**Table A-1. Expanded Payload MEL**

| Item | Per Assy | Flight Set | QUAL | EDU | ETU | Spares | CBE Unit Mass (kg) | CBE Total Flight Mass (kg) | AIAA Maturity Code | Mass Growth Allow (%) | Max Exp. Mass (kg) | Science Average (W) | Safehold (W) | Peak (W) | Comments, Material, Heritage |
|---|---|---|---|---|---|---|---|---|---|---|---|---|---|---|---|
| Stray Light Baffle, Ring 1 | 1 | 12 | 1 | 1 | 1 | 1 | 1.8 | 2 | 1 | 25% | 2.3 | n/a | n/a | n/a | Structural: Aluminum, 6061-T651 |
| SLB Thermal Pad, Ring 1 | 1 | 12 | 1 | 1 | 1 | 1 | 0.01 | 0.01 | 1 | 25% | 0.01 | n/a | n/a | n/a | Thermal: CHO-THERM |
| Thermal Shield | 1 | 12 | 1 | 1 | 1 | 1 | 0.01 | 0.01 | 1 | 25% | 0.01 | n/a | n/a | n/a | Thermal: Aluminum/Mylar |
| Ring 2 Modules | 24 | 24 | 24 | 0 | 0 | 2 | 11 | 393 | 2 | - | 454 | n/a | n/a | n/a | - |
| Module Structure Assy, Ring 2 | 1 | 24 | 1 | 0 | 0 | 2 | 0.005 | 0.005 | 2 | - | 5.8 | n/a | n/a | n/a | - |
| Module Front, Ring 2 | 1 | 24 | 1 | 0 | 0 | 2 | 0.001 | 0.001 | 2 | 15% | 0.9 | n/a | n/a | n/a | Structural: Titanium, Ti15Mo |
| Module Side, Ring 2 | 2 | 48 | 2 | 0 | 0 | 5 | 0.001 | 0.002 | 2 | 15% | 2.6 | n/a | n/a | n/a | Structural: Titanium, Ti15Mo |
| Module Back, Ring 2 | 1 | 24 | 1 | 0 | 0 | 2 | 0.001 | 0.001 | 2 | 15% | 0.6 | n/a | n/a | n/a | Structural: Titanium, Ti15Mo |
| Module Back Close-out, Ring 2 | 1 | 24 | 1 | 0 | 0 | 2 | 0.001 | 0.001 | 2 | 15% | 1.2 | n/a | n/a | n/a | Structural: Titanium, Ti15Mo |
| Radial Rail Supports, Ring 2 | 8 | 192 | 8 | 0 | 0 | 19 | 0.0001 | 0.001 | 2 | 15% | 0.6 | n/a | n/a | n/a | Structural: Titanium, Ti15Mo |
| Kinematic Mount Assy | 3 | 72 | 3 | 0 | 0 | 7 | 0.02 | 0.054 | 2 | - | 0.1 | n/a | n/a | n/a | - |
| Shear Pin | 1 | 72 | 3 | 0 | 0 | 7 | 0.01 | 0.010 | 2 | 15% | 0.01 | n/a | n/a | n/a | Structural: Titanium, Ti6Al4V |
| Shear Pin Bushing | 1 | 72 | 3 | 0 | 0 | 7 | 0.01 | 0.008 | 2 | 15% | 0.01 | n/a | n/a | n/a | Structural: Titanium, Ti6Al4V |
| Deployment Screw | 2 | 144 | 6 | 0 | 0 | 14 | 0.0002 | 0.0004 | 2 | 15% | 0.0005 | n/a | n/a | n/a | Structural: Stainless Steel, Inconel 718 |
| Segments 1479-2220 (P and S pair) | 115 | 2760 | 115 | 0 | 0 | 391 | Various | 10 | 2 | 15% | 11 | n/a | n/a | n/a | Optics: D263 Glass |
| Mirror Attachment Rails | 16 | 384 | 16 | 0 | 0 | 38 | 0.01 | 0.2 | 2 | 15% | 0.2 | n/a | n/a | n/a | Structural: Titanium, Ti15Mo |
| Bonding Tabs | 1840 | 44160 | 1840 | 0 | 0 | 4416 | 0.0002 | 0.4 | 2 | 15% | 0.4 | n/a | n/a | n/a | Structural: Titanium, Ti15Mo |
| Stray Light Baffle, Ring 2 | 1 | 24 | 1 | 0 | 0 | 2 | 0.5 | 1 | 1 | 25% | 0.6 | n/a | n/a | n/a | Structural: Aluminum, 6061-T651 |
| SLB Thermal Pad, Ring 2 | 1 | 24 | 1 | 0 | 0 | 2 | 0.01 | 0.01 | 1 | 25% | 0.01 | n/a | n/a | n/a | Thermal: CHO-THERM |
| Thermal Pre-Collimator, Ring 2 | 1 | 24 | 1 | 0 | 0 | 2 | 0.4 | 0.4 | 1 | 25% | 0.5 | n/a | n/a | n/a | Thermal: Fiberglass, G10 |
| Ring 3 Modules | 24 | 24 | 24 | 0 | 0 | 2 | 35.0 | 536 | 2 | - | 619 | n/a | n/a | n/a | - |
| Module Structure Assy, Ring 3 | 1 | 24 | 1 | 0 | 0 | 2 | 19.8 | 20 | 2 | - | 8.1 | n/a | n/a | n/a | - |
| Module Front, Ring 3 | 1 | 24 | 1 | 0 | 0 | 2 | 2.9 | 3 | 2 | 15% | 1.2 | n/a | n/a | n/a | Structural: Titanium, Ti6Al4V |
| Module Side, Ring 3 | 2 | 48 | 2 | 0 | 0 | 5 | 4.4 | 9 | 2 | 15% | 3.6 | n/a | n/a | n/a | Structural: Titanium, Ti6Al4V |
| Module Back, Ring 3 | 1 | 24 | 1 | 0 | 0 | 2 | 2.0 | 2 | 2 | 15% | 0.8 | n/a | n/a | n/a | Structural: Titanium, Ti15Mo |
| Module Back Close-out, Ring 3 | 1 | 24 | 1 | 0 | 0 | 2 | 4.1 | 4 | 2 | 15% | 1.7 | n/a | n/a | n/a | Structural: Titanium, Ti15Mo |
| Radial Rail Supports, Ring 3 | 8 | 192 | 8 | 0 | 0 | 19 | 0.2 | 2 | 2 | 15% | 0.8 | n/a | n/a | n/a | Structural: Titanium, Ti15Mo |
| Kinematic Mount Assy | 3 | 72 | 3 | 0 | 0 | 7 | 0.02 | 0.05 | 2 | - | 0.1 | n/a | n/a | n/a | - |
| Shear Pin | 1 | 72 | 3 | 0 | 0 | 7 | 0.01 | 0.01 | 2 | 15% | 0.01 | n/a | n/a | n/a | Structural: Titanium, Ti6Al4V |
| Shear Pin Bushing | 1 | 72 | 3 | 0 | 0 | 7 | 0.01 | 0.01 | 2 | 15% | 0.01 | n/a | n/a | n/a | Structural: Titanium, Ti6Al4V |
| Deployment Screw | 2 | 144 | 6 | 0 | 0 | 14 | 0.0002 | 0.0004 | 2 | 15% | 0.0005 | n/a | n/a | n/a | Structural: Stainless Steel, Inconel 718 |
| Segments 2312-3210 (P and S pair) | 103 | 2472 | 103 | 0 | 0 | 350 | Various | 14 | 2 | 15% | 16 | n/a | n/a | n/a | Optics: D263 Glass |
| Mirror Attachment Rails | 16 | 384 | 16 | 0 | 0 | 38 | 0.01 | 0.2 | 2 | 15% | 0.2 | n/a | n/a | n/a | Structural: Titanium, Ti15Mo |
| Bonding Tabs | 1648 | 39552 | 1648 | 0 | 0 | 3955 | 0.0002 | 0.3 | 2 | 15% | 0.4 | n/a | n/a | n/a | Structural: Titanium, Ti15Mo |
| Stray Light Baffle, Ring 3 | 1 | 24 | 1 | 0 | 0 | 2 | 0.5 | 1 | 1 | 25% | 0.6 | n/a | n/a | n/a | Structural: Aluminum, 6061-T651 |
| SLB Thermal Pad, Ring 3 | 1 | 24 | 1 | 0 | 0 | 2 | 0.0 | 0.01 | 1 | 25% | 0.01 | n/a | n/a | n/a | Thermal: CHO-THERM |
| Thermal Pre-Collimator, Ring 3 | 1 | 24 | 1 | 0 | 0 | 2 | 0.5 | 1 | 1 | 25% | 0.6 | n/a | n/a | n/a | Thermal: Fiberglass, G10 |
| HXMM Assembly | 1 | 1 | 1 | 0 | 1 | 0 | 51 | 51 | 1 | - | 59 | 10 | 8 | 10 | - |
| Module Structure Assy | 1 | 1 | 1 | 0 | 1 | 0 | 7.3 | 7 | 2 | - | 8.4 | n/a | n/a | n/a | - |
| Inner support Cylinder | 1 | 1 | 1 | 0 | 1 | 0 | 2.0 | 2 | 2 | 15% | 2.3 | n/a | n/a | n/a | Structural: Titanium, Ti15Mo |
| Outer Support Cylinder | 2 | 2 | 2 | 0 | 2 | 0 | 1.5 | 3 | 2 | 15% | 3.5 | n/a | n/a | n/a | Structural: Titanium, Ti15Mo |
| Front mounting spider | 1 | 1 | 1 | 0 | 1 | 0 | 1.2 | 1 | 2 | 15% | 1.3 | n/a | n/a | n/a | Structural: Titanium, Ti15Mo |
| Rear mounting spider | 1 | 1 | 1 | 0 | 1 | 0 | 1.2 | 1 | 2 | 15% | 1.3 | n/a | n/a | n/a | Structural: Titanium, Ti15Mo |
| Glass Segments | 2040 | 2040 | 0 | 0 | 2040 | 204 | 0.02 | 37 | 2 | 15% | 43 | n/a | n/a | n/a | Optics: D263 Glass |

# Table A-1. Expanded Payload MEL

| Item | Per Assy | Flight Set | QUAL | EDU | ETU | Spares | CBE Unit Mass (kg) | CBE Total Flight Mass (kg) | AIAA Maturity Code | Mass Growth Allow (%) | Max Exp. Mass (kg) | Science Average (W) | Safehold (W) | Peak (W) | Comments, Material, Heritage |
|---|---|---|---|---|---|---|---|---|---|---|---|---|---|---|---|
| Mirror Support bars | 6120 | 6120 | 0 | 0 | 6120 | 612 | 0.0007 | 4 | 2 | 15% | 4.6 | n/a | n/a | n/a | Structural: Titanium, Ti15Mo |
| Stray Light Baffle | 1 | 1 | 0 | 0 | 0 | 0 | 1.8 | 2 | 1 | 25% | 2.3 | n/a | n/a | n/a | Structural: Aluminum, 6061-T651 |
| Kinematic Mount Assy | 3 | 3 | 0 | 0 | 3 | 0 | 0.05 | 0.16 | 2 | 15% | 0.2 | n/a | n/a | n/a | - |
| Shear Pin | 3 | 3 | 0 | 0 | 3 | 0 | 0.01 | 0.03 | 2 | 15% | 0.03 | n/a | n/a | n/a | Structural: Titanium, Ti6Al4V |
| Shear Pin Bushing | 3 | 3 | 0 | 0 | 3 | 0 | 0.01 | 0.02 | 2 | 15% | 0.03 | n/a | n/a | n/a | Structural: Titanium, Ti6Al4V |
| Deployment Screw | 2 | 2 | 0 | 0 | 1 | 0 | 0.0002 | 0.0004 | 2 | 15% | 0.0005 | n/a | n/a | n/a | Structural: Stainless Steel, Inconel 718 |
| Thermal Shield/docking filter | 1 | 1 | 0 | 0 | 1 | 0 | 0.5 | 1 | 1 | 25% | 0.6 | n/a | n/a | n/a | Thermal: Fiberglass, G10 |
| HXMM Thermal Hardware | 1 | 1 | 0 | 0 | 0 | 0 | 0.9 | 1 | 1 | - | 1.6 | 10 | 8 | 10 | - |
| Heaters (Primary and Redundant) | 8 | 1 | 0 | 0 | 0 | 1 | 0.05 | 0.2 | 1 | 25% | 0.5 | 10 | 8 | 10 | - |
| Controllers | 1 | 1 | 0 | 0 | 0 | 0 | 0.22 | 0.2 | 1 | 25% | 0.3 | n/a | n/a | n/a | - |
| Thermal Harness | 1 | 1 | 0 | 0 | 0 | 0 | 0.5 | 1 | 1 | 55% | 0.8 | n/a | n/a | n/a | - |
| Thermisters | 8 | 1 | 0 | 0 | 0 | 0 | 0.0002 | 0.002 | 1 | 25% | 0.002 | n/a | n/a | n/a | - |
| Thermal Hardware | 1 | 1 | 0 | 0 | 0 | 0 | 119 | 119 | 2 | - | 148 | 1110 | 965 | 1110 | - |
| Heaters (internally redundant) | 1500 | 1 | 0 | 0 | 0 | 45 | 0.012 | 18 | 2 | 20% | 22 | 1110 | 965 | 1110 | - |
| Controllers | 400 | 1 | 0 | 0 | 0 | 12 | 0.22 | 88 | 2 | 20% | 106 | n/a | n/a | n/a | - |
| Thermal Harness | 1 | 1 | 0 | 0 | 0 | 0 | 13 | 13 | 1 | 55% | 20 | n/a | n/a | n/a | - |
| Thermisters | 3080 | 1 | 0 | 0 | 0 | 92 | 0.0002 | 1 | 1 | 20% | 0.7 | n/a | n/a | n/a | - |
| Fasteners | 1 | 1 | 0 | 0 | 0 | 0 | 5.2 | 5 | 2 | 20% | 6.2 | n/a | n/a | n/a | - |
| **XMS** | **1** | **1** | **0** | **0** | **0** | **0** | **189** | **263** | **-** | **24%** | **327** | **649** | **0** | **703** | |
| Filter Wheel Mechanism | 1 | 1 | 0 | 1 | 1 | 1 | 6.8 | 6.8 | - | - | 8.5 | n/a | n/a | n/a | - |
| Filters + Wheel -6 X filters, Support Wheel | 1 | 1 | 0 | 0 | 0 | 0 | 3.0 | 3.0 | 1 | 25% | 3.8 | n/a | n/a | n/a | Al |
| Stepper Motor, Planetary Gear Drive | 1 | 1 | 0 | 0 | 0 | 0 | 1.5 | 1.5 | 1 | 25% | 1.9 | n/a | n/a | n/a | Motor- n/a Gear - Al & STL |
| Encoder - RVDT, Potentiometer | 1 | 1 | 0 | 0 | 0 | 0 | 0.5 | 0.5 | 1 | 25% | 0.6 | n/a | n/a | n/a | - |
| Housing | 1 | 1 | 0 | 0 | 0 | 0 | 1.5 | 1.5 | 1 | 25% | 1.9 | n/a | n/a | n/a | Al |
| 5% Misc Hardware | 1 | 1 | 0 | 1 | 1 | 1 | 0.3 | 0.3 | 1 | 25% | 0.4 | n/a | n/a | n/a | - |
| Gate Valve | 1 | 1 | 0 | 0 | 0 | 1 | 0.03 | 0.03 | 4 | 5% | 0.03 | n/a | n/a | n/a | - |
| Pyro Devices for Gate Valve | 2 | 1 | 0 | 0 | 0 | 10 | 0.1 | 0.2 | 4 | 5% | 0.2 | n/a | n/a | n/a | - |
| XMS Cryostat Assembly | 1 | 1 | 0 | 1 | 1 | 0 | 72 | 87 | - | - | 104 | n/a | n/a | n/a | - |
| Filter Windows | 4 | 1 | 0 | 0 | 0 | 4 | 0.001 | 0.004 | 1 | 55% | 0.01 | n/a | n/a | n/a | Aluminized polyimide |
| Dewar Assembly | 1 | 1 | 0 | 0 | 0 | 0 | 72 | 87 | 1 | 20% | 104 | n/a | n/a | n/a | - |
| 300K Stage (Dewar mainshell) | 1 | 1 | 0 | 0 | 0 | 0 | 27 | 32 | - | - | 41.9 | n/a | n/a | n/a | - |
| Magnet Leads from 300K to 150K | 20 | 1 | 0 | 0 | 0 | 2 | 0.2 | 4.5 | 1 | 55% | 7 | n/a | n/a | n/a | Cu |
| Detector Leads to 300K to 150K | 1088 | 1 | 0 | 0 | 0 | 2 | 0.001 | 1.1 | 1 | 55% | 1.7 | n/a | n/a | n/a | Cu; Manganin |
| Top Cap | 1 | 1 | 0 | 0 | 0 | 0 | 1.2 | 1.2 | 1 | 25% | 1 | n/a | n/a | n/a | Al |
| Top Shell | 1 | 1 | 0 | 0 | 0 | 0 | 4.4 | 4.4 | 1 | 25% | 5.5 | n/a | n/a | n/a | Al |
| Upper Cylinder | 1 | 1 | 0 | 0 | 0 | 0 | 14 | 13.6 | 1 | 25% | 17 | n/a | n/a | n/a | Al |
| Lower Cylinder | 1 | 1 | 0 | 0 | 0 | 0 | 6.7 | 6.7 | 1 | 25% | 8.4 | n/a | n/a | n/a | Al |
| Bottom Cap | 1 | 1 | 0 | 0 | 0 | 0 | 0.7 | 0.7 | 1 | 25% | 1 | n/a | n/a | n/a | Al |
| 300K to 150K G-10 Spacer | 1 | 1 | 0 | 0 | 0 | 0 | 0.6 | 0.6 | 1 | 25% | 0.7 | n/a | n/a | n/a | G-10 |
| 150K Stage | 1 | 1 | 0 | 0 | 0 | 0 | 5.5 | 10 | - | - | 14 | n/a | n/a | n/a | - |
| Magnet Leads to 150K to 45K | 20 | 1 | 0 | 0 | 0 | 2 | 0.2 | 4.5 | 1 | 55% | 7.0 | n/a | n/a | n/a | Cu |
| Detector Leads from 150K to 45K | 1088 | 1 | 0 | 0 | 0 | 2 | 0.00002 | 0.02 | 3 | 55% | 0 | n/a | n/a | n/a | Cu; Manganin |
| Heat sinks | 16 | 1 | 0 | 0 | 0 | 0 | 0.02 | 0.4 | 3 | 10% | 0.39 | n/a | n/a | n/a | - |
| Top Shell | 1 | 1 | 0 | 0 | 0 | 0 | 0.4 | 0.4 | 1 | 25% | 0.5 | n/a | n/a | n/a | Al |
| Cylinder | 1 | 1 | 0 | 0 | 0 | 0 | 4.4 | 4.4 | 1 | 25% | 5.4 | n/a | n/a | n/a | Al |

# Table A-1. Expanded Payload MEL

| Item | Per Assy | Flight Set | QUAL | EDU | ETU | Spares | CBE Unit Mass (kg) | CBE Total Flight Mass (kg) | AIAA Maturity Code | Mass Growth Allow (%) | Max Exp. Mass (kg) | Science Average (W) | Safehold (W) | Peak (W) | Comments, Material, Heritage |
|---|---|---|---|---|---|---|---|---|---|---|---|---|---|---|---|
| Lower Shell | 1 | 1 | 0 | 0 | 0 | 0 | 0.5 | 0.5 | 1 | 25% | 0.6 | n/a | n/a | n/a | Al |
| 150K to 45K G-10 Spacer | 1 | 1 | 0 | 0 | 0 | 0 | 1.0 | 1.0 | 1 | 25% | 1.2 | n/a | n/a | n/a | G-10 |
| 45K Stage | 1 | 1 | 0 | 0 | 0 | 0 | 4.6 | 5.9 | - | - | 7.6 | n/a | n/a | n/a | - |
| Magnet Leads from 45K to 15K | 20 | 1 | 0 | 0 | 0 | 2 | 0.1 | 1.0 | 1 | 55% | 1.6 | n/a | n/a | n/a | - |
| Detector Leads to 45K to 15K | 1088 | 1 | 0 | 0 | 0 | 2 | 0.00002 | 0.02 | 1 | 55% | 0.04 | n/a | n/a | n/a | Cu, Manganin |
| Heat sinks | 16 | 1 | 0 | 0 | 0 | 0 | 0.02 | 0.4 | 3 | 10% | 0.4 | n/a | n/a | n/a | - |
| Top Shell | 1 | 1 | 0 | 0 | 0 | 0 | 0.3 | 0.3 | 1 | 25% | 0.4 | n/a | n/a | n/a | Al |
| Cylinder | 1 | 1 | 0 | 0 | 0 | 0 | 3.8 | 3.8 | 1 | 25% | 4.8 | n/a | n/a | n/a | Al |
| Lower shell | 1 | 1 | 0 | 0 | 0 | 0 | 0.4 | 0.4 | 1 | 25% | 0.5 | n/a | n/a | n/a | Al |
| 45K to 15K G-10 Spacer | 1 | 1 | 0 | 0 | 0 | 0 | 0.5 | 0.5 | 1 | 25% | 0.6 | n/a | n/a | n/a | G-10 |
| 15K Stage | 1 | 1 | 0 | 0 | 0 | 0 | 2.0 | 3.3 | - | - | 4.6 | n/a | n/a | n/a | - |
| Magnet Leads from 15K to 4.5K | 20 | 1 | 0 | 0 | 0 | 2 | 0.1 | 1.0 | 1 | 55% | 1.6 | n/a | n/a | n/a | YBCO |
| Detector Leads from 15K to 4.5K | 1088 | 1 | 0 | 0 | 0 | 2 | 0.00002 | 0.02 | 1 | 55% | 0.04 | n/a | n/a | n/a | Cu, Manganin |
| Heat sinks | 16 | 1 | 0 | 0 | 0 | 0 | 0.02 | 0.4 | 1 | 55% | 0.5 | n/a | n/a | n/a | - |
| Top Shell | 1 | 1 | 0 | 0 | 0 | 0 | 0.2 | 0.2 | 1 | 25% | 0.3 | n/a | n/a | n/a | AL |
| Cylinder | 1 | 1 | 0 | 0 | 0 | 0 | 1.4 | 1.4 | 1 | 25% | 2 | n/a | n/a | n/a | Al |
| Lower Shell | 1 | 1 | 0 | 0 | 0 | 0 | 0.3 | 0.3 | 1 | 25% | 0.4 | n/a | n/a | n/a | Al |
| 15K to 4.5 K G-10 Spacer | 1 | 1 | 0 | 0 | 0 | 0 | 0.4 | 0.4 | 1 | 25% | 0.5 | n/a | n/a | n/a | G-10 |
| 15K to 4.5K Aluminum Spacer | 1 | 1 | 0 | 0 | 0 | 0 | 0.5 | 0.5 | 1 | 25% | 0.6 | n/a | n/a | n/a | Al |
| 4.5K Stage | 1 | 1 | 0 | 0 | 0 | 0 | 2.0 | 3.3 | - | - | 5 | n/a | n/a | n/a | - |
| Magnet Leads from 4.5K to 1.0K | 20 | 1 | 0 | 0 | 0 | 1 | 0.1 | 1.0 | 1 | 55% | 2 | n/a | n/a | n/a | YBCO |
| Detector Leads to from 4.5K to 1.0K | 1088 | 1 | 0 | 0 | 0 | 2 | 0.00002 | 0.02 | 1 | 55% | 0.04 | n/a | n/a | n/a | Cu, Manganin |
| Heat sinks | 16 | 1 | 0 | 0 | 0 | 1 | 0.02 | 0.4 | 1 | 55% | 1 | n/a | n/a | n/a | AL |
| Top Shell | 1 | 1 | 0 | 0 | 0 | 1 | 0.2 | 0.2 | 1 | 25% | 0.3 | n/a | n/a | n/a | AL |
| Cylinder | 1 | 1 | 0 | 0 | 0 | 1 | 1.4 | 1.4 | 1 | 25% | 2 | n/a | n/a | n/a | Al |
| Lower shell | 1 | 1 | 0 | 0 | 0 | 1 | 0.3 | 0.3 | 1 | 25% | 0.4 | n/a | n/a | n/a | Al |
| ADR 1 assembly & 1K shell | 1 | 1 | 0 | 0 | 0 | 0 | 23.3 | 2.5 | 1 | 55% | 33 | n/a | n/a | n/a | - |
| ADR 3 + Temperature Sensor | 1 | 1 | 0 | 0 | 0 | 1 | 2.6 | 2.6 | 1 | 55% | 4 | n/a | n/a | n/a | - |
| ADR 5 + Temperature Sensor | 1 | 1 | 0 | 0 | 0 | 1 | 2.5 | 2.5 | 1 | 55% | 4 | n/a | n/a | n/a | - |
| ADR Suspension System (6 Kevlar Straps + | 1 | 1 | 0 | 0 | 0 | 1 | 1.0 | 1.0 | 1 | 25% | 1.3 | n/a | n/a | n/a | Kevlar straps, Cu |
| Resistive Network | 1 | 1 | 0 | 0 | 0 | 1 | 0.1 | 0.1 | 1 | 55% | 0.2 | n/a | n/a | n/a | Surface mount resistors |
| ADR 4 + Temperature Sensor | 1 | 1 | 0 | 0 | 0 | 1 | 3.0 | 3.0 | 1 | 25% | 4 | n/a | n/a | n/a | - |
| Detector Leads to from 1K to 50 mK | 1088 | 1 | 0 | 0 | 0 | 1 | 0.00002 | 0.02 | 1 | 55% | 0.004 | n/a | n/a | n/a | Cu, Manganin |
| Magnet Leads from 1.0K to ADRs | 20 | 1 | 0 | 0 | 0 | 1 | 1.0 | 1.0 | 1 | 55% | 2 | n/a | n/a | n/a | YBCO |
| Heat sinks on 1K shield | 16 | 1 | 0 | 0 | 0 | 0 | 0.02 | 0.4 | 3 | 10% | 0.4 | n/a | n/a | n/a | - |
| ADR 1 + Temperature Sensor | 1 | 1 | 0 | 0 | 0 | 1 | 0.6 | 0.6 | 1 | 55% | 0.9 | n/a | n/a | n/a | - |
| ADR 2 + Temperature Sensor | 1 | 1 | 0 | 0 | 0 | 1 | 1.7 | 1.7 | 1 | 55% | 2.6 | n/a | n/a | n/a | - |
| Top Cap | 1 | 1 | 0 | 0 | 0 | 0 | 3.0 | 3.0 | 1 | 25% | 3.8 | n/a | n/a | n/a | Cu |
| Cylinder | 1 | 1 | 0 | 0 | 0 | 0 | 6.8 | 6.8 | 1 | 25% | 8.5 | n/a | n/a | n/a | Cu |
| Bottom Shell | 1 | 1 | 0 | 0 | 0 | 2 | 2.0 | 2.0 | 1 | 25% | 2.5 | n/a | n/a | n/a | Cu |
| 50mK Detector Assembly | 1 | 1 | 0 | 0 | 0 | 0 | 1.0 | 1.0 | - | - | 1.6 | n/a | n/a | n/a | Cu |
| Microcalorimeter | 1 | 1 | 0 | 0 | 0 | 0 | 0.0 | 0.001 | 1 | 55% | 0.002 | n/a | n/a | n/a | Si |
| Microcalorimeter Bias Resistors | 1 | 1 | 0 | 0 | 0 | 0 | 0.003 | 0.003 | 1 | 55% | 0.005 | n/a | n/a | n/a | Si |
| Anticoincident detector | 1 | 1 | 0 | 0 | 0 | 0 | 0.0 | 0.004 | 1 | 55% | 0.006 | n/a | n/a | n/a | Si |
| Residual Amplifier (SQUID) | 1 | 1 | 0 | 0 | 0 | 0 | 0.003 | 0.003 | 1 | 55% | 0.005 | n/a | n/a | n/a | Si |
| Detector Suspension Systems - Braided Kevlar | 1 | 1 | 0 | 0 | 0 | 0 | 0.0 | 0.002 | 1 | 55% | 0.003 | n/a | n/a | n/a | Keval, Fibreglass, Al |
| Cover of detector enclosure with filtered aperture | 1 | 1 | 0 | 0 | 0 | 0 | 0.1 | 0.1 | 1 | 55% | 0.2 | n/a | n/a | n/a | Cu |
| Interface to cover of assembly | 1 | 1 | 0 | 0 | 0 | 0 | 0.0 | 0.001 | 1 | 55% | 0.002 | n/a | n/a | n/a | Cu |
| Detector enclosure (50 mK stage houses calorimeter | 1 | 1 | 0 | 0 | 0 | 0 | 0.9 | 0.9 | 1 | 55% | 1 | n/a | n/a | n/a | Cu |
| ADR Interface access | 1 | 1 | 0 | 0 | 0 | 0 | 0.0 | 0.02 | 1 | 55% | 0.03 | n/a | n/a | n/a | Vacuum |
| Main shell vent valve | 1 | 1 | 0 | 0 | 0 | 0 | 0.3 | 0.3 | 1 | 25% | 0.3 | n/a | n/a | n/a | - |
| 5% Misc Hardware | 1 | 1 | 0 | 0 | 0 | 1 | 3.1 | 3.1 | 1 | 25% | 4 | n/a | n/a | n/a | - |

# Table A-1. Expanded Payload MEL

| Item | Per Assy | Flight Set | QUAL | EDU | ETU | Spares | CBE Unit Mass (kg) | CBE Total Flight Mass (kg) | AIAA Maturity Code | Mass Growth Allow (%) | Max Exp. Mass (kg) | Science Average (W) | Safehold (W) | Peak (W) | Comments, Material, Heritage |
|---|---|---|---|---|---|---|---|---|---|---|---|---|---|---|---|
| Dewar Bipod Assembly | | | | | | | | | | | | | | | |
| Struts | 2 | 3 | 0 | 0 | 0 | 0 | 2.3 | 9.0 | - | 15% | 10 | n/a | n/a | n/a | Alum |
| Mounting Foot | 2 | 3 | 0 | 0 | 0 | 0 | 0.3 | 1.5 | 2 | 15% | 1.7 | n/a | n/a | n/a | Alum |
| Dewar Foot | 2 | 3 | 0 | 0 | 0 | 0 | 0.7 | 4.2 | 2 | 15% | 4.8 | n/a | n/a | n/a | Alum |
| Dewar interface | 1 | 1 | 0 | 0 | 0 | 0 | 1.0 | 3.0 | 2 | 15% | 3.5 | n/a | n/a | n/a | Alum |
| 5% Misc Hardware | 1 | 1 | 0 | 0 | 0 | 0 | 0.3 | 0.3 | 2 | 15% | 0.4 | n/a | n/a | n/a | - |
| XMS Electronics Boxes | | | | | | | | | | | | | | | |
| PreAmpBias (PBB) Box | 1 | 1 | 0 | 1 | 1 | 0 | 64 | 109 | - | - | 144 | 434 | 0 | 488 | - |
| PreAmp/Bias Voltages Card | 1 | 1 | 0 | 0 | 0 | 1 | 2.5 | 3.8 | - | - | 4.8 | 25 | 0 | 25 | - |
| Pre Amplifier/Bias Voltages Card | 4 | 1 | 0 | 0 | 0 | 1 | 0.5 | 2.0 | 1 | 25% | 2.5 | 6 | 0 | 6 | 10% connectors, 90% analog |
| Enclosure | 1 | 1 | 0 | 0 | 0 | 1 | 1.8 | 1.8 | 1 | 25% | 2.3 | 19 | 0 | 19 | Al |
| 5% Misc Hardware | 1 | 1 | 0 | 0 | 0 | 0 | 0.2 | 0.001 | 1 | 30% | 0.001 | n/a | n/a | n/a | - |
| Feedback/Controller Box | 1 | 4 | 0 | 0 | 0 | 0 | 5.0 | 12 | - | - | 16 | 180 | 0 | 220 | - |
| Digital Feedback cards, 2 FB columns per cards -2 FPGA's/card | 8 | 1 | 0 | 0 | 0 | 1 | 1.0 | 8.0 | 1 | 30% | 10 | n/a | n/a | n/a | 10% connectors, 60% digital, 30% analog |
| 1 row address/1 system Controller, card 2 - 2-FPGAs unique | 1 | 1 | 0 | 0 | 0 | 1 | 1.0 | 1.0 | 1 | 25% | 1.3 | n/a | n/a | n/a | 10% connectors, 60% digital, 30% analog |
| Enclosure | 1 | 1 | 0 | 0 | 0 | 0 | 2.8 | 2.8 | 1 | 25% | 3.5 | n/a | n/a | n/a | Al |
| 5% Misc Hardware | 1 | 1 | 0 | 0 | 0 | 0 | 0.2 | 0.2 | 1 | - | 0.3 | n/a | n/a | n/a | - |
| Pulse Processing Electronics ( PPE) Box | 1 | 1 | 0 | 0 | 0 | 0 | 6.2 | 14 | - | - | 18 | 62 | 0 | 62 | - |
| Cal Digital Pulse Processor Card - Maxwell SCS 750 | 4 | 1 | 0 | 0 | 0 | 1 | 1.5 | 6.0 | 1 | 30% | 7.8 | n/a | n/a | n/a | - |
| Pulse Detection - 2 FPGA Board | 4 | 1 | 0 | 0 | 0 | 1 | 1.0 | 4.0 | 1 | 30% | 5.2 | n/a | n/a | n/a | 10 connectors, 90% digital |
| Enclosure | 1 | 1 | 0 | 0 | 0 | 1 | 3.0 | 3.0 | 1 | 25% | 3.8 | n/a | n/a | n/a | Al |
| 5% Misc Hardware | 1 | 1 | 0 | 0 | 0 | 0 | 0.7 | 0.7 | 1 | 30% | 0.8 | n/a | n/a | n/a | - |
| Cryocooler Control Electronics (CCE) Box | 1 | 2 | 0 | 0 | 0 | 0 | 4.9 | 11 | - | - | 14.1 | 54 | 0 | 54 | - |
| Cryocooler Control Electronics Card | 4 | 1 | 0 | 0 | 0 | 1 | 2.0 | 8.0 | 1 | 30% | 10 | n/a | n/a | n/a | - |
| Enclosure | 1 | 1 | 0 | 0 | 0 | 0 | 2.7 | 2.7 | 1 | 25% | 3 | n/a | n/a | n/a | - |
| 5% Misc Hardware | 1 | 1 | 0 | 0 | 0 | 0 | 0.2 | 0.2 | 1 | 30% | 0.3 | n/a | n/a | n/a | - |
| ADR Control Electronics Box | 1 | 1 | 0 | 0 | 0 | 0 | 5.2 | 6.9 | 1 | - | 8.6 | 35 | 0 | 5 | - |
| ADR Control Electronics Card -1 FPGA PID controller | 8 | 1 | 0 | 0 | 0 | 1 | 0.3 | 2.0 | 1 | 20% | 2.4 | n/a | n/a | n/a | 10% connectors, 45% digital, 45% analog |
| Enclosure | 1 | 1 | 0 | 0 | 0 | 1 | 4.5 | 4.5 | 1 | 25% | 5.6 | n/a | n/a | n/a | Al |
| 5% Misc Hardware | 1 | 1 | 0 | 0 | 0 | 0 | 0.4 | 0.4 | 1 | 30% | 0.6 | n/a | n/a | n/a | - |
| Power Distribution Unit (PDU) Box | 1 | 1 | 0 | 0 | 0 | 0 | 16 | 38 | - | - | 48 | 108 | 0 | 122 | - |
| Main Transformer Card | 1 | 2 | 0 | 0 | 0 | 1 | 1.0 | 2.0 | 1 | 30% | 2.6 | n/a | n/a | n/a | 10%connectors, 80% analog, 10% digital |
| Mechanism Power Card | 1 | 2 | 0 | 0 | 0 | 1 | 1.0 | 2.0 | 1 | 30% | 2.6 | n/a | n/a | n/a | 10%connectors, 80% analog, 10% digital |
| Cryocooler Power Card | 4 | 2 | 0 | 0 | 0 | 1 | 1.0 | 8.0 | 1 | 30% | 10.4 | n/a | n/a | n/a | 10%connectors, 80% analog, 10% digital |
| Electronics Power -FCB Power Card | 2 | 2 | 0 | 0 | 0 | 1 | 1.0 | 4.0 | 1 | 30% | 5.2 | n/a | n/a | n/a | 10%connectors, 80% analog, 10% digital |
| PBB Card | 1 | 2 | 0 | 0 | 0 | 1 | 1.0 | 2.0 | 1 | 30% | 2.6 | n/a | n/a | n/a | 10%connectors, 80% analog, 10% digital |
| PPE Card | 1 | 2 | 0 | 0 | 0 | 1 | 1.0 | 2.0 | 1 | 30% | 2.6 | n/a | n/a | n/a | 10%connectors, 80% analog, 10% digital |
| ADR Card | 2 | 2 | 0 | 0 | 0 | 1 | 1.0 | 2.0 | 1 | 30% | 2.6 | n/a | n/a | n/a | 10%connectors, 80% analog, 10% digital |
| FWC Spacewire I/F Card | 1 | 2 | 0 | 0 | 0 | 1 | 6.0 | 12 | 1 | 25% | 15.0 | n/a | n/a | n/a | 10%connectors, 10% analog, 80% digita |
| Enclosure | 1 | 1 | 0 | 0 | 0 | 1 | 1.5 | 1.5 | 1 | 25% | 2.0 | n/a | n/a | n/a | Al |
| 5% Misc Hardware | 1 | 1 | 0 | 0 | 0 | 0 | 4.7 | 4.7 | - | 30% | 6.0 | n/a | n/a | n/a | - |
| Anticoincident detector pre-amp box | 1 | 1 | 0 | 0 | 0 | 0 | 2.7 | 2.7 | 1 | 25% | 3.4 | 14.6 | 0 | 0 | - |
| Enclosure | 1 | 1 | 0 | 0 | 0 | 1 | 1.0 | 1.0 | 1 | 30% | 1.3 | n/a | n/a | n/a | - |
| Anti-co pulse processing card | 1 | 1 | 0 | 0 | 0 | 1 | 1.0 | 1.0 | 1 | 30% | 1.3 | n/a | n/a | n/a | - |
| Digital FB card | 1 | 1 | 0 | 0 | 0 | 1 | 3.7 | 3.7 | - | 30% | 4.7 | n/a | n/a | n/a | - |
| Filter wheel electronics box | 1 | 1 | 0 | 0 | 0 | 0 | 2.7 | 2.7 | 1 | 25% | 3.4 | n/a | n/a | n/a | - |
| Enclosure | 1 | 1 | 0 | 0 | 0 | 1 | 1.0 | 1.0 | 1 | 30% | 1.3 | n/a | n/a | n/a | - |
| Filter wheel electronics card | 1 | 1 | 0 | 0 | 0 | 1 | 1.0 | 1.0 | 1 | 30% | 1.3 | n/a | n/a | n/a | - |
| Detector harness | 1 | 1 | 0 | 0 | 0 | 0 | 16 | 16 | - | 55% | 25 | n/a | n/a | n/a | - |
| XMS Thermal Subsystem | | | | | | | | | | | | | | | |
| Constant conductance heat pipe (CCHP) from Cryocooler to Radiator | 2 | 1 | 0 | 0 | 0 | 0 | 3.5 | 7.0 | 1 | 25% | 8.8 | n/a | n/a | n/a | Al, ammonia |
| Spreader Heat Pipes (also CCHP) for Cryocooler Radiator | 4 | 1 | 0 | 0 | 0 | 0 | 0.3 | 1.2 | 1 | 25% | 1.5 | n/a | n/a | n/a | Al, Ammonia |

Table A-1. Expanded Payload MEL.

| Item | Per Assy | Flight Set | QUAL | EDU | ETU | Spares | CBE Unit Mass (kg) | CBE Total Flight Mass (kg) | AIAA Maturity Code | Mass Growth Allow (%) | Max Exp. Mass [kg] | Science Average (W) | Safehold (W) | Peak (W) | Comments, Material, Heritage |
|---|---|---|---|---|---|---|---|---|---|---|---|---|---|---|---|
| MLI on Cryocooler CCHPs (15 layers) | 1 | 1 | 0 | 0 | 0 | 0 | 0.1 | 0.1 | 2 | 20% | 0.07 | n/a | n/a | n/a | - |
| MLI (15 layers) Enclosure for PDU & Cryocooler/ADR/Processing/Mechanism Electronics | 1 | 1 | 0 | 0 | 0 | 0 | 0.9 | 0.9 | 2 | 0.2 | 1.1 | n/a | n/a | n/a | - |
| MLI (15 layers) Enclosure for Calorimeter Amplifier & Controller Feedback | 1 | 1 | 0 | 0 | 0 | 0 | 0.4 | 0.4 | 2 | 20% | 0.4 | n/a | n/a | n/a | - |
| MLI (15 layers) Enclosure for Cryocooler/Dewar | 1 | 1 | 0 | 0 | 0 | 0 | 0.7 | 0.7 | 2 | 0.2 | 0.9 | n/a | n/a | n/a | - |
| MLI (15 layers) on Platform | 1 | 1 | 0 | 0 | 0 | 0 | 2.9 | 2.9 | 2 | 20% | 3.5 | n/a | n/a | n/a | - |
| Thermistors/Platinum RTDs | 60 | 1 | 0 | 0 | 0 | 0 | 0.0 | 0.1 | 4 | 0.1 | 0.1 | n/a | n/a | n/a | Honeywell 3100 Series |
| Thermostats for Mechanisms Op Heaters | 8 | 1 | 0 | 0 | 0 | 2 | 0.006 | 0.05 | 4 | 10% | 0.05 | n/a | n/a | n/a | Kapton Film |
| Mechanisms Op Heaters | 8 | 1 | 0 | 0 | 0 | 0 | 0.0 | 0.02 | 4 | 0.1 | 0.0 | n/a | n/a | n/a | - |
| Thermostats for Electronics, Mechnisms & Cryocooler Survival Heaters (8 each) | 8 | 13 | 0 | 0 | 0 | 20 | 0.006 | 0.6 | 4 | 10% | 0.7 | n/a | n/a | n/a | - |
| Survival Heaters (8 per electronics or cryocooler box) | 8 | 13 | 0 | 0 | 0 | 20 | 0.002 | 0.2 | 4 | 0.1 | 0.2 | n/a | n/a | n/a | - |
| MLI (20 Layers) Shields for Dewar | 1 | 1 | 0 | 0 | 0 | 0 | 1.6 | 1.6 | 2 | 20% | 1.9 | n/a | n/a | n/a | Kapton Film 5.5cmx6.4cm |
| Buttons, Velcro, and Tape for MLI | 1 | 1 | 0 | 0 | 0 | 0 | 0.7 | 0.7 | 4 | 0.1 | 0.7 | n/a | n/a | n/a | - |
| Cryocooler Harness | 1 | 1 | 0 | 0 | 0 | 0 | 2.0 | 2.0 | 2 | 30% | 2.6 | n/a | n/a | n/a | - |
| 5% Misc Hardware | 1 | 1 | 0 | 0 | 0 | 0 | 0.7 | 0.7 | 4 | 0.1 | 0.7 | n/a | n/a | n/a | - |
| Cryocooler | 1 | 1 | 0 | 0 | 0 | 0 | 31 | 32 | - | - | 37 | 215 | 0 | 215 | - |
| Pre Cooler Compressor | 1 | 1 | 0 | 0 | 0 | 1 | 12.0 | 12.0 | 2 | 0.15 | 14 | 215 | 0 | 215 | - |
| JT Compressor | 1 | 1 | 0 | 0 | 0 | 0 | 8 | 8.0 | 2 | 15% | 9.2 | n/a | n/a | n/a | - |
| Cold Head | 1 | 1 | 0 | 0 | 0 | 1 | 8.0 | 8.0 | 2 | 0.15 | 9.2 | n/a | n/a | n/a | Al, Stainless Steel |
| Cryocooler Deck | 1 | 1 | 0 | 0 | 0 | 0 | 0.8 | 0.8 | 2 | 15% | 0.9 | n/a | n/a | n/a | Alum h/c core, facesheet, t=1.0 |
| Cryocooler Deck Mounts | 3 | 1 | 0 | 0 | 0 | 0 | 0.3 | 0.9 | 2 | 0.15 | 1.0 | n/a | n/a | n/a | Alum |
| 5% Misc Hardware | 1 | 1 | 0 | 0 | 0 | 0 | 2.2 | 2.2 | 2 | 15% | 2.5 | n/a | n/a | n/a | - |
| **WFI/HXI** | **1** | **1** | **0** | **0** | **0** | **0** | | **89** | | **26%** | | **268** | **0** | **308** | - |
| WFI | 1 | 1 | | | | | | 65 | | 28% | 83 | 222* | 0 | 262* | - |
| Focal Plane Array (FPA) | 1 | 1 | 1 | 1 | 1 | 1 | 17.6 | 17.6 | | 46% | 26 | 25.1 | 0 | 43.1 | - |
| DEPFET based Active Pixel Sensor Array | 1 | 0 | 0 | 0 | 0 | 0 | 3.1 | 3.1 | | 32% | 4 | 25 | 0 | 43 | - |
| DEPFET Hybrid Board | 1 | 0 | 0 | 0 | 0 | 0 | 0.2 | 0.2 | 1 | 55% | 0.3 | 25 | 0 | 43 | - |
| DEPFET Sensor Wafer | 1 | 0 | 0 | 0 | 0 | 0 | - | - | | - | - | n/a | n/a | n/a | Simbol-X LED, BepiColombo MIXS |
| Front-End Ceramic Board | 1 | 0 | 0 | 0 | 0 | 0 | - | - | | - | - | 7 | 0 | 25 | - |
| Analog Frontend ASICs (Vela/Asteroid/CA/MEX) | 16 | 0 | 0 | 0 | 0 | 0 | - | - | | - | - | n/a | n/a | n/a | ASTEROID/VELA: TRL 3, heritage Simbol-X, BepiColombo |
| Digital Frontend ASICs (SWITCHER) | 16 | 0 | 0 | 0 | 0 | 0 | - | - | | - | - | n/a | n/a | n/a | XMM-Newton, SOHO, ABRIXAS, eROSITA |
| Heater assembly | 1 | 0 | 0 | 0 | 0 | 0 | - | - | | - | - | 7 | 0 | 25 | - |
| Hybrid Board Mounting Structure | 1 | 0 | 0 | 0 | 0 | 0 | 2.9 | 2.9 | 2 | 30% | 3.8 | n/a | n/a | n/a | - |
| FPA cases, incl thermal strap | 1 | 0 | 0 | 0 | 0 | 0 | 3.5 | 3.5 | 2 | 30% | 4.6 | n/a | n/a | n/a | - |
| Radiation shield | 1 | 0 | 0 | 0 | 0 | 0 | 11 | 11 | 1 | 55% | 17.1 | n/a | n/a | n/a | - |
| Hemisphere Preprocessor Boxes (HPP) | 2 | 2 | 2 | 1 | 1 | 1 | 6.7 | 13 | | 21% | 16.2 | 98 | 0 | 98 | - |
| ADC Cluster | 4 | - | - | - | - | - | 0.5 | 2.0 | 1 | 30% | 2.6 | n/a | n/a | n/a | - |
| Framelet Builder FPGA Board | 1 | - | - | - | - | - | 0.5 | 0.5 | 1 | 30% | 0.7 | n/a | n/a | n/a | - |
| HPP Case | 1 | - | - | - | - | - | 4.2 | 4.2 | 2 | 15% | 4.8 | n/a | n/a | n/a | - |
| Frame Builder / Brain Box (FBB) | 1 | 1 | 1 | 1 | 1 | 1 | 8.7 | 8.7 | - | 18% | 10.3 | 24 | 0 | 24 | eROSITA |
| Redundant Frame-Builder / Brain Modules | 2 | 0 | 0 | 0 | 0 | 0 | 1 | 2.0 | - | - | 2.6 | 12 | 0 | 24 | - |
| Image Controller Unit | 1 | 0 | 0 | 0 | 0 | 0 | 0.5 | 0.5 | 1 | 30% | 0.7 | 12 | 0 | 6 | - |
| Frame Builder Unit | 1 | 0 | 0 | 0 | 0 | 0 | 0.5 | 0.5 | 1 | 30% | 0.7 | 12 | 0 | 6 | - |
| FBB Case | 1 | 0 | 0 | 0 | 0 | 0 | 6.7 | 6.7 | 2 | 15% | 7.7 | n/a | n/a | n/a | - |
| Power Conditioner Units (PCU) | 2 | 2 | 2 | 1 | 1 | 2 | 6.7 | 13 | - | 21% | 16.2 | 6 | 0 | 6 | - |

Table A-1. Expanded Payload MEL

| Item | Per Assy | Flight Set | QUAL | EDU | ETU | Spares | CBE Unit Mass (kg) | CBE Total Flight Mass (kg) | AIAA Maturity Code | Mass Growth Allow (%) | Max Exp. Mass (kg) | Science Average (W) | Safehold (W) | Peak (W) | Comments, Material, Heritage |
|---|---|---|---|---|---|---|---|---|---|---|---|---|---|---|---|
| Power Conditioners | 1 | 0 | 0 | 0 | 0 | 0 | 2.5 | 2.5 | 1 | 30% | 3.3 | n/a | n/a | n/a | - |
| PCU case | 1 | 0 | 0 | 0 | 0 | 0 | 4.2 | 4.2 | 2 | 15% | 4.8 | n/a | n/a | n/a | eROSITA |
| Filter Sled (4 positions: open, closed, calibration, filter) | 1 | 1 | 1 | 1 | 1 | 1 | 11 | 11 | - | 26% | 14 | 3 | 0 | 12 | - |
| Filter Sled mechanics | 1 | 1 | 0 | 0 | 0 | 0 | 11 | 11 | 1 | 25% | 14 | 3 | 0 | 12 | XMM-Newton |
| Optical Blocking Filter (100nm Al, on film, mechanical support grid) | 1 | 0 | 0 | 0 | 0 | 0 | 0.2 | 0.2 | 1 | 55% | 0.3 | 3 | 0 | n/a | XMM-Newton |
| Calibration Sources | 2 | 0 | 0 | 0 | 0 | 0 | 0.2 | 0.2 | 1 | 55% | 0.3 | n/a | n/a | n/a |  |
| HXI | 1 | 1 |  |  |  |  | 24 | 24 |  | 19% | 28.6 | 46 | 0 | 46 |  |
| HXI Sensor (HXI-S) | 1 | 1 | 1 | 1 | 0 | 1 | 15 | 15 | - | 21% | 18.2 | 20 | 0 | 20 | ASTRO-H HXI |
| Si Imager | 1 | 0 | 0 | 0 | 0 | 0 | 0.2 | 0.2 | 1 | 55% | 0.3 | n/a | n/a | n/a | ASTRO-H HXI |
| CdTe Imager | 1 | 0 | 0 | 0 | 0 | 0 | 0.1 | 0.1 | 1 | 55% | 0.2 | n/a | n/a | n/a | ASTRO-H HXI |
| ASIC | 1 | 0 | 0 | 0 | 0 | 0 | 0.01 | 0.01 | 1 | 55% | 0.02 | n/a | n/a | n/a | ASTRO-H HXI (TRL 5)/Simbol-X (TRL 5) Ovesolescence risk of technology requires new developments for IXO era (TRL 3) |
| active shield | 1 | 0 | 0 | 0 | 0 | 0 | 5 | 5.0 | 2 | 30% | 6.5 | n/a | n/a | n/a | Suzaku HXD |
| HK, HV, heater | 1 | 0 | 0 | 0 | 0 | 0 | 0.2 | 0.2 | 1 | 25% | 0.3 | n/a | n/a | n/a | ASTRO-H HXI |
| housing | 1 | 0 | 0 | 0 | 0 | 0 | 9.5 | 10 | 2 | 15% | 10.9 | n/a | n/a | n/a | Suzaku HXD |
| HXI Analog Electronics (HXI-E) | 1 | 1 | 1 | 1 | 0 | 1 | 5 | 5.0 | - | 16% | 5.8 | 20 | 0 | 20 | ASTRO-H HXI |
| DIO board | 1 | 0 | 0 | 0 | 0 | 0 | 0.2 | 0.2 | 2 | 25% | 0.3 | n/a | n/a | n/a | ASTRO-H HXI |
| power supply | 1 | 0 | 0 | 0 | 0 | 0 | 0.2 | 0.2 | 1 | 30% | 0.3 | n/a | n/a | n/a | ASTRO-H HXI |
| housing | 1 | 0 | 0 | 0 | 0 | 0 | 4.6 | 4.6 | 2 | 15% | 5.3 | n/a | n/a | n/a | Suzaku HXD |
| HXI Digital Electronics (HXI-D) | 1 | 1 | - | 1 | 0 | 1 | 4 | 4 | - | 16% | 4.7 | 6 | 0 | 6 | ASTRO-H HXI |
| CPU board | 1 | 0 | 0 | 0 | 0 | 0 | 0.2 | 0.2 | 2 | 25% | 0.3 | n/a | n/a | n/a | ASTRO-H HXI |
| power supply | 1 | 0 | 0 | 0 | 0 | 0 | 0.2 | 0.2 | 1 | 30% | 0.3 | n/a | n/a | n/a | ASTRO-H HXI |
| housing | 1 | 0 | 0 | 0 | 0 | 0 | 3.6 | 3.6 | 2 | 15% | 4.1 | n/a | n/a | n/a | Suzaku HXD |
| **XGS** | **1** | **1** |  |  |  |  |  | **50** |  | **21%** | **61** | **77** | **0** | **83** |  |
| XGS Readout Camera Assembly | 1 | 1 | 0 | 1 | 1 | 0 | 41.2 | 41 | - | 19% | 49 | 77 | 0 | 83 |  |
| Focal Plane Assembly | 1 | 1 | 0 | 0 | 0 | 0 | 27.5 | 28 | - | 16% | 32 | 7 | 0 | 9 | - |
| Paddle and CCD Assembly | 1 | 1 | 0 | 0 | 0 | 1* | 2.76 | 2.8 | 2 | 25% | 3.45 | n/a | n/a | n/a | - |
| MIT-LL CCDID41B (CCD) | 32 | 32 | 0 | 0 | 0 | 0 | 0.0001 | 0.005 | 2 | 30% | 0.01 | n/a | n/a | n/a | 1024x1026 24um pixels; Framestore Array +2" 512x1026 (21umx13.5um) |
| Al2O3 Substrate (mount) | 32 | 32 | 0 | 0 | 0 | 0 | 0.023 | 0.7 | 2 | 30% | 0.96 | n/a | n/a | n/a | Al2O3 |
| Fanout board (interconnect between CCD and flex circuit) | 32 | 32 | 0 | 0 | 0 | 0 | 0.003 | 0.10 | 2 | 30% | 0.12 | n/a | n/a | n/a | Al2O3 |
| Frame Store Shield | 32 | 32 | 0 | 0 | 0 | 0 | 0.004 | 0.1 | 2 | 30% | 0.17 | n/a | n/a | n/a | Al |
| Isolators | 18 | 18 | 0 | 0 | 0 | 0 | 0.0015 | 0.03 | 2 | 30% | 0.03 | n/a | n/a | n/a | G-10, 1 cm O.D., 0.5 cm I.D. 1.4 cm tall |
| Paddle -Machined to Optical precision | 1 | 1 | 0 | 0 | 0 | 0 | 1.6 | 1.6 | 2 | 30% | 2.1 | n/a | n/a | n/a | Al |
| Mechanical Standoffs (for Paddle) | 18 | 18 | 0 | 0 | 0 | 0 | 0.002 | 0.04 | 2 | 30% | 0.05 | n/a | n/a | n/a | Al |
| 5% misc Hardware | 1 | 1 | 0 | 0 | 0 | 0 | 0.1 | 0.1 | 2 | 30% | 0.2 | n/a | n/a | n/a | - |
| Front Radiation Shielding | 1 | 1 | 0 | 0 | 0 | 0 | 4.2 | 4.2 | 2 | 15% | 4.8 | n/a | n/a | n/a | Al2O3 |
| Read Radiation Shielding | 1 | 1 | 0 | 0 | 0 | 0 | 7.6 | 7.6 | 2 | 15% | 8.7 | n/a | n/a | n/a | Al |
| Side Cover Insulator | 1 | 1 | 0 | 0 | 0 | 0 | 0.01 | 0.01 | 2 | 15% | 0.01 | n/a | n/a | n/a | G-10 |

Table A-1. Expanded Payload MEL

| Item | Per Assy | Flight Set | QUAL | EDU | ETU | Spares | CBE Unit Mass (kg) | CBE Total Fight Mass (kg) | AIAA Maturity Code | Mass Growth Allow (%) | Max Exp. Mass [kg] | Science Average (W) | Safehold (W) | Peak (W) | Comments, Material, Heritage |
|---|---|---|---|---|---|---|---|---|---|---|---|---|---|---|---|
| Top Cover Insulator | 1 | 1 | 0 | 0 | 0 | 0 | 0.07 | 0.1 | 2 | 15% | 0.08 | n/a | n/a | n/a | G-10 |
| Side Cover | 1 | 1 | 0 | 0 | 0 | 0 | 0.6 | 0.6 | 2 | 15% | 0.7 | n/a | n/a | n/a | Al |
| Top Cover | 1 | 1 | 0 | 0 | 0 | 0 | 0.8 | 0.8 | 2 | 15% | 0.9 | n/a | n/a | n/a | Al |
| Connector Panel (Detectors) | 1 | 1 | 0 | 0 | 0 | 0 | 0.14 | 0.1 | 2 | 15% | 0.2 | n/a | n/a | n/a | Al |
| Camera Housing Enclosure | 1 | 1 | 0 | 0 | 0 | 0 | 10 | 10 | 2 | 15% | 12 | n/a | n/a | n/a | Al |
| 5% misc Hardware | 1 | 1 | 0 | 0 | 0 | 0 | 1.2 | 1.2 | 2 | 25% | 1.5 | n/a | n/a | n/a | - |
| Camera Structure | 1 | 1 | 0 | 0 | 0 | 0 | 1.6 | 1.6 | - | 15% | 1.8 | n/a | n/a | n/a | - |
| Stray Light Shield | 1 | 1 | 0 | 0 | 0 | 0 | 1.5 | 1.5 | 2 | 15% | 1.7 | n/a | n/a | n/a | - |
| Camera Bellows | 1 | 1 | 0 | 0 | 0 | 0 | 1.5 | 1.5 | 2 | 15% | 2.0 | n/a | n/a | n/a | - |
| Lamp Holder -Fiducials | 1 | 4 | 0 | 0 | 0 | 0 | 0.01 | 0.02 | 2 | 15% | 0.02 | n/a | n/a | n/a | Al |
| 5% misc Hardware | 1 | 1 | 0 | 0 | 0 | 0 | 0.1 | 0.1 | 2 | 15% | 0.09 | n/a | n/a | n/a | - |
| Detector Electronics Assembly Box (DEA) | 1 | 1 | 0 | 0 | 0 | 1 | 6.3 | 6.3 | - | 25% | 7.9 | 50 | 0 | 50 | % analog/digital |
| Read out Control/Digitizer Boad + 1 FPGA | 8 | 8 | 0 | 0 | 0 | 0 | 0.3 | 2.0 | 1 | 25% | 2.5 | n/a | n/a | n/a | 7025 |
| Power Converter –Side A/B | 2 | 2 | 0 | 0 | 0 | 0 | 0.3 | 0.5 | 2 | 25% | 0.6 | n/a | n/a | n/a | 905 |
| Backplane | 1 | 1 | 0 | 0 | 0 | 0 | 0.5 | 0.5 | 2 | 25% | 0.6 | n/a | n/a | n/a | - |
| Housing | 1 | 1 | 0 | 0 | 0 | 0 | 3.0 | 3.0 | 2 | 25% | 3.8 | n/a | n/a | n/a | Al |
| 5% misc Hardware | 1 | 1 | 0 | 0 | 0 | 0 | 0.3 | 0.3 | 2 | 25% | 0.4 | n/a | n/a | n/a | - |
| Digital Processing Assy Box (DPA) | 1 | 1 | 0 | 0 | 1 | 1 | 5.4 | 5.4 | - | 25% | 6.7 | 20 | 0 | 24 | % analog/digital |
| SpaceCube Processor Board | 2 | 2 | 0 | 0 | 0 | 0 | 0.5 | 1.0 | 2 | 25% | 1.3 | n/a | n/a | n/a | 5/90 |
| Heater Control & H/K Board +1 FPGA | 2 | 2 | 0 | 0 | 0 | 0 | 0.5 | 1.0 | 2 | 25% | 1.3 | n/a | n/a | n/a | 7025 |
| Power Converter | 2 | 2 | 0 | 0 | 0 | 0 | 0.5 | 1.0 | 2 | 25% | 1.3 | n/a | n/a | n/a | 905 |
| Backplane | 1 | 1 | 0 | 0 | 0 | 0 | 0.6 | 0.6 | 2 | 25% | 0.8 | n/a | n/a | n/a | - |
| Housing | 1 | 1 | 0 | 0 | 0 | 0 | 1.5 | 1.5 | 2 | 25% | 1.9 | n/a | n/a | n/a | Al |
| 5% misc Hardware | 1 | 1 | 0 | 0 | 0 | 0 | 0.3 | 0.3 | 2 | 25% | 0.3 | n/a | n/a | n/a | - |
| Thermal Subsystem | 1 | 1 | 0 | 0 | 0 | 0 | 0.4 | 0.4 | - | 20% | 0.4 | n/a | n/a | n/a | - |
| Ethane CCHPs Attached to FPA Paddle | 1 | 2 | 0 | 0 | 0 | 0 | 0.2 | 0.4 | 2 | 20% | 0.4 | n/a | n/a | n/a | 1.5 m; 1.27 cm diameter; Aluminum; |
| CAT Grating Assembly | 1 | 1 | - | 0 | 0 | 1 | 9.1 | 9.1 | 1 | 32% | 12 | n/a | n/a | n/a | - |
| Grating Assembly A | 1 | 1 | 0 | 0 | 0 | 0 | 2.3 | 2.3 | 2 | 32% | 3.0 | n/a | n/a | n/a | - |
| Grating AssemblyStructure | 1 | 1 | 0 | 0 | 0 | 0 | 0.7 | 0.7 | 2 | 30% | 0.9 | n/a | n/a | n/a | Ti |
| Grating Facet Frames - 8 different frame sizes. | 48 | 48 | 0 | 0 | 0 | 0 | 0.03 | 1.4 | 2 | 30% | 1.9 | n/a | n/a | n/a | - |
| Silicon Membranes | 48 | 48 | 0 | 0 | 0 | 0 | 0.003 | 0.1 | 1 | 55% | 0.2 | n/a | n/a | n/a | - |
| Grating Assembly to FMA Module Flexures | 3 | 3 | 0 | 0 | 0 | 0 | 0.01 | 0.02 | 2 | 30% | 0.03 | n/a | n/a | n/a | - |
| Grating Assembly B | 1 | 1 | 0 | 0 | 0 | 0 | 2.3 | 2.3 | 1 | 32% | 3.0 | n/a | n/a | n/a | - |
| Grating AssemblyStructure | 1 | 1 | 0 | 0 | 0 | 0 | 0.7 | 0.7 | 2 | 30% | 0.9 | n/a | n/a | n/a | Ti |
| Grating Facet Frames - 8 different frame sizes. | 48 | 48 | 0 | 0 | 0 | 0 | 0.03 | 1.4 | 2 | 30% | 1.9 | n/a | n/a | n/a | - |
| Silicon Membranes | 48 | 48 | 0 | 0 | 0 | 0 | 0.003 | 0.1 | 1 | 55% | 0.2 | n/a | n/a | n/a | - |
| Grating Assembly to FMA Module Flexures | 3 | 3 | 0 | 0 | 0 | 0 | 0.01 | 0.02 | 2 | 30% | 0.03 | n/a | n/a | n/a | - |
| Grating Assembly C | 1 | 1 | 0 | 0 | 0 | 0 | 2.3 | 2.3 | 1 | 32% | 3.0 | n/a | n/a | n/a | - |
| Grating AssemblyStructure | 1 | 1 | 0 | 0 | 0 | 0 | 0.7 | 0.7 | 2 | 30% | 0.9 | n/a | n/a | n/a | Ti |
| Grating Facet Frames - 8 different frame sizes. | 48 | 48 | 0 | 0 | 0 | 0 | 0.03 | 1.4 | 2 | 30% | 1.9 | n/a | n/a | n/a | - |
| Silicon Membranes | 48 | 48 | 0 | 0 | 0 | 0 | 0.003 | 0.1 | 1 | 55% | 0.2 | n/a | n/a | n/a | - |
| Grating Assembly to FMA Module Flexures | 3 | 3 | 0 | 0 | 0 | 0 | 0.01 | 0.02 | 2 | 30% | 0.03 | n/a | n/a | n/a | - |
| Grating Assembly D | 1 | 1 | 0 | 0 | 0 | 0 | 2.3 | 2.3 | 1 | 32% | 3.0 | n/a | n/a | n/a | - |

# Table A-1. Expanded Payload MEL

| Item | Per Assy | Flight Set | QUAL | EDU | ETU | Spares | CBE Unit Mass (kg) | CBE Total Flight Mass (kg) | AIAA Maturity Code | Mass Growth Allow (%) | Max Exp. Mass [kg] | Science Average (W) | Safehold (W) | Peak (W) | Comments, Material, Heritage |
|---|---|---|---|---|---|---|---|---|---|---|---|---|---|---|---|
| Grating Assembly/Structure | 1 | 1 | 0 | 0 | 0 | 0 | 0.7 | 0.7 | 2 | 30% | 0.9 | n/a | n/a | n/a | - |
| Grating Facet Frames- 8 different frame sizes. | 1 | 48 | 0 | 0 | 0 | 0 | 0.03 | 1.4 | 2 | 30% | 1.9 | n/a | n/a | n/a | - |
| Silicon Membranes | 1 | 48 | 0 | 0 | 0 | 0 | 0.003 | 0.1 | 1 | 55% | 0.2 | n/a | n/a | n/a | - |
| Grating Assembly to FMA Module Flexures | 1 | 3 | 0 | 0 | 0 | 0 | 0.01 | 0.02 | 2 | 30% | 0.03 | n/a | n/a | n/a | - |
| HTRS | 1 | 1 | 1 | 0 | 0 | 1 | 10.5 | 23 | - | 22% | 27 | 109* | 0 | 109* | - |
| Detector Unit | 1 | 1 | 0 | 1 | 1 | 1 | 0.66 | 2.5 | - | - | 2.9 | 22 | 0 | 22 | Mars rovers |
| SDD array | 1 | 1 | 0 | 0 | 0 | 0 | 0.1 | 0.1 | 1 | 55% | 0.2 | n/a | n/a | n/a | - |
| Housing | 1 | 1 | 0 | 0 | 0 | 0 | 0.5 | 0.5 | 2 | 15% | 0.6 | n/a | n/a | n/a | - |
| Charge sensitive preamplifiers | 37 | 1 | 0 | 0 | 0 | 0 | 0.05 | 1.9 | 2 | 15% | 2.1 | n/a | n/a | n/a | - |
| Heaters | 4 | 1 | 0 | 0 | 0 | 0 | 0.01 | 0.04 | 1 | 25% | 0.1 | n/a | n/a | n/a | - |
| Filter wheel | 1 | 1 | 1 | 0 | 0 | 0 | 2.1 | 2.1 | 2 | 15% | 2.4 | 20 | 0 | 20 | ASTRO-H |
| Wheel - 5 positions | 1 | 1 | 0 | 0 | 0 | 0 | 0.2 | 0.2 | - | - | - | n/a | n/a | n/a | - |
| Wheel supporting plate | 1 | 1 | 0 | 0 | 0 | 0 | 0.6 | 0.6 | - | - | - | n/a | n/a | n/a | - |
| Motor | 1 | 1 | 0 | 0 | 0 | 1 | 0.5 | 0.5 | - | - | - | n/a | n/a | n/a | - |
| Mechanical structure | 1 | 1 | 0 | 0 | 0 | 0 | 0.8 | 0.8 | - | - | - | n/a | n/a | n/a | - |
| Detector Electronic Unit (DEU) | 1 | 1 | 1 | 1 | 1 | 1 | 0.6 | 6.4 | - | - | 7.8 | 20 | 0 | 20 | INTEGRAL SPI, ECLAIRs |
| ADC boards (shaping, sample and hold) | 37 | 1 | 0 | 0 | 0 | 0 | 0.1 | 3.7 | 2 | 25% | 4.6 | n/a | n/a | n/a | - |
| Power configuration | 37 | 1 | 0 | 0 | 0 | 0 | 0.05 | 1.9 | 3 | 20% | 2.2 | n/a | n/a | n/a | - |
| Configuration control | 2 | 1 | 0 | 0 | 0 | 0 | 0.4 | 0.8 | 3 | 20% | 1.0 | n/a | n/a | n/a | - |
| Housing | 1 | 1 | 0 | 0 | 0 | 0 | 1 | 1.0 | 2 | 15% | 1.2 | n/a | n/a | n/a | - |
| Central Electronic Unit (CEU) | 1 | 1 | 1 | 1 | 1 | 1 | 7.2 | 12 | - | - | 14 | 34 | 0 | 34 | INTEGRAL SPI, ECLAIRs |
| FPGA boards | 4 | 1 | 0 | 0 | 0 | 0 | 0.5 | 2 | 2 | 25% | 2.5 | n/a | n/a | n/a | - |
| DC/DC cards | 4 | 1 | 0 | 0 | 0 | 0 | 1.3 | 5.2 | 2 | 25% | 6.5 | n/a | n/a | n/a | - |
| Data Processing Unit | 1 | 1 | 0 | 0 | 0 | 0 | 2 | 2 | 3 | 20% | 2.4 | n/a | n/a | n/a | - |
| Configuration control | 1 | 1 | 0 | 0 | 0 | 0 | 0.4 | 0.4 | 3 | 20% | 0.5 | n/a | n/a | n/a | - |
| Power and filter wheel electronics | 1 | 1 | 0 | 0 | 0 | 0 | 2 | 2.0 | 3 | 20% | 2.4 | n/a | n/a | n/a | - |
| Housing | 1 | 1 | 0 | 0 | 0 | 0 | 1 | 1.0 | 2 | 15% | 1.2 | n/a | n/a | n/a | - |
| XPOL | 1 | 1 | 0 | 0 | 0 | 0 | 8.8 | 8.8 | - | 20% | 10.6 | 46 | 0 | 46 | - |
| Focal Plane Assembly | 1 | 1 | 0 | 0 | 0 | 0 | 4.8 | 4.8 | - | - | 5.7 | 12 | 0 | 12 | - |
| GPD+FW | 1 | 1 | 0 | 0 | 0 | 0 | 3.3 | 3.3 | - | - | 3.8 | 2 | 0 | 2 | - |
| GAS Pixel Detector (GPD) | 1 | 1 | 0 | 0 | 0 | 1 | 0.4 | 0.4 | 2 | 30% | 0.5 | 2 | 0 | 2 | - |
| Filter Wheel (FW) | 1 | 1 | 0 | 0 | 0 | 1 | 1.6 | 1.6 | 2 | 15% | 1.8 | 4 | 0 | 4 | - |
| Mech I/F (+ prebaffle) | 1 | 1 | 0 | 0 | 0 | 1 | 1.3 | 1.3 | 2 | 15% | 1.5 | n/a | n/a | n/a | - |
| Back End Electronics (BEE) | 1 | 1 | 0 | 0 | 0 | 0 | 1.6 | 1.6 | - | - | 1.9 | 10 | 10 | 10 | - |
| Mech I/F (+ Backplane) | 1 | 1 | 0 | 0 | 0 | 1 | 1.1 | 1.1 | 2 | 15% | 1.2 | n/a | n/a | n/a | - |
| #4 Cards cPCI (1 = High Voltage Power Supply) | 4 | 1 | 0 | 0 | 0 | 4 | 0.13 | 0.5 | 2 | 25% | 0.7 | n/a | n/a | n/a | - |
| Control Electronics (CE) | 1 | 1 | 0 | 0 | 0 | 0 | 4.0 | 4.0 | - | - | 4.9 | 34 | 0 | 34 | - |
| Mech I/F (+ Backplane) | 1 | 1 | 0 | 0 | 0 | 1 | 2.3 | 2.3 | 2 | 25% | 2.8 | n/a | n/a | n/a | - |
| #3 Cards Extended | 3 | 1 | 0 | 0 | 0 | 3 | 0.4 | 1.3 | - | 20% | 1.5 | n/a | n/a | n/a | - |
| Mass Memory (MM) Card Extended | 1 | 1 | 0 | 0 | 0 | 1 | 0.5 | 0.5 | 2 | 25% | 0.6 | n/a | n/a | n/a | - |

* Power total includes 70% dc-dc converter efficiency

Table A-2. Spacecraft MEL

| Item | # per Assy | # Flight Assy's | Qual Assy's | EDU | ETU | Spares | CBE Unit Mass per Assy (kg) | CBE Total Flight Mass (Kg) | AIAA Maturity Code | Mass Growth Allow. (%) | Max Exp. Mass [kg] | Description | TRL | Heritage |
|---|---|---|---|---|---|---|---|---|---|---|---|---|---|---|
| **INSTRUMENT MODULE** | | | | | | | | | | | | | | |
| **Mechanical Component List** | | | | | | | | **141** | | **16%** | **163** | | | |
| IM fixed deck (Fixed Instrument Platform, FIP) | 1 | 1 | 0 | 0 | 1 | 0 | 85 | 85 | 2 | 15% | 98 | honeycomb panel, .020" aluminum facesheets, 4" 3.1 pcf alum core, with 2 VCHPs | 8 | Honeycomb panel construction (common to many/most structures |
| MIP deck | 1 | 1 | 0 | 0 | 1 | 0 | 20 | 20 | 2 | 15% | 23 | honeycomb panel, .020" aluminum facesheets, 4" 3.1 pcf alum core, with 2 VCHPs | 8 | Honeycomb panel construction (common to many/most structures |
| MIP Drive Motor / Shaft Assy | 1 | 1 | 0 | 0 | 1 | 0 | 10 | 10 | 2 | 15% | 12 | 2 stepper motors with encoders, driving a gear train. Motors are primary & redundant via clutch | 7 | Chandra, Soho, Voyager |
| MIP launch locks | 1 | 3 | 0 | 0 | 1 | 0 | 1.0 | 3.0 | 3 | 10% | 3 | 1/2" diameter separation nut pyro-actuated with 2 NSI's. Bolt catcher and additional bracketry | 9 | Space Shuttle, Deep Impact, Messenger, many others |
| MIP radiator struts & fittings (not incl radiator) | 1 | 2 | 0 | 0 | 1 | 0 | 1 | 2 | 2 | 15% | 3 | CFRP round tubes bonded to Titanium end fittings | 8 | Swift-BAT, UARS, many others |
| FIP radiator struts & fittings (not incl radiator) | 1 | 4 | 0 | 0 | 1 | 0 | 1.6 | 6.4 | 2 | 15% | 7 | CFRP round tubes bonded to Titanium end fittings | 8 | Swift-BAT, UARS, many others |
| fasteners | 1 | 1 | 0 | 0 | 1 | 0 | 3 | 3 | 2 | 20% | 3 | -- | 9 | Common |
| Sunshade | 1 | 1 | 0 | 0 | 1 | 0 | 10.5 | 11 | 1 | 25% | 13 | composite tube frame with MLI. | 8 | Swift, many others |
| **GN&C Subsystem Component List** | | | | | | | | **2** | | **47%** | **3** | | | |
| Coarse Sun Sensor | 1 | 2 | 0 | 0 | 0 | 1 | 0.16 | 0.32 | 5 | 3% | 0 | Coarse solar aspect angle detector | 8 | TRMM, TRACE, MAP, QUICKSAT, EO-1, HESSI, Swift, SDO, LRO |
| Fiducial Light Assy (ea) | 1 | 18 | 1 | 0 | 2 | 2 | 0.1 | 1.80 | 2 | 55% | 3 | | 8 | Chandra |
| **Thermal Component List** | | | | | | | | **51** | | **15%** | **58** | | | |
| XMS Cryocooler Radiator(.71 sq.m) | 1 | 1 | 0 | 0 | 0 | 0 | 4 | 4 | 3 | 15% | 5 | honeycomb panel | 8 | Common |
| WFI Cold Finger Radiator(.38 sq.m) | 1 | 1 | 0 | 0 | 0 | 0 | 2.6 | 2.6 | 3 | 15% | 3 | -- | 8 | Common |
| MIP Radiator(.86 sq. m) | 1 | 1 | 0 | 0 | 0 | 0 | 5 | 5 | 3 | 15% | 6 | -- | 8 | Common |
| FIP Radiator(1.25 sq.m) | 1 | 1 | 0 | 0 | 0 | 0 | 7.4 | 7.4 | 3 | 15% | 9 | -- | 8 | Common |
| XGS CCD Camera Radiator (.3 sq. m) | 1 | 1 | 0 | 0 | 0 | 0 | 1 | 1 | 3 | 15% | 1 | -- | 8 | Common |
| Heat Pipes on MIP | 1 | 10 | 0 | 0 | 0 | 0 | 0.50 | 5.0 | 3 | 15% | 6 | -- | 8 | Common |

**Table A-2. Spacecraft MEL**

| Item | # per Assy | # Flight Assy's | Qual | EDU | ETU | Spares | CBE Unit Mass (kg) | CBE Total Flight Mass (kg) | AIAA Maturity Code | Mass Growth Allow. (%) | Max Exp. Mass [kg] | Description | TRL | Heritage |
|---|---|---|---|---|---|---|---|---|---|---|---|---|---|---|
| Heat Pipes on FIP | 1 | 20 | 0 | 0 | 0 | 0 | 1 | 10 | 3 | 15% | 12 | -- | 8 | Landsat-5 ETM |
| MLI (15) on Cryocooler Rad Backside | 1 | 1 | 0 | 0 | 0 | 0 | 0.36 | 0.36 | 3 | 15% | 0.41 | -- | 8 | Common |
| MLI (15) on MIP Radiator Backside | 1 | 1 | 0 | 0 | 0 | 0 | 0.4 | 0.4 | 3 | 15% | 1 | -- | 8 | Common |
| MLI (15) Tent for MIP Electronics | 1 | 1 | 0 | 0 | 0 | 0 | 2.5 | 2.5 | 3 | 15% | 3 | -- | 8 | Common |
| Aft Fixed Sunshield (15 Layer MLI) | 1 | 1 | 0 | 0 | 0 | 0 | 1 | 1 | 1 | 25% | 1 | -- | 8 | Common |
| Buttons, Velcro and tapes ML1 | 1 | 100 | 0 | 0 | 0 | 0 | 0.01 | 1.0 | 3 | 15% | 1 | -- | 8 | Common |
| Buttons, Velcro,Tape for Sunshield | 1 | 50 | 0 | 0 | 0 | 0 | 0.01 | 1 | 3 | 15% | 1 | -- | 8 | TRMM, ICESat/GLASS |
| Survival Heaters (redundancy incl) | 1 | 40 | 0 | 0 | 0 | 1 | 0.05 | 2.0 | 3 | 15% | 2 | -- | 8 | Common |
| MIP Heaters | 1 | 8 | 0 | 0 | 0 | 1 | 0.1 | 1 | 3 | 15% | 1 | -- | 8 | Common |
| MLI (15) XGS CCD Camera Radiator | 1 | 1 | 0 | 0 | 0 | 0 | 0.42 | 0.42 | 3 | 15% | 0 | -- | 8 | Common |
| MLI (15 layers) Enclosure XMS | 1 | 1 | 0 | 0 | 0 | 0 | 1 | 1 | 3 | 15% | 1 | -- | 8 | Common |
| Thermostats for Mech Op Heaters | 1 | 10 | 0 | 0 | 0 | 0 | 0.10 | 1.0 | 3 | 15% | 1 | -- | 8 | Common |
| Mechanisms Op Heaters | 1 | 10 | 0 | 0 | 0 | 0 | 0.1 | 1 | 3 | 15% | 1 | -- | 8 | Common |
| XGS Camera Radiator Standoff | 1 | 3 | 0 | 0 | 0 | 0 | 0.01 | 0.04 | 2 | 20% | 0.05 | Ti | 6 | Common |
| XGS Radiator Bracket | 1 | 1 | 0 | 0 | 0 | 0 | 0.1 | 0.1 | 2 | 20% | 0 | Alum | 6 | Common |
| Thermostats | 1 | 50 | 0 | 0 | 0 | 0 | 0.10 | 5.0 | 3 | 15% | 6 | -- | 8 | Common |
| **Avionics Component List** | | | | | | | | 14 | | 47% | 21 | | | |
| Instrument Module RIU | 1 | 1 | 0 | 1 | 0 | 0 | 14 | 14 | -- | -- | 21 | -- | 7 | Common |
| Instrument Module PDU | 1 | 1 | 0 | 0 | 0 | 0 | 1.8 | 1.8 | 2 | 15% | 2.1 | -- | 7 | Common |
| Low Voltage Power Converter | 1 | 1 | 0 | 0 | 0 | 0 | 1.8 | 1.8 | 2 | 15% | 2.1 | -- | 7 | Common |
| Focus Driver | 1 | 2 | 0 | 0 | 0 | 0 | 0.8 | 1.6 | 2 | 15% | 1.8 | -- | 7 | Common |
| Analog I/O Card | 1 | 2 | 0 | 0 | 0 | 0 | 0.8 | 1.6 | 2 | 15% | 1.8 | -- | 7 | Common |
| Analog I/O Card | 1 | 2 | 0 | 0 | 0 | 0 | 0.8 | 1.6 | 2 | 15% | 1.8 | -- | 7 | Common |
| SpaceWire Router | 1 | 2 | 0 | 0 | 0 | 0 | 0.8 | 1.6 | 2 | 15% | 1.8 | -- | 7 | WMAP, SDO, Swift, GOES-R |
| Backplane | 1 | 1 | 0 | 0 | 0 | 0 | 0.25 | 0.25 | 2 | 15% | 0.3 | -- | 7 | Common |
| Chassis | 1 | 1 | 0 | 0 | 0 | 0 | 4.0 | 4.0 | 2 | 15% | 4.5 | -- | 8 | Common |
| **Power and Data Harnesses** | 1 | | | | | | **0.59** | **59** | | **30%** | **76** | | | |
| Power and Data Harnesses | 1 | 100 | 0 | 1 | 0 | 0 | 0.59 | 59 | 2 | 30% | 76 | -- | 6 | Common |

## Table A-2. Spacecraft MEL

| Item | # per Assy | # Flight Assy's | Qual | EDU | ETU | Spares | CBE Unit Mass (kg) | CBE Total Flight Mass (kg) | AIAA Maturity Code | Mass Growth Allow. (%) | Max Exp. Mass [kg] | Description | TRL | Heritage |
|---|---|---|---|---|---|---|---|---|---|---|---|---|---|---|
| **Payload Accommodations Equip** | | | | | | | | 46 | | 21% | 56 | | | |
| Linear Actuator Mount (Cylinder) | 1 | 5 | 0 | 0 | 1 | 1 | 0.2 | 1 | 2 | 15% | 1 | Actuator/metrology assembly, dual redundant motors, 2 for XGS Camera Focus, 2 for XGS Camera Translation, 3 for XMS, 3 for WFI | 6 | JWST |
| Linear Actuator Mount (gusseted) | 1 | 5 | 0 | 0 | 1 | 1 | 1.0 | 5.0 | 2 | 15% | 6 | Actuator/metrology assembly, dual redundant motors, 2 for XGS Camera Focus, 2 for XGS Camera Translation, 3 for XMS, 3 for WFI | 6 | JWST |
| Mount and Baffle for XGS | 1 | 1 | 0 | 0 | 0 | 0 | 13 | 13 | 1 | 25% | 17 | CFRP stand with baffles. .060" wall thickness | 6 | typical CFRP construction |
| Radiation Monitor | 1 | 1 | 0 | 0 | 0 | 1 | 3.6 | 3.6 | 5 | 5% | 4 | EPHIN - Electron Proton Helium Instrument | 9 | Soho , chandra |
| Magnetic Broom | 1 | 1 | 0 | 0 | 0 | 1 | 7 | 7 | 2 | 25% | 8 | _ | 6 | Chandra |
| Calibration Source Shutter Assembly | 1 | 1 | 0 | 0 | 0 | 1 | 0.5 | 0.5 | 2 | 25% | 1 | Low precision blocking mechanism | 8 | Spitzer IRAC Calibration Source Blocker |
| Purge System | 1 | 1 | 0 | 0 | 0 | 0 | 1 | 1 | 1 | 25% | 1 | _ | 9 | Cassini CIRS |
| Door Assembly | 1 | 1 | 0 | 0 | 0 | 0 | 5.2 | 5.2 | 1 | 25% | 6 | Al, Stainless Steel | 6 | Common |
| Door | 1 | 1 | 0 | 0 | 0 | 0 | 2.50 | 2.5 | 1 | 25% | 3.13 | Alum | 6 | Common |
| Frame | 1 | 1 | 0 | 0 | 0 | 0 | 1.4 | 1.4 | 1 | 25% | 2 | Alum | 6 | Common |
| Travel Sensor - CME it | 1 | 1 | 0 | 0 | 0 | 0 | 0.01 | 0.01 | 1 | 25% | 0.01 | _ | 6 | Common |
| Hinge + Pin puller | 1 | 1 | 0 | 0 | 0 | 0 | 1.0 | 1.00 | 1 | 25% | 1 | Alum | 6 | Common |
| 5% misc Hardware | 1 | 1 | 0 | 0 | 0 | 0 | 0.25 | 0.25 | 1 | 30% | 0.32 | _ | 6 | Common |
| Common "Snout" Baffle for XMS and WFI | 1 | 1 | 0 | 0 | 0 | 0 | 5.0 | 5.0 | 2 | 15% | 6 | CFRP hollow cone shape. .060" wall thickness | 6 | Chandra |
| **DEPLOYMENT MODULE** | | | | | | | | | | | | | | |
| **Mechanical Component List** | | | | | | | | 316 | | 12% | 355 | | | |
| ADAM mast | 1 | 3 | 0 | 0 | 0 | 0 | 56 | 169 | 3 | 10% | 186 | Articulated Deployable mast in a canister. 12.1m deployment length | 7 | SRTM mission on STS-99, Int'l Space Station, Nustar |
| top deck (previously listed in Bus Module section) | 1 | 1 | 0 | 0 | 1 | 0 | 16 | 16 | 2 | 15% | 19 | _ | 8 | Honeycomb panel construction (common to many/most structures) |
| FIP to Bus Launch Locks (both sides of entire system) | 1 | 6 | 0 | 0 | 0 | 0 | 1 | 6 | 3 | 10% | 7 | 1/2" diameter separation nut pyro-actuated with 2 NSI's. Bolt catcher and additional bracketry | 9 | Space Shuttle, Deep Impact, Messenger, many others |
| Shroud Stowage Channel (was: "MLI sleeve canister") | 1 | 1 | 0 | 0 | 0 | 0 | 15 | 15 | 2 | 15% | 17 | CFRP C-channel | 8 | Common |

# Table A-2. Spacecraft MEL

| Item | # per Assy | # Flight Assy's | Qual | EDU | ETU | Spares | CBE Unit Mass (kg) | CBE Total Flight Mass (kg) | AIAA Maturity Code | Mass Growth Allow. (%) | Max Exp. Mass [kg] | Description | TRL | Heritage |
|---|---|---|---|---|---|---|---|---|---|---|---|---|---|---|
| MLI Shroud (Whipple Shield type) | 1 | 1 | 0 | 0 | 0 | 0 | 100 | 100 | 2 | 15% | 115 | two concentric collapsible 5-layer MLI blanket cylinders (2-mil outer kapton, 5-layers 1/4-mil inner + Dacron scrim, 1-mil inner kapton), tailored to an accordion shape | 6 | accordian structure used on HST Electronic Support Module on STS-109 (March 2002). |
| MLI baffles (10 layer MLI) | 1 | 2 | 0 | 0 | 0 | 0 | 4.1 | 8.3 | 2 | 15% | 9 | 1-mil Tantalum foil with 2-mil Kapton backers on both sides | 8 | Swift-BAT Graded-Z shield |
| MLI attachment to IM | 1 | 72 | 0 | 0 | 0 | 0 | 0.004 | 0.29 | 3 | 10% | 0 | 72 NAS1351N3-10 screws and #10 washers | 9 | Common |
| fasteners | 1 | 1 | 0 | 0 | 0 | 1 | 2.0 | 2.0 | 3 | 15% | 2 | standard NAS1351 series fasteners | 9 | Common |
| **Thermal Component List** | | | | | | | **4** | **4** | | **15%** | **5** | | | |
| thermofoil heaters (canister mechanism) | 1 | 12 | 0 | 0 | 0 | 1 | 0.1 | 1 | 3 | 15% | 1 | Tayco Heaters | 8 | Common |
| thermostats for heaters (canister mechanism) | 1 | 12 | 0 | 0 | 0 | 1 | 0.10 | 1.20 | 3 | 15% | 1 | Honeywell 701 series | 8 | COBE, TRMM, WMAP, and many others |
| thermistors for Metering Truss | 1 | 80 | 0 | 0 | 0 | 4 | 0.02 | 2 | 3 | 15% | 2 | YSI | 8 | Common |
| **Power and Data Harnesses** | | | | | | | **1** | **118** | | **30%** | **154** | | | |
| Power and Data Harness | 1 | 200 | 0 | 1 | 0 | 0 | 1 | 118 | 2 | 30% | 154 | — | 6 | Common |
| **SPACECRAFT MODULE** | | | | | | | | **586** | | | **677** | | | |
| **Mechanical Component List** | | | | | | | | **586** | | **15%** | **677** | | | |
| bottom deck | 1 | 1 | 0 | 0 | 0 | 0 | 47 | 47 | 2 | 15% | 54 | honeycomb panel, .020" aluminum facesheets, 2" 3.1 pcf CFRP core, with 2 CCHPs | 8 | Honeycomb panel construction (common to many/most structures). Swift-BAT |
| bus frame | 1 | 1 | 0 | 0 | 0 | 0 | 36 | 36 | 2 | 15% | 41 | CFRP longerons and L-shaped brackets | 8 | innumerable. Chandra, etc |
| component panels (9) | 1 | 9 | 0 | 0 | 0 | 0 | 6.1 | 55.3 | 2 | 15% | 64 | honeycomb panel, .020" CFRP facesheets M55J/954-3, 1.5" 3.1 pcf core | 8 | Honeycomb panel construction (common to many/most structures). SDO |
| secondary structure (RW brackets, etc) | 1 | 1 | 0 | 0 | 0 | 0 | 20 | 20 | 2 | 15% | 22 | aluminum riveted structure to hold wheel | 8 | SDO |
| fasteners | 1 | 1 | 0 | 0 | 0 | 0 | 9.8 | 9.8 | 1 | 25% | 12.3 | Various NAS1351N-series screws & washers. | 9 | Common |
| isogrid metering tube and end fittings | 1 | 1 | 0 | 0 | 0 | 0 | 312 | 312 | 2 | 15% | 358 | .040" thick CFRP tube with 1.375" tall, .188" wide isogrid stiffeners. Includes Titanium fittings top and bottom rings | 8 | Boeing 787 Dreamliner, Minotaur AGS rocket fairing |
| Venting assembly | 1 | 14 | 0 | 0 | 0 | 1 | 0.5 | 7.0 | 5 | 3% | 7.21 | | 9 | SDO |

Table A-2. Spacecraft MEL

| Item | # per Assy | # Flight Assy's | Qual Assy's | EDU | ETU | Spares | CBE Unit Mass (kg) | CBE Total Flight Mass (kg) | AIAA Maturity Code | Mass Growth Allow. (%) | Max Exp. Mass [kg] | Description | TRL | Heritage |
|---|---|---|---|---|---|---|---|---|---|---|---|---|---|---|
| propulsion structure, valve & plumbing mounting plates | 1 | 2 | 0 | 0 | 0 | 0 | 10 | 20 | 2 | 15% | 23 | aluminum plate 60 x 91 x 6 cm | 8 | innumerable. Following SDO propulsion system layout. |
| Mounting brackets for propulsion tanks | 1 | 5 | 0 | 0 | 0 | 0 | 4.1 | 20.6 | 1 | 25% | 25.8 | aluminum frustum interfaces to prop tank equatorial mount flange | 8 | innumerable. Following SDO propulsion system layout. |
| Baffles | 1 | 1 | 0 | 0 | 0 | 0 | 2 | 2 | 2 | 15% | 2 | 1-mil Tantalum foil with 2-mil Kapton backers on both sides | 8 | Swift-BAT Graded-Z shield |
| body mounted solar array substrate w/mounts | 1 | 1 | 0 | 0 | 0 | 0 | 49 | 49 | 2 | 15% | 56.6 | .015" CFRP facesheets, 1.5" aluminum honeycomb core. | 8 | Construction typical to most previous spacecraft. SDO |
| ultraflex solar array mounts (release mechanism) | 1 | 2 | 0 | 0 | 0 | 1 | 3 | 6 | 1 | 25% | 8 | pyro activated capture latch | 8 | Composite & machined titanium al ally |
| high-gain antenna mount | 1 | 1 | 0 | 0 | 0 | 0 | 1.6 | 1.6 | 1 | 15% | 1.89 | aluminum bracket to hold the antenna assy | 8 | Common |
| **GN&C Subsystem Component List** | | | | | | | 80 | 80 | | 3% | 83 | | | |
| Reaction Wheel | 1 | 5 | 0 | 0 | 0 | 1 | 15 | 75 | 5 | 3% | 77 | Fine balance option | 7 | NPOESS |
| Coarse Sun Sensor | 1 | 6 | 0 | 0 | 0 | 0 | 0.16 | 0.96 | 5 | 3% | 0.99 | Coarse solar aspect angle detector | 8 | TRMM, TRACE, MAP, QUICKSAT, EO-1, HESSI, Swift, SDO, LRO |
| Gyro | 1 | 1 | 1 | 0 | 0 | 1 | 5 | 5 | 5 | 3% | 5 | HRG technology Internally redundant (4 for 3 gyros) | 8 | GE 1 through 4, Echostar 3 & 4, Chinastar, NEAR, AcES 1 & 2, Cassini, ICO Constellation, EOS, TDRSS, GALAXY XI, GOES N Series |
| **Propulsion Component List** | | | | | | | 56 | 56 | | 3% | 57 | | | |
| Hz Tank | 1 | 2 | 1 | 0 | 0 | 0 | 5.7 | 11.3 | 5 | 3% | 11.7 | Ti PMD tank | 8 | INMARSAT-3 |
| NTO Tank | 1 | 2 | 1 | 0 | 0 | 0 | 4 | 8 | 5 | 3% | 8 | Ti PMD tank | 8 | SPACENET |
| He Tank | 1 | 1 | 1 | 0 | 0 | 0 | 7.0 | 7.0 | 5 | 3% | 7.20 | Composite Over-wrapped Pressurant Vessel | 8 | ETS8 Xenon |
| 22N Hz/NTO Thruster | 1 | 12 | 1 | 0 | 0 | 1 | 0.74 | 8.9 | 5 | 3% | 9.15 | ISP >300 Fee press 80-400 psia | 8 | OSC Wild Geese |
| 0.9N Hz Thruster | 1 | 4 | 1 | 0 | 0 | 1 | 0 | 1 | 5 | 3% | 1 | ISP 224-209s Feed press 400-90 pia (27.6-6.2 bar) | 8 | DS-1, Skynet 4, ADEOS3, MSTI |
| 3/8", dual coil, LP Latch Valve | 1 | 6 | 1 | 0 | 0 | 1 | 0.73 | 4.38 | 5 | 3% | 4.51 | 5,000 cycle life | 8 | MUSES-C, ASTRO-F, classified |
| 3/8", dual coil, HP Latch Valve | 1 | 1 | 1 | 0 | 0 | 1 | 1 | 1 | 5 | 3% | 1 | 5,000 cycle life | 8 | MUSES-C, ASTRO-F, classified |
| Pyro Valves (NO/NC) | 1 | 12 | 1 | 0 | 0 | 1 | 0.16 | 1.9 | 5 | 3% | 1.98 | Zero leak metal seal | 8 | Atlas, Rosetta, WINDS, many more |
| Check Valve (dual seat) | 1 | 1 | 1 | 0 | 0 | 1 | 0.1 | 0.2 | 5 | 3% | 0 | 1000 operation cycles | 8 | Many programs, first flew for LM in 1994 |

**Table A-2. Spacecraft MEL**

| Item | # per Assy | # Flight Assy's | Qual | EDU | ETU | Spares | CBE Unit Mass (kg) | CBE Total Flight Mass (kg) | AIAA Maturity Code | Mass Growth Allow. (%) | Max Exp. Mass [kg] | Description | TRL | Heritage |
|---|---|---|---|---|---|---|---|---|---|---|---|---|---|---|
| He Regulator (Series Redundant) | 1 | 1 | 1 | 0 | 0 | 1 | 1.3 | 1.3 | 5 | 3% | 1.29 | Set regulated output press between | 8 | Mars Odyssey, Mars Orbiter, Cluster II, Messenger |
| HP Filters (10μ) | 1 | 1 | 1 | 1 | 0 | 1 | 0 | 0 | 5 | 3% | 0 | 10 micron filtration | 8 | HS-601, HS-702, Cassini, Chandra, many other S/C |
| LP Filters (10μ) | 1 | 4 | 1 | 1 | 0 | 1 | 0.30 | 1.20 | 5 | 3% | 1.2 | 10 micron filtration | 8 | HS-601, HS-702, Cassini, Chandra, many other S/C |
| Fill/Drain Valves | 1 | 11 | 1 | 1 | 0 | 1 | 0 | 1 | 5 | 3% | 1 | Load fuel and pressurant as well as test ports | 8 | Many flight programs |
| Pressure Transducer | 1 | 5 | 1 | 1 | 0 | 1 | 0.27 | 1.35 | 5 | 3% | 1.4 | Reads in voltage the pressure of the tanks | 8 | X-34, ASTRIUM, NEXT |
| Venturi Orifices | 1 | 4 | 1 | 1 | 0 | 1 | 0.1 | 0.2 | 5 | 3% | 0 | Eliminate effects of water hammer | 8 | SDO, LRO, many other S/C |
| Manifold, etc. | 1 | 1 | 1 | 1 | 0 | 0 | 6.8 | 6.8 | 5 | 3% | 7.0 | Tubes, manifold stands, etc | 8 | Many S/C |
| **Thermal Component List** | | | | | | | | 86 | , | 23% | 105.02 | | | |
| bus heaters | 1 | 50 | 0 | 0 | 0 | 2 | 0.1 | 3 | 2 | 20% | 3 | Tayco Heaters | 8 | COBE, WMAP, TRMM, SMEX, SDO, LRO |
| bus thermistors, standard | 1 | 100 | 0 | 0 | 0 | 2 | 0.02 | 2.00 | 2 | 20% | 2.40 | YSI | 8 | COBE, WMAP, TRMM, SMEX, SDO, LRO |
| bus thermostats | 1 | 50 | 0 | 0 | 0 | 2 | 0.1 | 5 | 2 | 20% | 6 | Honeywell 701 series | 8 | COBE, WMAP, TRMM, SMEX, SDO, LRO |
| Bus MLI | 1 | 1 | 0 | 0 | 0 | 0 | 5.5 | 5.5 | 2 | 20% | 6.60 | GBK | 8 | COBE, WMAP, TRMM, SMEX, SDO, LRO |
| Bus thermal interface material | 1 | 15 | 0 | 0 | 0 | 0 | 0.02 | 0.3 | 2 | 20% | 0 | Choseal | 8 | COBE, WMAP, TRMM, SMEX, SDO, LRO |
| Isogrid MLI | 1 | 1 | 0 | 0 | 0 | 0 | 50 | 50 | 2 | 20% | 60 | GBK | 8 | COBE, WMAP, TRMM, SMEX, SDO, LRO |
| ISOgrid Heaters | 1 | 200 | 0 | 0 | 0 | 6 | 0 | 10 | 2 | 20% | 12 | -- | 8 | Common |
| Controllers | 1 | 20 | 0 | 0 | 0 | 1 | 0.22 | 4.40 | 2 | 20% | 5.3 | -- | 8 | SWIFT |
| Thermal Harness | 1 | 1 | 0 | 0 | 0 | 0 | 6 | 6 | 1 | 55% | 10 | -- | 8 | Common |
| ISOgrid Thermistors | 1 | 160 | 0 | 0 | 0 | 5 | 0.0002 | 0.032 | 2 | 20% | 0.04 | -- | 8 | Common |
| **EPS Component List** | | | | | | | | 119 | | 8% | 128 | | | |
| Flex Arrays (incl mast, wing structure & mech) | 1 | 2 | 0 | 0 | 0 | 0 | 14 | 27 | 5 | 3% | 28 | -- | 7 | Common |
| Body Mounted Array (cells & glass | 1 | 1 | 0 | 0 | 0 | 0 | 21 | 21 | 3 | 10% | 23 | -- | 7 | Common |
| 100 Ah LiIon Battery | 1 | 1 | 1 | 0 | 1 | 1 | 30 | 30 | 5 | 3% | 31 | -- | 8 | Common |

Table A-2. Spacecraft MEL

| Item | # per Assy | # Flight Assy's | Qual Assy | EDU | ETU | Spares | CBE Unit Mass (kg) | CBE Total Flight Mass (kg) | AIAA Maturity Code | Mass Growth Allow. (%) | Max Exp. Mass [kg] | Description | TRL | Heritage |
|---|---|---|---|---|---|---|---|---|---|---|---|---|---|---|
| PSE | 1 | 1 | 0 | 1 | 0 | 0 | 41 | 41 | 2 | 15% | 47 | – | 7 | Common |
| **RF Comm Component List** | | | | | | | **30** | **30** | | **4%** | **32** | | | |
| S/Ka Transponder | 1 | 2 | 0 | 1 | 0 | 1 | 3 | 6 | 5 | 3% | 6 | S-band  TT&C  Ka transmit at 26GHz | 7 | flown on DSN missions at 32 GHz |
| S/Ka High Gain Antenna (0.7 Meter), incl. Gimbal assy | 1 | 1 | 0 | 1 | 0 | 1 | 8.5 | 8.5 | 5 | 3% | 8.8 | – | 8 | LRO, JWST |
| 10 watt Ka TWTA | 1 | 2 | 0 | 1 | 0 | 1 | 4 | 7 | 5 | 3% | 7 | 10 watts RF | 8 | LRO |
| 5 watt S-band PA | 1 | 2 | 0 | 1 | 0 | 1 | 0.5 | 1.0 | 5 | 3% | 1.0 | 5 watts RF | 8 | Many missions |
| S-band omni | 1 | 2 | 0 | 1 | 0 | 1 | 1 | 1 | 5 | 3% | 1 | omnis | 8 | XTE, TRMM., TERRA |
| Diplexer | 1 | 2 | 0 | 1 | 0 | 1 | 0.6 | 1.2 | 5 | 3% | 1.2 | dual frequency | 8 | GLAST |
| Triplexer | 1 | 1 | 0 | 1 | 0 | 1 | 1 | 1 | 5 | 3% | 1 | Tri frequency | 8 | Many DSN missions |
| Hybrids | 1 | 2 | 0 | 1 | 0 | 1 | 0.2 | 0.4 | 5 | 3% | 0.4 | splits signal | 8 | Most mission flown |
| Switches | 1 | 4 | 0 | 1 | 0 | 1 | 0.3 | 1 | 5 | 3% | 1 | Ka and s-Band | 8 | SDO |
| Isolator and cabling, misc | 1 | 1 | 0 | 0 | 0 | 0 | 3.0 | 3.0 | 3 | 15% | 3.5 | – | 8 | SDO, JWST, LRO |
| **Avionics Component List** | | | | | | | **35** | **35** | | **15%** | **40** | | | |
| Command and Data Handling Unit | 1 | 1 | 0 | 1 | 0 | 0 | 15 | 15 | | | 18 | – | 7 | – |
| Low Voltage Power Converter - | 1 | 1 | 0 | 0 | 0 | 0 | 1.80 | 1.80 | 2 | 15% | 2.1 | – | 7 | Common |
| Fault Detection and Recovery | 1 | 1 | 0 | 0 | 0 | 0 | 0.80 | 0.80 | 2 | 15% | 0.9 | – | 7 | Common |
| Comm Driver Card | 1 | 2 | 0 | 0 | 0 | 0 | 0.80 | 1.60 | 2 | 15% | 1.8 | – | 7 | Common |
| Memory(SMEM) | 1 | 3 | 0 | 0 | 0 | 0 | 0.80 | 2.40 | 2 | 15% | 2.8 | – | 7 | Common |
| Solid State Data Recorder Card | 1 | 2 | 0 | 0 | 0 | 0 | 0.80 | 1.60 | 2 | 15% | 1.8 | – | 7 | Common |
| Spacewire router - identical card | 1 | 2 | 0 | 0 | 0 | 0 | 0.80 | 1.60 | 2 | 15% | 1.8 | – | 7 | WMAP, SDO, Swift, GOES-R |
| Backplane | 1 | 1 | 0 | 0 | 0 | 0 | 0.28 | 0.28 | 2 | 15% | 0.3 | – | 7 | Common |
| Chassis | 1 | 1 | 0 | 0 | 0 | 0 | 5.17 | 5.17 | 2 | 15% | 5.9 | – | 8 | Common |
| Integrated Avionics Unit | 1 | 1 | 0 | 1 | 0 | 0 | 19.3 | 19.3 | | | 22.2 | – | 7 | – |
| Low Voltage Power Converter - | 1 | 1 | 0 | 0 | 0 | 0 | 1.80 | 1.80 | 2 | 15% | 2.1 | – | 7 | Common |
| Attitude Interface Card | 1 | 2 | 0 | 0 | 0 | 0 | 0.80 | 1.60 | 2 | 15% | 1.8 | – | 7 | Common |
| Deployment Control Card | 1 | 1 | 0 | 0 | 0 | 0 | 0.80 | 1.60 | 2 | 15% | 1.8 | – | 7 | Common |
| Thruster Valve Drive Card | 1 | 4 | 0 | 0 | 0 | 0 | 0.80 | 3.20 | 2 | 15% | 3.7 | – | 7 | Common |
| Analog I/O Card - identical card | 1 | 2 | 0 | 0 | 0 | 0 | 0.80 | 1.60 | 2 | 15% | 1.8 | – | 7 | Common |
| RAD 750 SBC | 1 | 1 | 0 | 0 | 0 | 0 | 0.80 | 1.60 | 2 | 15% | 1.8 | – | 7 | Common |
| Spacewire router - identical card | 1 | 2 | 0 | 0 | 0 | 0 | 0.80 | 1.60 | 2 | 15% | 1.8 | – | 7 | WMAP, SDO, Swift, GOES-R |
| Backplane | 1 | 1 | 0 | 0 | 0 | 0 | 1.75 | 1.75 | 2 | 15% | 2.0 | – | 7 | Common |
| Chassis | 1 | 1 | 0 | 0 | 0 | 0 | 4.58 | 4.58 | 2 | 15% | 5.3 | – | 8 | Common |
| Ultra Stable Oscillator | 1 | 1 | 0 | 0 | 0 | 1 | 0.50 | 0.50 | 2 | 15% | 0.6 | – | 6 | Chandra |

# Table A-2. Spacecraft MEL

| Item | # per Assy | # Flight Assy's | Qual | EDU | ETU | Spares | CBE Unit Mass (kg) | CBE Total Flight Mass (kg) | AIAA Maturity Code | Mass Growth Allow. (%) | Max Exp. Mass [kg] | Description | TRL | Heritage |
|---|---|---|---|---|---|---|---|---|---|---|---|---|---|---|
| **Power and Data Harnesses** | | | | | | | | **92** | | **30%** | **120** | | | |
| Power and Data Harnesses | 1 | 100 | 0 | 1 | 0 | 0 | 0.92 | 92 | 2 | 30% | 120 | — | 6 | Common |
| **OPTICS MODULE** | | | | | | | | | | | | | | |
| **Mechanical Component List** | | | | | | | | **65** | | **15%** | **75** | | | |
| FMA mount (bolts only) | 1 | 24 | 0 | 0 | 0 | 0 | 0.1 | 2 | 3 | 10% | 2 | NAS6708J16 bolts with washers | 9 | Common |
| Fore Sunshade Deploy Mechanisms | 1 | 1 | 0 | 0 | 1 | 0 | 1.6 | 1.6 | 1 | 30% | 2.07 | either an inflatable frame or fiberglass rod frame with restraint latch | 7 | various space missions have used restraint latches. Inflatable technology demonstrated on STS-77 |
| Fore Sunshield | 1 | 1 | 0 | 0 | 0 | 0 | 0.22 | 0.22 | 2 | 30% | 0 | MLI blanket with either an inflatable frame or inflatable frame of fiberglass rod frame | 8 | MLI used on most spacecraft. Inflatables per STS-77 Spartan mission. |
| Spacecraft Adapter (S/C Side of Separation System) | 1 | 1 | 0 | 0 | 1 | 0 | 62 | 62 | 2 | 15% | 71 | aluminum cylinder with flanges | 8 | Machined/Sheetmetal al aly |
| **GN&C Subsystem Component List** | | | | | | | | **24** | | **14%** | **28** | | | |
| Star Tracker | 1 | 2 | 0 | 0 | 0 | 0 | 7 | 14 | 5 | 3% | 15 | Autonomous star tracker outputs attitude quaternion | 8 | Spitzer |
| TADS Retroreflector/Periscope Assembly | 1 | 1 | 1 | 1 | 0 | 0 | 7 | 7.0 | 2 | 30% | 9.1 | — | 6 | Chandra |
| Star Tracker Mount Assembly | 1 | 2 | 0 | 0 | 0 | 0 | 2 | 3 | 2 | 30% | 4 | — | 6 | Custom |
| **Thermal Component List** | | | | | | | | **18** | | **17%** | **21** | | | |
| FMA MLI | 1 | 1 | 1 | 0 | 0 | 0 | 8 | 8 | 2 | 20% | 9 | — | 8 | Common |
| Heat Pipes on Spacecraft Adapter | 1 | 8 | 0 | 0 | 1 | 1 | 1.3 | 10.2 | 3 | 15% | 12 | — | 8 | Common |
| **Avionics Component List** | | | | | | | | **12** | | **15%** | **14** | | | |
| Optics Module RIU | 1 | 1 | 0 | 1 | 0 | 0 | 12 | 12 | 2 | 15% | 14 | — | 7 | Common |
| Optic Module PDU | 1 | 1 | 0 | 0 | 0 | 0 | 1.8 | 1.8 | 2 | 15% | 2.1 | — | 7 | Common |
| Low Voltage Power Converter | 1 | 1 | 0 | 0 | 0 | 0 | 1.8 | 1.80 | 2 | 15% | 2.1 | — | 7 | Common |
| Analog I/O Card | 2 | 2 | 0 | 0 | 0 | 0 | 0.8 | 1.6 | 2 | 15% | 1.8 | — | 7 | Common |
| Deployment Control Card | 2 | 2 | 0 | 0 | 0 | 0 | 0.8 | 1.60 | 2 | 15% | 1.8 | — | 7 | Common |
| SpaceWire Router | 2 | 2 | 0 | 0 | 0 | 0 | 0.8 | 1.6 | 2 | 15% | 1.8 | — | 7 | WMAP, SDO, Swift, GOES-R |
| Backplane | 1 | 1 | 0 | 0 | 0 | 0 | 0.20 | 0.20 | 2 | 15% | 0.2 | — | 7 | Common |
| Chassis | 1 | 1 | 0 | 0 | 0 | 0 | 3.6 | 3.6 | 2 | 15% | 4.17 | — | 8 | Common |
| **Power and Data Harnesses** | | | | | | | | **4** | | **30%** | **5** | | | |
| Power and Data Harnesses | 1 | 10 | 0 | 1 | 0 | 0 | 0.4 | 4 | 2 | 30% | 5 | — | 6 | Common |
| **PAF & Separation System (LV Side)** | | | | | | | | **78** | | **15%** | **90** | | | |
| Separation Ring (LV Side of Separation System, stays with LV) | 1 | 1 | 1 | 0 | 0 | 0 | 46 | 46 | 2 | 15% | 53 | Machined/Sheetmetal al aly | 8 | Common |
| Separation System (stays with LV) | 1 | 1 | 1 | 0 | 0 | 0 | 32 | 32 | 2 | 15% | 37 | Machined/Sheetmetal al aly | 6 | Common |

## Table A-2. Spacecraft MEL

| Item | # per Assy | # Flight Assy's | Qual | EDU | ETU | Spares | CBE Unit Mass (kg) | CBE Total Flight Mass (kg) | AIAA Maturity Code | Mass Growth Allow. (%) | Max Exp. Mass [kg] | Description | TRL | Heritage |
|---|---|---|---|---|---|---|---|---|---|---|---|---|---|---|
| **Payload Accomodations Equipment** | | | | | | | | 88 | | 22% | 108 | | | |
| FMA Outer Mirror Cover | 1 | 1 | 0 | 0 | 0 | 0 | 23 | 23 | 2 | 15% | 26 | CFRP isogrid lid | 6 | Cassini CIRS |
| FMA Inner Mirror Cover | 1 | 1 | 0 | 0 | 0 | 0 | 3.0 | 3.0 | 1 | 25% | 4 | 2-mil Kapton sheet with G10 fiberglass hoop which is preloaded and hinged to retract after restraint latch is opened. | 6 | Cassini CIRS |
| FMA Outer Cover Jettison Mechanism | 1 | 3 | 0 | 0 | 1 | 0 | 2 | 6 | 1 | 25% | 8 | paraffin wax release system | 6 | Multiple |
| FMA Inner Cover Deploy Mechanism | 1 | 3 | 0 | 0 | 1 | 0 | 19 | 57 | 1 | 25% | 71 | paraffin wax release system, with a preloaded spring and hinge system | 6 | Cassini CIRS |

# Appendix B. Acronyms

| | |
|---|---|
| AANM | Astronomy and Astrophysics in the New Millennium Survey |
| ACE | Advanced Composition Explorer |
| ACEIT | Automated Cost Estimating Integrated Tools |
| ACIS | AXAF CCD Imaging Spectrometer |
| ACS | Attitude Control System |
| ASIC | Application Specific Integrated Circuit |
| ADAM™ | Able Deployable Articulated Mast |
| ADR | Adiabatic Demagnetization Refrigerator |
| ADRC | ADR Controller |
| AGILE | Astro-rivelatore Gamma a Immagini LEggero |
| AGN | Active Galactic Nucleus |
| AIAA | American Institute of Aeronautics and Astronautics |
| AO | Announcement of Opportunity |
| APS | Active Pixel Sensor |
| ASCA | Advanced Satellite for Cosmology & Astrophysics |
| ASIC | Application-Specific Integrated Circuit |
| ASIST | Advanced System for Integration and Spacecraft Testing |
| AST | Autonomous Star Tracker |
| ASTEROID | Active current Switching Technique ReadOut In X-ray spectroscopy with DEPFET |
| ATP | Authorization to proceed |
| AXAF | Advanced X-ray Astrophysics Facility |
| BBXRT | Broad Band X-ray Telescope |
| BEPAC | Beyond Einstein Program Assessment Committee |
| BH | Black Hole |
| BHC | Black Hole Candidate |
| BI | Back-Illumination |
| BOL | Beginning of Life |
| C&DH | Command and Data Handling |
| CAD | Computer-Aided Design |
| CAD | Cost Analysis Division |
| CADR | Continuous Adiabatic Demagnetization Refrigerator |
| CALDB | Calibration Database |
| CASG | Cost Analysis Steering Group |
| CAT | Critical Angle Transmission |
| CBE | Current Best Estimate |
| CCD | Charge-Coupled Device |
| CCE | Cryocooler Control Electronics |
| CCHPs | Constant Conductance Heat Pipe |
| CCSDS | Consultative Committee for Space Data System |
| CDF | Chandra Deep Field |
| CDR | Critical Design Review |
| CE | Control Electronics |
| CEH | Cost Estimating Handbook |
| CESR | Centre d'Etude Spatiale des Rayonnements |
| CEU | Central Electronic Unit |
| CfA | Center for Astrophysics |
| CFRP | Carbon Fiber Reinforced Plastic |
| CIAO | Chandra Interactive Analysis of Observations |
| CL | Confidence-level |
| CMOS | Complementary Metal–oOxide–Semiconductor |
| CNES | Centre National d'Etudes Spatiales |
| COBE | Cosmic Background Explorer |
| Con-X | Constellation-X |
| COTS | Commercial Off-the-Shelf |
| CPU | Central Processing Unit |
| DA | Data Analysis |
| DC | Direct Current |
| DE | Dark Energy |
| DEA | Detector Electronics Assembly |
| DEPFET | Depleted P-channel Field Effect Transistor |
| DEU | Detector Electronics Unit |
| DM | Dark Matter |
| DM | Deployment Module |
| DS-1 | Deep Space 1 |
| DS-CdTe | Double-sided Strip Cadmium Telluride |
| DSN | Deep-Space Network |
| DSSD | Double-sided Si Strip Detector |
| EA | Effective Area |
| ECLAIRs | Not an acronym; Instrument on SVOM mission |
| EDU | Engineering Demonstration Unit |
| EELV | Evolved Expendable Launch Vehicle |
| EGSE | Electrical Ground Support Equipment |
| ELV | Expendable Launch Vehicle |
| EO-1 | Earth Observing-1 |
| EOB | Extendable Optical Bench |
| EOL | End of Life |
| EOS | Earth Observing System |
| EoS | Equation of State |
| EPIC | European Photon Imaging Camera |
| EPO | Education and Public Outreach |
| EPS | Electrical and Power Subsystem |
| ESA | European Space Agency |

| | | | |
|---|---|---|---|
| ESLOC | Equivalent Software Lines Of Code | HXMM | Hard X-ray Mirror Module |
| ETU | Engineering Test Unit | HXT | Hard X-ray Telescope |
| EURECA | European Retrievable Carrier | HZ | Hydrazine |
| FCB | Feedback/Controller Box | I&T | Integration and Test |
| FEDS | Front End Data System | I/F | Interface |
| FEM | Finite Element Modeling | I/O | Input/Output |
| FIP | Fixed Instrument Platform | IASF | Instituto di Astrofisica Spaziale e Fisica Cosmica |
| FITS | Flexible Image Transport System | | |
| FMA | Flight Mirror Assembly | ICE | Independent Cost Estimate |
| FMS | Fixed Metering Structure | IDL | Instrument Design Lab |
| FOV | Field of View | IGM | Intergalactic Medium |
| FPGA | Field Programmable Gate Array | IIRT | Integrated Independent Review Team |
| FRR | Flight Readiness Review | IM | Instrument Manager |
| FSW | Flight Software | IM | Instrument Module |
| FTE | Full Time Equivalent | IMAGE | Imager for Magnetopause-to-Aurora Global Imaging |
| FTOOLS | FITS Tools | | |
| FWC | Filter Wheel Control Electronics | INAF | Istituto Nazionale di Astrofisica Nucleare |
| FWHM | Full Width Half Maximum | | |
| FY | Fiscal Year | IP | Internet Protocol |
| GAS | Grating Assembly Structure | IR | Infrared |
| GEMS | Gravity and Extreme Magnetism SMEX | ISAS | Institute of Space and Astronautical Sciences |
| GMSEC | GSFC Mission Services Evolution Center | ISOC | IXO Science and Operations Center |
| | | ISS | International Space Station |
| GN&C | Guidance, Navigation & Control | ITAR | International Traffic in Arms Regulations |
| GNC | Guidance, Navigation & Control | | |
| GOES | Geostationary Operational Environmental Satellites | ITOS | Integrated Test and Operations System |
| GOES-R | Geostationary Operational Environmental Satellite-R Series | IXO | International X-ray Observatory |
| | | JAXA | Japan Aerospace Exploration Agency |
| GOLD | Goddard Open Learning Design | JPL | Jet Propulsion Laboratory |
| GOTS | Government Off-The-Shelf | JT | Joule-Thomson |
| GPD | Graphical Pilot Display | JWST | James Webb Space Telescope |
| GPM | Global Precipitation Measurement mission | KDP | Key Decision Point |
| | | KSC | Kennedy Space Center |
| GR | General Relativity | LL | Lincoln Laboratory |
| GS | Ground System | LLNL | Lawrence Livermore National Laboratory |
| GSE | Ground Support Equipment | | |
| GSFC | Goddard Space Flight Center | LM | Lockheed-Martin |
| HEASARC | High Energy Astrophysics Science Archive Research Center | LMXB | Low Mass X-ray Binary |
| | | LRD | Launch Readiness Date |
| HEFT | High Energy Focusing Telescope | LRO | Lunar Reconnaissance Orbiter |
| HESSI | High Energy Solar Spectroscopic Imager | LV | Launch Vehicle |
| | | LVPC | Low Voltage Power Converter |
| HETG | High Energy Transmission Grating | MAR | Mission Assurance Requirements |
| HGA | High Gain Antenna | MCR | Mission Confirmation Review |
| HK | Housekeeping data | MDL | Mission Design Lab |
| HPD | Half Power Diameter | MDP | Minimum Detectable Polarization |
| HQ | Headquarters | MDR | Mission Definition Review |
| HST | Hubble Space Telescope | MDS | Mission Data System |
| HTRS | High Time Resolution Spectrometer | MEL | Master Equipment List |
| HUDF | Hubble Ultra-Deep Field | MEOP | Maximum Expected Operating Pressure |
| HV | High Voltage | | |
| HXI | Hard X-ray Imager | | |

| | |
|---|---|
| MESSENGER | MErcury Surface, Space ENvironment, GEochemistry, and Ranging |
| MIP | Movable Instrument Platform |
| MIRI | Mid-InfraRed Instrument |
| MIXS | Mercury Imaging X-ray Spectrometer |
| MLI | Multi-Layer Insulation |
| MM | Mass Memory |
| MMS | Magnetosphere Multi-Scale satellite constellation |
| MO | Mission Operations |
| MO&DA | Mission Operations and Data Analysis |
| MOCM | Mission Operations Cost Model |
| MODA | Mission Operations & Data Analysis |
| MOR | Missions Operation Review |
| MPE | Max-Planck-Institut für Extraterrestrische Physik |
| MS | Mission Simulation |
| MSFC | Marshall Space Flight Center |
| N/A | Not Available |
| NAR | Non-Advocate Review |
| NASA | National Aeronautics and Space Administration |
| NEAR | Near Earth Asteroid Rendezvous |
| NGST | Next Generation Space Telescope |
| NIST | National Institute of Standards and Technology |
| NPOESS | National Polar-orbiting Operational Environmental Satellite System |
| NRC | National Research Council |
| NS | Neutron Star |
| NTO | Nitrogen tetroxide (or dinitrogen tetroxide), rocket fuel |
| OB | Optical Bench |
| OGS | Objective Grating Spectrometer |
| OM | Observatory Manager |
| OM | Optics Module |
| OP | Optical path |
| ORR | Operations Readiness Review |
| P | parabolic |
| PA | Power Amplifier |
| PA&E | Program Analysis and Evaluation |
| PBB | Pre-Amplifier Bias Box |
| PDD | Payload Definition Document |
| PDR | Preliminary Design Review |
| PDU | Power Distribution Unit |
| PER | Pre-Environmental Review |
| PLF | Payload Fairing |
| PM | Project Manager |
| PPE | Pulse Processing Electronics |
| PRT | Positive Resistance Thermistors |
| PRICE-H | Parametric Review of Information for Costing and Evaluation Hardware |
| PSE | Power Supply Electronics |
| PSF | Point Spread Function |
| PSR | Pre-Shipment Review |
| QA | Quality Assurance |
| QE | Quantum Efficiency |
| QSO | Quasi - stellar Objects |
| RASS | ROSAT All-Sky Survey |
| RF | Radio Frequency |
| RFI | Request for Information |
| RFP | Request for Proposals |
| RGS | Reflection Grating Spectrometer |
| RIU | Remote Interface Units |
| ROM | Rough Order of Magnitude |
| ROSAT | Röntgensatellit - a German X-ray satellite telescope |
| RVDT | Rotary Variable Differential Transformer |
| RXTE | Rossi X-ray Timing Explorer |
| RY | Real Year |
| S/A | Solar Array |
| S/C | Spacecraft |
| S/N | Signal to Noise |
| SAO | Smithsonian Astrophysical Observatory |
| SAX | Satellite per Astronomia X-ray |
| SBIL | Scanning Beam Interference |
| SCG | Study Coordination Group |
| SDD | Silicon Drift Diodes |
| SDO | Solar Dynamics Observatory |
| SDS | Science Data System |
| SDT | Science Definition Team |
| SE | Systems Engineer |
| SEER-H | Software Estimation and Evaluation of Resources for Hardware |
| SIRU | Scalable Inertial Reference Unit |
| SIRu | Spacecraft Intertial Reference unit |
| SLOC | Software Line of Code |
| SM | Spacecraft Module |
| SMEX | Small Explorer mission |
| S&MA | Safety and Mission Assurance |
| SMBH | Super Massive Black Hole |
| SMEX | Small Explorer |
| SMILES | Superconducting Submilimeter-Wave Limb-Emission Sounder |
| SNR | Signal-to-Noise Ratio |
| SOC | Science Operations Center |
| SOHO | Solar and Heliospheric Observatory |
| SPO | Silicon Pore Optic |
| SQUID | Superconducting Quantum Interference Device |
| SR&T | Supporting Research and Technology |
| SRR | Systems Requirements Review |
| SRTMM | Shuttle Radar Topography Mapper |
| ST | Star Tracker |
| STEREO | Solar Terrestrial Relations Observatory |
| SVOM | Space multi-band Variable Object Monitor |
| SWG | Science Working Group |
| SXS | Soft X-ray Spectrometer |
| SXT | Spectroscopy X-ray Telescope |
| TADS | Telescope Aspect Determination System |
| TB | Thermal Balance |
| TBD | To Be Determined |
| TBR | To Be Revised |
| TCP/IP | Transmission Control Protocol/ Internet Protocol |
| TDM | Time-Division-Multiplexing |
| TDRSS | Tracking and Data Relay Satellite System |
| TES | Transition Edge Sensor |
| TM | Telescope Module |
| TMCO | Technical, Management, Cost and Other |
| TOPEX | Topography Experiment for Ocean Circulation |
| TRACE | Transition Region and Coronal Explorer |
| TRL | Technology Readiness Level |
| TRMM | Tropical Rainfall Monitoring Mission |
| TT&C | Tracking, Telemetry and Command |
| TV | Thermal Vacuum |
| TWINS | Two Wide- Angle Imaging Neutral-Atom Spectrometers mission |
| TWG | Telescope Working Group |
| UV | Ultraviolet |
| V&V | Verification and Validation |
| VELA | VLSI ELectronic for Astronomy |
| VCHP | Variable Conductance Heat Pipe |
| WBS | Work Breakdown Structure |
| WFI | Wide Field Imager |
| WHIM | Warm-Hot Intergalactic Medium |
| WMAP | Wilkinson Microwave Anisotropy Probe |
| WP | White Paper |
| XEUS | X-ray Evolving Universe Spectroscopy |
| XGS | X-ray Grating Spectrometer |
| XIS | X-ray Imaging Spectrometer |
| XMM | X-ray Multi-Mirror Mission |
| XMS | X-ray Microcalorimeter Spectrometer |
| XPOL | X-ray Polarimeter |
| XQC | X-ray Quantum Calorimeter |
| XRB | X-ray Binary |
| XRCF | X-Ray Calibration Facility |
| XRS | X-ray Spectrometer |
| XRT | X-ray Telescope |
| YTD | year-to-date |

## Appendix C. References

### Science References

### Technical References

Appendix C References

# Appendix D. Supplementary Document Description

## D.1 Overview of the Supplemental Documents

As a supplement to the IXO response to Astro2010 RFI#2, we have provided the following documents at **http://ixo.gsfc.nasa.gov/RFI2/Supplemental/**.

The IXO Systems Definition Document (D.2) is the controlling document, followed by the other supplemental documents in order of presentation below.

## D.2 IXO Systems Definition Document

The IXO Systems Definition Document describes NASA's engineering concept of the IXO mission. This document includes the baseline configuration, main functions, key performance metrics (including pointing error budgets and resource budgets), launch and flight dynamics parameters, and the operations concept. An overview of all subsystems is included. The systems engineering process, requirements flowdown, integrated modeling, mission operations and integration and test are also documented.

## D.3 IXO Segmented Glass FMA Concept Study

This document describes the Flight Mirror Assembly (FMA) Concept Study, which derived the FMA requirements from the IXO mission level requirements. This study consisted of the development of a reference/preliminary design of the FMA that met all requirements, including mass and power allocations, optical design, launch environment, and structural and thermal distortion. Detailed I&T plans and schedules are provided. This study defined the technical areas that require development to meet the angular resolution requirement.

## D.4 NASA XMS Reference Concept

This document is a technical overview of the XMS reference design that was used to determine the XMS cost, interfaces, mass, and power estimates that are provided in the response to RFI#2. This is a streamlined complement to the XMS description in the ESA IXO PDD, which covers the ESA XMS reference concept in detail, in addition to introducing a variety of alternative approaches for the various subsystems.

## D.5 ESA Payload Definition Document (PDD) with Corrigendum

The ESA PDD compiles the IXO instrument requirements and their related reference designs. The PDD contains a description of the IXO instruments and a summary of instrument accommodation and interfaces. It plays a key role in defining the technical resources required by the IXO instruments and provides the information necessary to conduct the ESA mission assessment study and the ESA IXO spacecraft design. The ESA PDD therefore is not fully commensurate with the NASA IXO Concept.

The PDD corrigendum notes specific instances where the ESA and NASA concepts differ and provides updates to instrument performance and interface parameters that occurred since the PDD release date of April 23, 2009.

## D.6 Technology Development Plans

These Technology Development Plans provide, for both optics technologies and each IXO instrument, a description including performance requirements, current technology status, and a technology development plan with a detailed schedule.

### D.6.1.  Mirror Technology Development Roadmap for the International X-ray Observatory [Segmented Glass]

This document outlines the work being done to develop the glass mirror technology to TRL 6 by 2012. It specifies requirements, summarizes current status, identifies problems, and details their solutions. In particular, technical risks and the strategies adopted to mitigate them are identified.

### D.6.2.  IXO Silicon Pore Mirror Technology Development Plan

This document outlines the work being done to develop the glass mirror technology to TRL 6 by 2012. It specifies requirements, summarizes current status, identifies problems, and details their solutions. In particular, technical risks and the strategies adopted to mitigate them are identified.

### D.6.3.  Technology Roadmap for the X-ray Microcalorimeter Spectrometer of the IXO

The technology development and demonstration needed for the NASA XMS reference concept. This describes how the 50 mK cooler (the CADR) and the detector system will be advanced to TRL 6. This document does not cover the technology development needed for the ESA XMS reference concept.

### D.6.4.  WFI/HXI Technology Development Roadmaps

The WFI/HXI instrument consists of two separate detectors. A development roadmap is provided for each detector.

#### D.6.4.1.  WFI Technology Development Roadmap

Specific tasks listed include component development, performance characterization, thermal modeling and environmental testing of the detector, front-end electronics, and data acquisition systems.

#### D.6.4.2.  IXO-HXI Technology Development Roadmap

Specific tasks listed include component development, performance characterization, thermal modeling and environmental testing of the detector, front-end electronics, and data acquisition systems.

### D.6.5.  The Off-Plane X-ray Grating Spectrometer Technology Development Roadmap

Tasks described include laboratory testing of an existing prototype grating, environmental and X-ray testing of a flight-like grating mounted in a flight-like module, and the fabrication, performance testing, and environmental testing of a high-fidelity partial grating array.

### D.6.6.  Critical Angle Transmission X-ray Grating Spectrometer Technology Development Plan

Tasks described include component fabrication, performance testing followed by environmental and X-ray testing of the gratings and readout system.

### D.6.7.  XPOL Technology Development Roadmap

This document summarizes the instrument concept and performance requirements, and describes the approach for the high count rate ASIC that will bring XPOL to TRL 6. Development activities that will be undertaken to improve the XPOL beyond its minimum requirements are also outlined.

# Appendix E. WBS Dictionary

**IXO WBS Dictionary and Mapping to RFI#2 Cost Elements**

**Summary Work Breakdown Structure (WBS) Dictionary**

1.0  **Project Management**:  This includes the management, business and administrative planning, organizing, directing, coordinating, controlling, and approval processes used to accomplish overall project objectives not associated with specific hardware or software elements.  This element includes project reviews and documentation, non-project owned facilities, and project reserves. For Phase E, Project Management for Operations, including NASA Project Office effort, is carried under WBS 4.6.

2.0  **Systems Engineering**:  This includes the technical management of directing and controlling an integrated systems engineering effort for the mission. This element includes the efforts to define flight and ground system, conducting trade studies, the integrated planning and control of the technical program efforts of design engineering, software engineering, reliability engineering, system architecture development and integrated test planning, system requirements definition, configuration control, technical oversight, control and monitoring of the technical program, and risk management activities.

3.0  **Safety and Mission Assurance:** This includes the efforts of directing and controlling the safety and mission assurance elements of the project, including verification of practices and procedures. This element includes safety and mission assurance management, reliability analysis, quality assurance, safety, materials assurance, and electronic parts control.

4.0  **Science & Technology:**  This includes managing and directing the science investigation aspects, as well as leading, managing, and performing the technology demonstration elements of the project. Sub-elements 4.1 – 4.3 are FMA, XMS, WFI/HXI, XGS, and XPOL technology development (through TRL 6). Other sub-elements cover pre-launch science support (including calibration) and management, post-launch science support and management, and the science grants program.

5.0  **Payload:** This includes the following for the FMA, science instrument, and Instrument Module (IM):  management, system engineering, design, and development. Includes integration and test of the FMA to the OM, and instruments to the IM.  Integration and test of the FMA to the OM is included. Includes integration and test of the instruments to the IM. The completed IM and OM are delivered to the Observatory I&T (in WBS 10.0).

6.0  **Spacecraft:**  This includes management, system engineering, design and development of the Spacecraft Module (SM), Deployment Module (DM), and Optical Module (OM).  The OM is delivered to the FMA for I&T (WBS 5.0). This WBS includes I&T of each of the SM and DM prior their delivery to observatory I&T in WBS 10.0.

7.0  **Mission Operations:** This includes the management, development, and implementation of procedures, documentation, and training required to conduct mission operations. This element includes all aspects of flight operations (commanding, receiving/processing telemetry, analyses of system status, flight dynamics, etc.); mission planning, sustaining engineering for the hardware and software for the spacecraft and payload elements. DSN, TDRSS, and related services are included.

8.0  **Launch Vehicle/Services:**  Includes launch vehicle and efforts required to support launch campaign.

9.0 **Ground System:** This element includes the management, design, development, implementation, integration, and test of mission unique facilities required to conduct mission and science operations, including all ground system hardware and software required for processing, archiving, and distributing telemetry and science data. Ground system maintenance is included. DSN, TDRSS, and related services are included in WBS 7.0.

10.0 **System Integration and Testing:** Includes integration and testing of IM, DM&SM, and OM at the observatory level. Environmental testing and performance verification at the observatory level. Includes shipping the observatory to the launch site and support of observatory activities at the launch site. Includes sustaining science and engineering support for instruments and observatory subsystems through on-orbit check-out (which are carried in WBS 4.0 and 7.0 for Phase E.)

11.0 **Education and Public Outreach:** Provides for the education and public outreach (EPO) responsibilities in alignment with the Strategic Plan for Education. Includes management and coordinated activities, formal education, informal education, public outreach, media support, and website development.

**IXO Work Breakdown Structure (WBS) Crosswalk to RFI#2-Specified Cost Elements**
**RFI-2 Required WBS Structure (Tables 6-5 & 6-6 of IXO RFI#2 Response)**

| RFI#2-Specified Cost Elements I | Mapping to IXO Project WBS Element |
|---|---|
| Concept Studies | All Pre-Phase A and Phase A except 4.1- 4..3 |
| Technology Development | WBS 4.1 -  4.3 Technology Development |
| PM/SE/MA | Portion of WBS 1.0, 2.0 and 3.0 for Phases B - D |
| Instrument PM/SE | Portion of WBS 1.0, 2.0 and 3.0 for Phases B - D |
| FMA | WBS 5.1 FMA (Phases B-D) |
| XMS | WBS 5.2 XMS (Phases B-D) |
| WFI/HXI | WBS 5.3 WFI/HXI (Phases B-D) |
| XGS | WBS 5.4 XGS (Phases B-D) |
| XPOL | WBS 5.5 XPOL (Phases B-D) |
| HTRS | WBS 5.6 HTRS (Phases B-D) |
| Spacecraft incl. MSI&T | WBS 5.7, 6.0, 7.0, 10.0 (all Phases B – D) |
| Pre-launch Science | WBS 4.4 – 4.5 Pre-Launch Science (Phases B – D) |
| Ground System Dev | WBS 9.0 (Phases B – D) |
| Launch Services | WBS 8.0 |
| MODA | WBS 4.5 – 4.9, , 9.0 and 7.0 (all for Phase E) |
| Education/Outreach | WBS 11.0 |
| Reserves | Reserves* |

\* Reserves estimate developed by WBS based on point estimate versus 70% CL estimate.

# Appendix F.  Flight Mirror Assembly (FMA) Grassroots Estimate and Schedule

## F.1   FMA Grassroots Cost Estimate Overview

### Appendix F.1
### Flight Mirror Assembly (FMA) Grassroots Cost Estimate Overview

The results of the FMA grassroots estimate are provided in the Appendix G (restricted data). These include the summary of the FMA cost by WBS as well as summary by year.  For each sub-element of the WBS, tables are provided that include the task description, estimating rationale, assumptions. In addition, labor hours are summarized by year, as are procurements, materials, and other associated costs.   The FMA cost estimate was based on the initial FMA design and production concept as defined in the *IXO Segmented Glass FMA Concept Study* (referenced in Appendix D.3),

The grassroots estimate was generated by a team of experienced managers, scientists and engineers from GSFC and SAO with extensive relevant expertise including with NuSTAR, IXO, SDO, and Chandra. The estimate includes all efforts and other costs required to fabricate, integrate and test, verify performance, qualify, calibrate and deliver the FMA. The costs are phased by WBS element consistent with the FMA development schedule provided in Appendix F.2. The estimate includes the effort for Phase B – D, assuming the FMA is developed by a major industry contractor; with appropriate fully loaded rates (with fee) as verified with a major aerospace optics contractor. An appropriate mix of personnel capabilities, from technicians to systems engineers and managers, has been factored in for each task. Phase A estimates for technology development and industry studies are not included here, but are included elsewhere in the RFI#2 response (Section 6). This estimate also does not include contingency, which has been allocated separately. Costs for management, systems engineering, and safety and mission assurance efforts have been estimated based on projected labor for sub-elements informed by experience on the Chandra mirror development.

GSFC's recent and ongoing experience in producing the NuSTAR mirror segments was factored into estimate the costs for the IXO FMA mirror segments.  Costs for process and other adjustments to meet IXO requirements, such as improved optics performance, additional metrology, longer slumping time, and increased scale (segment size and number) have been factored into the estimate. Schedule factors have been accounted for: the facilities for mirror segment production require 44 ovens, 10 metrology set-ups, and 4 cutting stations with associated labor support. A conservative assumption was made that four mandrels are processed per oven (at a time), and one slumping technician operates four ovens.  Mirror segment production yield is accounted for, as is an additional 10% for spare segments. The mandrel costs were based on actual costs of industry mandrel procurement, extrapolated for the FMA based on the required polished surface area of 60 sq-meters (including 20% contingency), and adjusted for inflation and improved figure requirements (including the additional metrology required). Cost for mandrel preparation and treatment for mirror segment slumping has also been included.

The cost estimates for the FMA primary and module structures were made based on experience with similar structures such as that for the recently developed GSFC's Solar Dynamics Observatory (SDO).  Both the primary and module structure cost include building and testing Engineering Test Units and qualification units.  Alignment and mounting of the mirror segments into modules assumes 24 optical alignment stations operating in parallel as described in the *IXO Segmented Glass FMA Concept Study*.  Fully-integrated modules will be vibrated (four at a time) and X-ray tested (two at a time) to verify performance.   The cost of the Hard X-ray Mirror Module was estimated based on NuSTAR actual costs, factoring in improve angular resolution and the slightly larger size.  Costs for qualification, testing and calibration of HXMM to verify performance are included.

Costs are included for FMA I&T the effort to integrate the modules into the FMA structure and perform complete system testing and verification. Optical testing is assumed for each module as it is installed into the FMA and aligned, followed by X-ray tests to calibrate the effective area and provide an initial assessment of the point spread function. Pre- and post-environmental X-ray testing will then be performed to validate the fully-integrated FMA, and provide final calibration and verification data.

# E2  FMA Schedules

## Top-Level FMA Schedule

| Task Name | 2013 | 2014 | 2015 | 2016 | 2017 | 2018 | 2019 |
|---|---|---|---|---|---|---|---|

**FMA  Major Milestone** — FMA Award 09/13, PDR 09/14, CDR 09/15, Deliver to Obs. 08/19

**Mandrel and Mirror Segment Production (WBS 5.1.4, 5.1.5)**
- 09/13
- Facilitization 09/14
- 09/14
- Mandrel Production 02/17
- 09/14
- Mirror Segment Production 08/17

**Structures Design, Fabrication, Assembly, and Test (WBS 5.1.6)**
- 09/13
- Structure Design (Primary and Module Structures) 09/15
- Module Fabrication and Assembly 09/15 – 05/17
- Primary Structure Fabrication, Assembly, and Test 09/15 – 10/16
- Module Qualification Unit Testing 07/16 – 02/17
- Primary Structure Qualification Unit Testing 02/17 – 07/17

**Flight Module Integration and Test (WBS 5.1.8)**
- Inner Module Segment Integration 02/16 – 09/16
- Middle Module Segment Integration 06/16 – 03/17
- Outer Module Segment Integration 02/17 – 10/17
- Integrated Module Testing 09/16 – 05/18

**FMA/Optics Module Integration and Test (WBS 5.1.9)**
- 09/15
- Facilitization, Chamber Hardware Design and Build
- Module Integration into FMA w/ Optical and Pencil Beam Testing
- Inner Modules 11/16
- HXMM 02/17
- Middle Modules 11/16
- Outer Modules 06/18
- Integrated FMA Pencil Beam Testing / Calibration 11/16 – 06/18
- Optics Module Acoustic Testing / Thermal Testing / Post-Test Pencil Beam Testing / Calibration 06/18 – 10/18
- Install FMA Covers, TADS - Optics Module Configuration 10/18
- Schedule Reserve 05/19 – 08/19

# Appendix F.2: FMA Schedule (cont.)

## FMA Mandrel and Mirror Segment Production Schedule (WBS 5.1.4 & WBS 5.1.5)

Revision - 6

| Task Name | 2013 | 2014 | 2015 | 2016 | 2017 | 2018 |
|-----------|------|------|------|------|------|------|

**FMA Major Milestone**

- FMA Award ▼ 09/13
- PDR ▼ 09/14
- CDR ▼ 09/15

**Mandrel Production (WBS 5.1.4)**

- 09/13 — Mandrel Fabrication Facilitization — 09/14
- 09/14 — Inner Mandrel Production — 09/15
- 09/15 — Middle Mandrel Production — 06/16
- 06/16 — Outer Mandrel Production — 02/17

**Mirror Segment Production (WBS 5.1.5)**

- 09/13 — Mirror Fabrication Facilitization — 09/14
- 09/14 — Inner Mandrel Treatment & Segment Production — 01/16
- Middle Mandrel Treatment & Segment Production 09/15 — 12/16
- Outer Mandrel Treatment & Segment Production 06/16 — 08/17

# Appendix F.2: FMA Schedule (cont.)

## FMA Structures Schedule (WBS 5.1.6)

| Task Name | 2013 | 2014 | 2015 | 2016 | 2017 | 2018 |
|---|---|---|---|---|---|---|
| **FMA Major Milestone** | FMA Award ▼ 09/13 | PDR ▼ 09/14 | CDR ▼ 09/15 | | | |
| **Design and Development** | 09/13 | 04/14 / 09/14 | 09/15 / 09/15 / 09/15 | 02/16 / 02/16 | | |
| | | | FMA Structure and Module Design | Module ETU | | |
| | | | | FMA Structure ETU | | |
| **FMA Structure and Module Fabrication and Asssembly** | | | 09/15 | Inner Module Fab & Assy 02/16 / 10/16 | Middle Module Fab & Assy 10/16 / Outer Module Fab & Assy 05/17 | |
| | | | 09/15 | FMA Flight Structure 09/16 / FMA Structure Acceptance 09/16 10/16 | | |
| **Qualification Unit Design and Test** | | | Inner and Middle Module Qual Unit Testing | 07/16 / 02/17 | Outer Module Qual Unit Testing 09/17 / FMA Qual Unit Fabrication 12/17 | |
| | | | | FMA Qual Unit Test 09/16 / 02/17 | 02/17 / 07/17 | |

# Appendix F.2: FMA Schedule (cont.)

FMA Flight Module Integration and Test Schedule (WBS 5.1.8)

Revision - 6

| Task Name | 2015 | 2016 | 2017 | 2018 | 2019 | 2020 |
|-----------|------|------|------|------|------|------|
| FMA Major Milestone | CDR ▼ 09/15 | | | | Deliver to Obs. ▼ 08/19 | |
| Module I&T Facilitization | ▽ 09/15 | ▽ 02/16 Design and Build 24 Alignment Stations and Prepare Facility | | | | |
| MSFC X-Ray Facility Modifications | | ▽ 03/16 Modify the XRCF and Prepare for Module X-Ray Testing ▽ 08/16 | | | | |
| Inner Module Integration and Test | | ▽ 02/16 Inner Module Segment Integration ▽ 09/16 09/16 Inner Module Testing (Optical, Vibe, X-ray) 12/16 | | | | |
| Middle Module Integration and Test | | ▽ 06/16 Middle Module Segment Integration ▽ 03/17 12/16 Middle Module Testing (Optical, Vibe, X-ray) ▽ 09/17 | | | | |
| Outer Module Integration and Test | | ▽ 02/17 Outer Module Segment Integration ▽ 10/17 09/17 End of Vibe Tests 01/18 Outer Module Testing (Optical, Vibe, X-ray) ▽ 05/18 | | | | |

# Appendix F.2: FMA Schedule (cont.)

Flight FMA/Optics Module Integration and Test Schedule (WBS 5.1.9)

Revision - 7

| Task Name | 2015 | 2016 | 2017 | 2018 | 2019 | 2020 |
|---|---|---|---|---|---|---|
| **FMA Major Milestone** | CDR 09/15 | | | | Deliver to Obs. 08/19 | |
| FMA I&T Facilitization | 09/15 | 11/16 Prepare Facility and Test Equipment for I&T | | | | |
| Inner Module Integration | | 11/16 Mate FMA w/ S/C Adapter and Separation Ring | 02/17 HXMM Integration / Inner Module Integration into FMA w/ Optical and Pencil Beam Testing | | | |
| Middle Module Integration | | | 03/17 10/17 Middle Module Integration into FMA w/ Optical and Pencil Beam Testing | | | |
| Outer Module Integration | | | Outer Module Integration into FMA w/ Optical and Pencil Beam Testing 11/17 06/18 | | | |
| Integrated FMA/Optics Module Test and Calibration | | | Integrated FMA Pencil Beam Testing / Calibration 06/18 10/18 Optics Module Acoustic Test/ Pencil Beam Testing 10/18 12/18 Optics Module Thermal Test/ Pencil Beam Testing/Calibration 12/18 | Install Forward and Aft FMA Covers, TADS, Thermal Control System - Optics Module Configuration 05/19 Install Sunshade, Separation Mechanism Schedule Reserve 05/19 08/19 | | |

# Appendix F.2: FMA Schedule (cont.)

## FMA Hard X-Ray Mirror Module (HXMM) Integration and Test Schedule (WBS 5.1.10)

Revision - 6

| Task Name | 2013 | 2014 | 2015 | 2016 | 2017 |
|---|---|---|---|---|---|
| **FMA Major Milestone** | FMA Award ▼ 09/13 | PDR ▼ 09/14 | CDR ▼ 09/15 | | Deliver to FMA ▼ 02/17 |
| **HXMM Design and Development** | ▽ 09/13 | | HXMM Design ▽ 09/15 / HXMM ETU ▽ 09/15 / 08/14 | | |
| **HXMM Mandrel Production** | ▽ 09/13 | HXMM Mandrel Fabrication Facilitization ▽ 05/14 / ▽ 05/14 | HXMM Mandrel Production ▽ 05/15 | | |
| **HXMM Mirror Segment Production** | ▽ 09/13 | HXMM Mirror Fabrication Facilitization ▽ 05/14 / ▽ 05/14 | | HXMM Mandrel Treatment & Segment Production ▽ 02/16 | |
| **HXMM Integration and Test** | | | HXMM Testing/Calibration (Optical, Vibe, X-ray) ▽ 09/15 | HXMM Segment Integration ▽ 09/16 / ▽ 09/16 | 02/17 |